\documentclass{aa}  
\usepackage{graphicx}
\usepackage{caption}
\usepackage{subcaption}
\usepackage{booktabs}
\usepackage{tablefootnote}
\usepackage{footnote}
\usepackage{txfonts}
\usepackage[colorlinks=true, linkcolor=black, citecolor=blue]{hyperref}
\usepackage{ulem}
\usepackage{color}  
\usepackage{amsmath}
\usepackage{amssymb}

\def\aliii{Al{\sc iii}$\lambda$1860\/}
\def\al{Al{\sc iii}}

\def\chisq{$\chi^{2}$}
\def\ciii{{\sc{Ciii]}}$\lambda$1909\/}
\def\ciiip{{[\sc{Ciii]}}$\lambda$1906\/}
\def\cnl{{\sc{Ciii]}}}
\def\civ{{\sc{Civ}}$\lambda$1549\/}

\def\civbc{{\sc{Civ}}$\lambda$1549$_{\rm BC}$\/}

\def\cm3{cm$^{-3}$\/}

\def\ergss{ergs s$^{-1}$\/}
\def\fe{{\sc{Fe}}\/}
\def\fe6087{{\sc [Fe vii]}$\lambda$6087\/}

\def\feii{{Fe\sc{ii}}\/}
\def\feUV{{Fe\sc ii~}UV191}

\def\feiii{{Fe\sc{iii}}\/}
\def\feiiiuv{Fe{\sc iii}(UV34)}

\def\gtsima{$\; \buildrel > \over \sim \;$}
\def\gtsim{\lower.5ex\hbox{\gtsima}} 
\def\Gsoft{$\Gamma_{\rm soft}$}

\def\hb{{\sc{H}}$\beta$\/}
\def\hbbc{{\sc{H}}$\beta_{\rm BC}$\/}

\def\heiiuv{He{\sc{ii}}$\lambda$1640}

\def\civonly{C{\sc iv}\/}

\def\kms{km~s$^{-1}$}

\def\lbol{L$_{\rm bol}$\/}

\def\ledd{L$_{\rm Edd}$\/}
\def\lvir{L$_{\rm Vir}$\/}

\def\ltsima{$\; \buildrel < \over \sim \;$}
\def\ltsim{\lower.5dex\hbox{\ltsima}}  
\def\lya{{ Ly}$\alpha$}

\def\mbh{M$_{\rm BH}$\/}
\def\mgii{{Mg\sc{ii}}$\lambda$2800\/}

\def\nc{$N_{\rm c}$\/}

\def\nh{$n_{\mathrm{H}}$\/}

\def\niii{{\sc{Niii]}}$\lambda$1750} 

\def\o4363{{\sc{[Oiii]}}$\lambda$4363\/}
\def\oiiD{{\sc [Oii]}$\lambda\lambda$3727,3729}
\def\oii{{\sc [Oii]}$\lambda$3728}
\def\o2{{\sc [Oii]}}

\def\oiiiopt{{\sc{[Oiii]}}$\lambda\lambda$4959,5007\/}
\def\oiiia{{\sc [Oiii]}$\lambda$5007}

\def\rblr{$r_{\rm BLR}$\/}
\def\rfe{$R_{\rm FeII}$}
\def\redd{$R_{\rm Edd}$}

\def\siii{Si{\sc ii}$\lambda$1816}

\def\siiii{Si{\sc iii]}$\lambda$1892\/}
\def\si{Si{\sc iii]}}

\def\siblue{{\sc{[Sii]}}$\lambda\lambda$6716,6731}

\defcitealias{MS14}{MS14}
\defcitealias{vp06}{VP06}
\defcitealias{Marzianietal2022}{M22}

\begin{document} 

\titlerunning{Statistics of \aliii}
\authorrunning{Buendia-Rios et al.}
   \title{Statistical analysis of \ion{Al}{{\sc iii}} and \ion{C}{\sc iii]} emission lines as virial black hole mass estimators in quasars}

   \author{T. M. Buendia-Rios\inst{1},
         C. A. Negrete\inst{2}
          \and
          P. Marziani\inst{3}\fnmsep
          \and
          D. Dultzin\inst{1}
          }
    \institute{
    Instituto de Astronomía, UNAM, México DF. 04510, Mexico
    \and
    CONACyT Research Fellow - Instituto de Astronomía, UNAM
    \and
    INAF, Osservatorio astronomico di Padova, 35122, Padova, Italy\\
            \email{tbuendia@astro.unam.mx}
             }

   \date{Received ...; accepted ...}

  \abstract 
   {}
   { We test the usefulness of the intermediate ionisation lines \aliii\ and \ciii\ as reliable virial mass estimators for quasars. 
   }
   { We identify a sample of 309 quasars from the SDSS DR16 in the redshift range 1.2$\leq$z$\leq$1.4 to have \oii\ recorded on the same spectrum of \aliii, \siiii, and \ciii.
   We set the systemic quasar redshift using careful measurements of \oii.  We then classified the sources as Population A, extreme Population A (xA) and Population B, and  analysed the 1900 \AA\ blend using multi-component models to look for systematic line shifts of the \aliii\ and \ciii \ along the quasar main sequence. 
   }
   {We do not find significant shifts of the \aliii\ line peak in Pop. B and the wide majority of Pop. A. For Pop. xA, a small median blueshift of -250 \kms\ was observed, motivating a decomposition of the \al\ line profile into a virialized component centred at rest-frame and a blueshifted component for an outflow emission. For Pop. B objects, we proved the empirical necessity to fit a redshifted very broad component (VBC), clearly seen in \cnl, and analysed the physical implications on a Pop. B  composite spectrum using CLOUDY simulations.  
    We find consistent black hole mass estimations using \al\ and \cnl\ as virial estimators for the bulk of Population A.
    }
   {\al\ (and even \cnl) is a reliable virial black hole mass estimator for Pop. A and B objects. xA sources deserve special attention due to the significant blueshifted excess observed in the line profile of \al, although not as large as those observed in \civ. 
   }
   

   \maketitle
%

\section{Introduction}\label{sec:introduction}
    
    In \citeyear{ByG}, \citeauthor{ByG} carried out a principal component analysis (PCA) on a sample of $\sim$80 Palomar-Green quasars. Their analysis    identified a first eigenvector dominated by an anticorrelation between the \oiiia\ peak intensity and the strength of optical \feii\ emission. In this first eigenvector (from now on "Eigenvector 1"; E1), two dimensions, (1) the full width at half maximum of \hb\ (FWHM(\hb)) and (2) the \feii\ emission parameterised by the ratio of the equivalent widths of the \feii\ emission at 4750\AA\ and \hb, \rfe\ = EW(FeII$\lambda$4750)/EW(\hb) define what is known as the optical plane of the main sequence of quasars \citep[MS,][]{M18}.  In this way  the MS is an empirical sequence based on optical parameters, easy to measure in single-epoch spectra. The spectroscopic trends led \citet{sulentic00a} to distinguish two populations: quasars with FWHM(\hb) < 4000 \kms\ belongs to Population A (Pop. A), while objects with FWHM(\hb) > 4000 \kms are Population B (Pop. B).
     
    The E1 gained two more dimensions over the years, becoming the 4DE1 \citep{sulentic00a,sulentic00b,sulentic07}. The 4DE1 involves optical, UV and X-ray data, and its additional dimensions are  (3) the photon index in the soft X-rays domain (below 2 keV), \Gsoft\ \citep{wbb96}, and (4) the systematic blueshift of the high-ionisation line \civ\ \citep{sulenticetal00c,sulentic07}, specially prominent among Pop. A objects. The soft X-ray excess \citep{singh85} is a dominant component of the X-ray spectra of many AGN. It was adopted as a critical parameter of the 4DE1 that correlates opposite extremes of populations A and B \citep{bensch15}. Sources with higher values of soft X-ray excess (corresponding to a   value of the soft-X photon index \Gsoft $\approx 3 - 4$) concentrate among the highly accreting Pop. A quasars \citep{grupe04,sulentic08}, while Pop. B quasars typically have \Gsoft $\approx$ 2. The most widely-accepted interpretation of the excess detected in soft X-rays is a measure of comptonized emission in a corona connected with the innermost accretion disk (\citealt{WF93, petrucci20} and references therein). The systematic high amplitude of the \civ\ blueshift of quasars with high Eddington ratio may indicate the presence of strong outflows most likely originating in a disc wind \citep{netzer15, coatman16,coatman17}. \cite{sulentic07} introduced the {line centroid velocity shift at half-maximum \citep[see][for a more detailed description]{zamfir10}} of \civonly\ as the UV E1 measurement in the 4DE1 parameter space. The observational definition of the accretion and outflow processes is the motivation behind the two additional dimensions.  In the following, we shall  consider the  \civonly\ only, as it is available for most quasars at $z \gtrsim 1.4$\ surveyed by the  Sloan Digital Sky Survey (SDSS). 
    The division between the two populations is not enough to reflect the spectral diversity \citep[e.g.][]{marziani10}.   \citet{sulentic02} made sub-divisions of $\Delta$FWHM(\hb)=4000 \kms\ and $\Delta$\rfe=0.5 to emphasise the trends in spectral properties especially seen in Pop. A sources {as a function of \rfe\ \citep[e.g.,]{duetal16,shenho14,sun15}}. This division defines the A1, A2, A3 and A4 bins as \rfe\ increases, and B1, B1+, B1++ (as well as B2, B2+ in the range of \rfe\ 0.5-1) as FWHM(\hb) increases. Spectra belonging to the same bin should have similar characteristics concerning the line profiles and optical and UV line ratios \citep{sulentic07,zamfir10}. 

    A quasar spectrum can be characterised using two physical parameters: the Eddington ratio (defined as the ratio of the bolometric and Eddington luminosities, \redd\ = \lbol/\ledd) and the black hole mass (\mbh) which can be only coarsely estimated employing the MS of quasars \citep{P19}. This is why it is necessary to accurately obtain \mbh\ either with the reverberation mapping technique \citep{NP97, panda19, dallabontaetal20} or with methods  based on single-epoch spectra \citep{shen13}. The \mbh\ relates the evolutionary stage of the quasar with the accretion process \citep{small92,dimatteo03,fraix-burnetetal17}. The knowledge of the \mbh\ allows us to assess the strength of the gravitational forces  and gain inferences on the dynamics of the region surrounding the black hole \citep[e.g., ][]{ferlandetal09,marconietal09,NM10}. The definition of the virial mass as used in this paper is: 
    
    \begin{equation}
        M_{BH} = f \frac{r \delta v^2_r}{G}
        \label{eq:mbh}
    \end{equation}
    
    \noindent where $r$ is the distance of the line emitting gas from the central black hole, $\delta v_{r}$ is the line broadening due to virial motions, and $G$ is the gravitational constant. {$f$ is the virial factor dependent on the geometry, orientation and kinematics of the emitting region \citep[e.g., ][]{petersonetal93,Liuetal2017,mejia-restrepoetal18a}.} {The parameter $f$ is intended to take into account phenomena that affects the measure of the line broadening (usually FWHM or velocity dispersion) and that may include radiation-pressure effects \citep{NM10,Liuetal2017}, as well non-virial kinematical components due to outflow or inflow.}  
    All methods using optical and UV broad lines are based on estimating the distance of the broad line region (BLR) from the central continuum source, \rblr. At low redshift (z $\leq$ 0.8), one can estimate \mbh\ using FWHM(\hb) as the $\delta v_r$, but as further in redshift we go, the less practical this measure becomes. So few options are left: (1) following the \hb\ line into the infrared, a feat requiring large telescopes and IR spectrometers or (2) adopting other broad lines (e.g. \mgii, \aliii\ or \civ) as surrogate virial estimators if $z > 1$. 
    
    The use of \civ\ emission line in the UV as a virial mass estimator is problematic due to the strong blueshift with respect to the rest frame frequently observed \citep[][and references therein]{richards11}, associating the high ionisation line (HIL $\geq$ 40 eV) to outflowing winds \citep{gaskell82,marziani96}. This outflow indicates that the emitting gas is not in virial equilibrium and therefore is not a reliable mass indicator due to the systematic biases. Other HILs such as \oiiiopt\ could also present blueshifts arising from an outflowing gas \citep{zamanov02}, possibly associated with a disc wind. 
    Hence, the narrow-line region (NLR) in sources showing spectroscopic blue shifts is not likely to be dynamically related to the gravitational potential of the host galaxy. To avoid systematic shifts in the velocity dispersion estimates (as well as in the proper rest frame), the width of \siblue\ or the doublet \oiiD\ (hereafter \oii) has been proposed as an alternative to \oiiiopt\ \citep{komossa07,salviander07}. Narrow low ionisation lines (LILs) serve as better rest frame references, as they provide a value closest to the one of the host galaxy \citep{ms12,bon20}.  

    To avoid the HILs in the UV spectral range we propose to work with the LILs or the intermediate ionisation lines (IILs $\sim $ 20-40 eV).     A possible strategy is to use high signal-to-noise ratio (S/N) of \hb\ as a template to model the IILs \aliii\ and \siiii\ (among others).  
    {Velocity-resolved reverberation mapping \citep{bentz09,denney10,feng21} detect inflow and outflow motions, although the effects seem to be relatively minor as far as the width of the LILs is concerned. We emphasise the importance of finding a reliable virial mass estimator in the UV range that is paired to LIL H$\beta$\ and we propose the surrogate lines \al\ and \cnl\ present in the 1900\AA\ blend.}
    The blend includes the IILs: \aliii, \siiii\ and \ciii. \aliii\ is a resonant doublet ($^2P^o_{3/2,1/2} \rightarrow ^2S_{1/2}$) in the sodium isoelectronic configuration, while \siiii\ and \ciii\ are due to the inter-combination transitions ($^2P^o_{1} \rightarrow   ^1S_0$) with widely different critical densities ($\approx 2 \times 10^{12}$ cm$^{-3}$ and $3.2 \times 10^9$ cm$^{-3}$, respectively; \citealt{negrete12,m20}). The parent ionic species imply ionisation potentials intermediate between LILs and HILs. 
    
    The usefulness of \si\ and \al\ as virial estimators is suggested by their profiles showing  consistent width and shape with the one of \hbbc\   \citep{marziani10,Marzianietal2022}.  They are symmetric and are usually not affected by strong outflows often observed in the \civ\ profile \citep{marziani17,mla18}. 
    The rest-frame of the 1900\AA\ blend based on the quasar redshift derived from the \oii\ line \citep{bon20} would prove the effectiveness of \al\ and \cnl\ as virial broadening estimators in the absence of systematic blueshifts.

    The main objective of the present work is (1) { to test the consistency of the \al\ and \cnl\ emission lines with the systemic redshift derived from the \o2\ line and in the case of a systematic shift is found, to propose a correction}; and (2) to probe the usefulness of the IILs \al\ and \cnl\ as reliable virial mass estimators. The outline of the paper is as follows. Section \ref{samp_descrip} is a description of the sample and  employed selection criteria. Section \ref{data_an} describes the analysis of the redshift estimation as well as the multi-component fitting of the 1900\AA\ blend and \oii\ region. In Section \ref{results} we present our results for the entire sample and by populations, analysing the trends and correlations obtained along with the physical parameters. Section \ref{sec:discussion} discusses the virial \mbh\ obtained from \al\ and \cnl,  virial luminosity estimates for the extreme Population A sub-sample, and the inter-comparison between \civonly\ and \al, along with a schematic interpretation of the $\lambda 1900$\ blend in Population B.   Section \ref{conclusions} provides the summary and conclusions. 

\section{Sample description}\label{samp_descrip}

    The spectroscopic quasar sample used in this work was selected from the  SDSS data release 16 (DR16, \citealt{Lykeetal2020}), limited with the following filters: (1) 1.2 < z < 1.4 to cover the 1900\AA\ blend and the \oii\ doublet line simultaneously; and (2) high S/N> 20 (measured in the continuum range around 1700\AA) to be able to decompose the blend at 1900\AA. These criteria give us a sample of 1379 objects. 
    
    Not all spectra have a visible \o2\ emission. To determine the visibility of \o2\ in the selected objects, we use a third criterion applied to all spectra normalised by a continuum region around the line: (3) the  ratio f$_{[\mathrm{OII}]}$ defined as the intensity ratio of \o2\ in the range 3722-3734\AA\ ($F([\mathrm{OII}]$), and the observed  continuum over the range 3670-3700\AA\ ($F(cont)$, composed of  the AGN continuum and   strong \feii\ emission contaminating the region),  f$_{[\mathrm{OII}]}=[F([\mathrm{OII}])+F(cont)]/F(cont)$, with the constriction of f$_{[\mathrm{OII}]}>1.3$. Only 309 spectra satisfied this last condition. 
    
    \oii\ is a relatively weak line that is also affected by the emission of the \feii\ multiplet (e.g. \citealt{vanden01}. The fact that \oii\ was detected in only 22\% of the objects in the initial sample (with S/N $>$ 20) may be due to two reasons. The first one is that it may be severely contaminated by sky subtraction residuals, whose emission is strong at the red end of the spectra. The second reason is attributed to the intrinsic \oii\ emission. We already know the trends of the MS in regards to oxygen, in \citet[][right panel of their Figure 2]{sulentic00a} it is very clear that for Pop. A quasars the oxygen is fainter than in Pop. B, and decreasing along the sequence, indicating that the higher the accretion (bins A3-A4), the less prominent oxygen lines we are observing \citep{sulentic00a,shenho14}. So, discarding spectra with no detectable \oii, creates a bias against highly accreting sources.
    
    The next step was to separate Pops. A and B (Sec. \ref{ssec:spec_types}). For this purpose, it was necessary to apply the luminosity dependent criterion of \citet{sulentic17} which brings the limit at FWHM $\approx$ 4000 \kms\  to significantly higher values for sources of bolometric luminosity $\log L_\mathrm{bol} \gtrsim 46$ \ergss:  FWHM$_\mathrm{AB} \approx 3500 + 500 (L_\mathrm{bol}/3.69\times 10^{44})^{0.15}$ \kms (applied to the \cnl\ line) to separate Pop. A and B.
    In our sample, Pop. A FWHM(\aliii) goes from $\sim$ 2500 \kms to almost 4500 \kms, and only one source had a value less than 2000 \kms\ (Sec. \ref{ssec:lwidths}). 
    
    Afterwards, we looked for extreme Pop. A sources (xA or highly accreting quasars) using the UV line ratios from \citet[][hereafter \citetalias{MS14}]{MS14}: \aliii/\siiii\ $\geq$ 0.5 and \ciii/\siiii\ $\leq$ 1 (see Section \ref{ssec:spec_types} for a more detailed description), finding 11 xA quasar candidates. 
    
    \begin{figure}
        \includegraphics[width=\hsize]{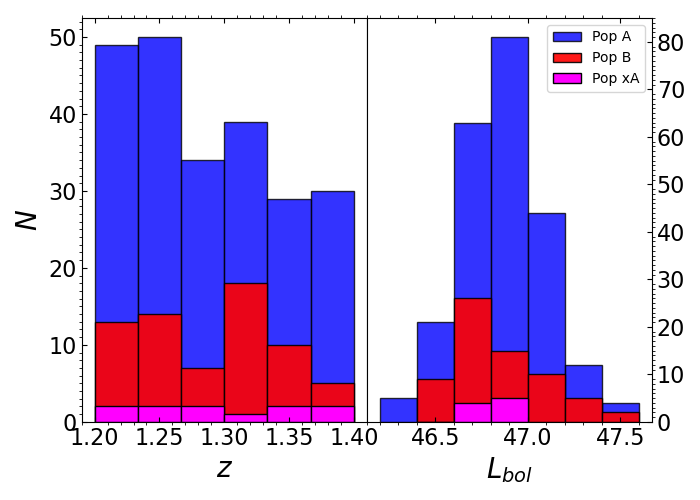}
        \caption{Distribution of redshift (left) and of bolometric luminosity (right) for the total sample (309 objects). Colour code: Pops. A, xA and B in blue, magenta and red, respectively.}
        \label{fig:Lbol_hist}
    \end{figure}

    The median value of the S/N distribution of our final sample is $\approx$31. The redshift distribution is reported in Figure \ref{fig:Lbol_hist} (left). The $z$-range is small, due to the values of the line wavelength range needed to have both \al\ and \o2\ recorded on the same spectrum: \al\ is at the blue edge and \o2\ at the red edge of each spectrum (see Figure \ref{fig:dzvsz}, upper panel, for an example). This $z$\ range is the most relevant condition to ensure that the systemic redshift of the quasar is measured precisely (see also Section \ref{ssec:redshift_est}). In Figure \ref{fig:Lbol_hist} (right), the \lbol\  distribution of our sample shows that our sample is made of luminous AGN. 
    The luminosity median values are log \lbol\ = 46.8 \ergss\ for Pop. A objects (including xA quasars), and log \lbol\ = 46.6 \ergss\ for Pop B. 
    Previous works usually found 50\%\ Pop. A and 50 \%\ Pop. B in flux-limited samples \citep{zamfir10,marziani13}. The larger sample size of Pop. A might be due to the flux limit of the Sloan survey along with the relatively high redshift $z \sim 1$ \citep{sulentic14} that might have caused  a Malmquist-type bias yielding an excess of Pop. A sources (i.e., radiating at relatively high Eddington ratio) with respect to Population B (radiating at lower Eddington ratio).

\section{Data analysis}\label{data_an}

    Optical spectral data used in this work were wavelength- and flux-calibrated by the SDSS DR16 pipeline. For the Galactic dust extinction, we use the reddening estimates from \cite{SanF11} assuming the value of the R$_V$ coefficient as 3.1. The Galactic absorption median value was $\mu_\frac{1}{2}(A_B) 
    \approx 0.14$, ranging from 0.03 up to $\sim$ 0.6. We choose to apply this correction only in correspondence of the redshifted 1900\AA\ blend, where the median $A_B$\ implies a flux increase of 14\%. The chosen value only affected the luminosity computation, not the spectral slope between the blue and red edge of the 1900\AA\ blend. Redshift and flux corrections of the spectra were first done using the $z$\ values provided by the SDSS DR16. An additional $z$\ correction was applied using the rest-frame estimated with the \oii\ line, as described below.

\subsection{\oii\ redshift estimation}\label{ssec:redshift_est} 

\begin{figure}
        \centering
        \includegraphics[width=\hsize]{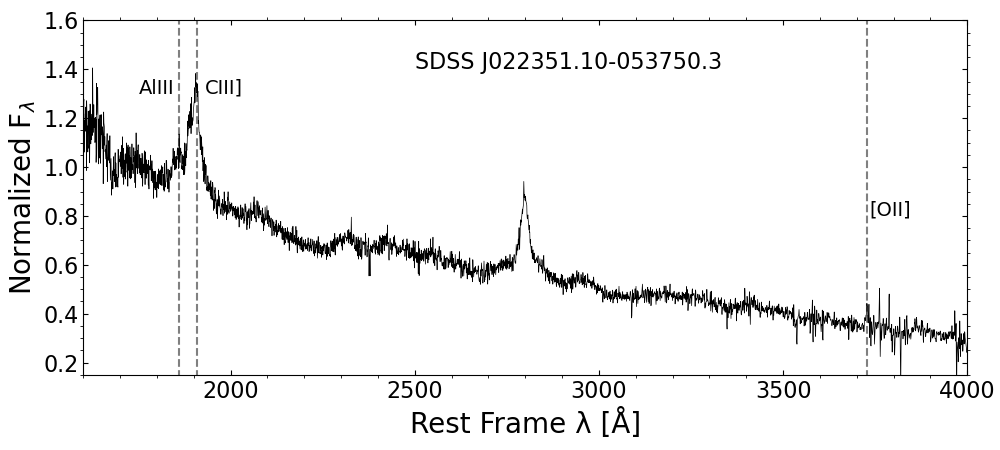}\\
        \includegraphics[width=\hsize]{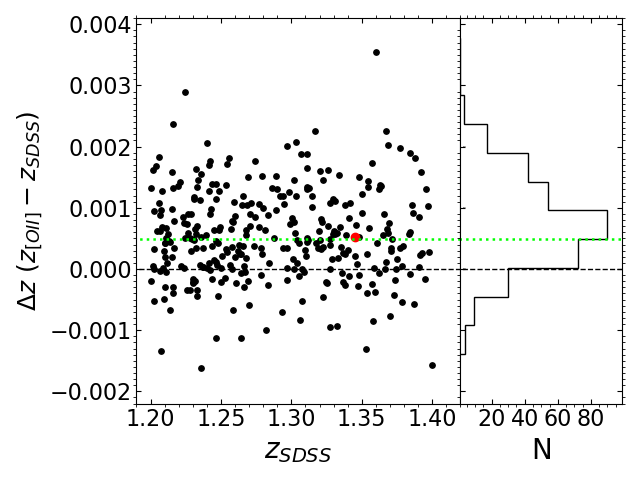}
        \caption{\textit{Upper panel}: Example of a spectrum from our sample. Vertical lines are the prominent emission lines of our work: \aliii, \ciii\ and \oii. Abscissa scale is rest-frame wavelength in \AA. Ordinate scale is the normalised flux at 1700 \AA. 
        \textit{Lower left panel}: $z_\mathrm{SDSS}$ vs. $\Delta$z for the sample. The red point corresponds to the values of $\Delta$z and $z$\ of the upper panel spectrum. \textit{Lower right panel}: Distribution of $\Delta z$. The lime line shows the shift $\Delta z$ median value: 4.918$\times10^{-4}$\ equivalent to $\sim$65 \kms. }
        \label{fig:dzvsz}
    \end{figure}

    The SDSS redshift estimates can be biased \citep{hewett10}. We observed discrepancies between the peak and rest-frame wavelength of \o2\ (as seen in Figure \ref{fig:dzvsz}) after the SDSS-based $z$ correction. We applied an additional redshift correction using the peak intensity wavelength of \oii\ as described in Section \ref{sssec:OII_fit} with a more carefully fitting using the {\tt specfit} task \citep{kriss94} from Image Reduction and Analysis Facility ({\fontfamily{lmtt}\selectfont IRAF}, \citealt{tody86}).
 
    We compare the differences between the SDSS DR16 redshift and the $z$\ values obtained from the narrow line \o2\ in Figure \ref{fig:dzvsz}. The median value of $\Delta z = z_\mathrm{[OII]} - z_\mathrm{SDSS} = 4.918 \times 10^{-4}$  (roughly $\sim$70 \kms in the rest frame, green dotted line of the Figure \ref{fig:dzvsz}) indicates that the SDSS-$z$ values were underestimated. A fraction of the objects, $\sim$ 25\% of the sample, showed a difference in the rest-frame $z$\ higher than 150 \kms\ (up to $\sim$450 \kms). The distribution is not symmetric around the median value. The main reason of these systematic differences is probably a bias of the SDSS due to the usage of HILs.  

\subsection{Multicomponent fitting}\label{sec:multifitting}

    \begin{figure}
        \centering
        \includegraphics[trim={1cm 0 0 0},clip,width=\hsize]{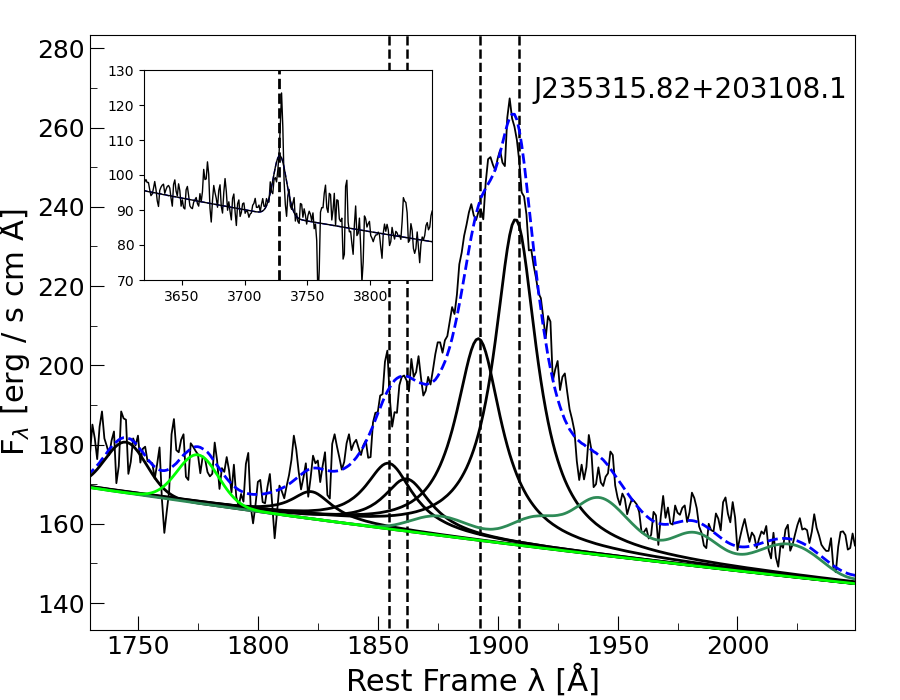}
        \includegraphics[trim={1cm 0 0 0},clip,width=\hsize]{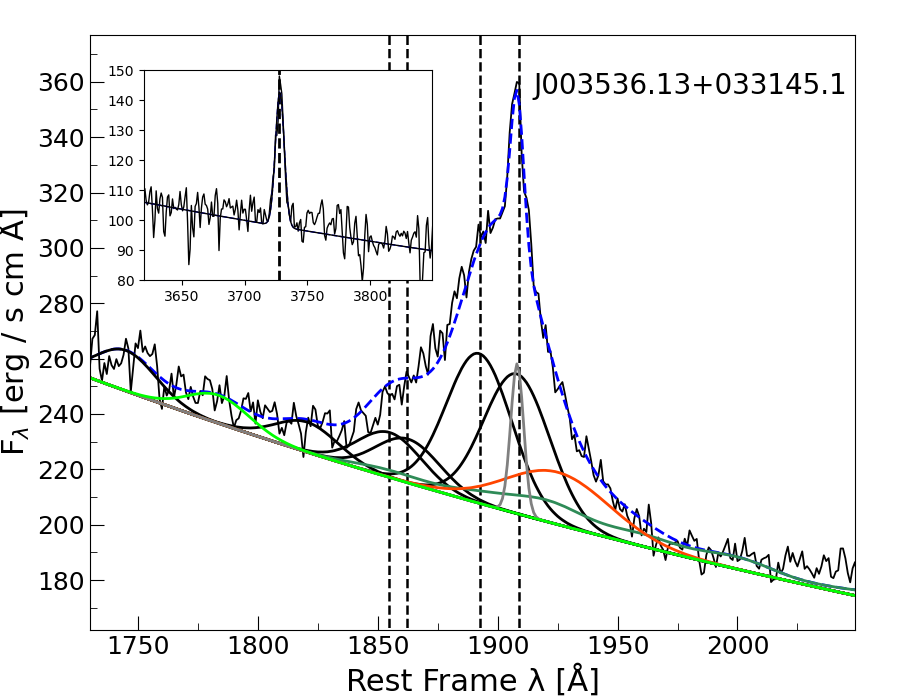}
        \caption{Analysis of the 1900\AA\ blend and the \oii\ region (small box within the Figure) as described in section \ref{sec:multifitting}. Abscissa scale is rest-frame wavelength in \AA. Ordinate scale is the specific flux in units of 10$^{-17}$\ergss\ cm$^{-2}$ \AA$^{-1}$ \textbf{. \textit{Top:}} Example of a Pop. A source fitted with a Lorentzian profile. \textit{Bottom:} Example of a Pop. B source fitted with a Gaussian profile. The black lines identify the BC of \aliii, \siiii\ and \ciii\ (red and grey are the VBC and NC of \ciii\, respectively if present). The dashed blue line is the multi-component model obtained by \texttt{specfit}. Green lines trace the adopted \feii\ (pale) and \feiii\ template (dark).}
        \label{fig:spec_fits}
    \end{figure}

    The UV range covered in the sample is populated by blended, intermediate ionisation lines. To analyse the emission lines of the spectra,   multi-component fits were done using the task \texttt{specfit}. This routine allows us to simultaneously fit all components present in the spectrum:  continuum, \feii\ features, and emission lines, computing the \chisq\ parameter that measures the difference between the original spectra and the fitted one. The task \texttt{specfit} minimises the \chisq\ to find the best fit.  Intensity measures of \oii\ were carried out with the \texttt{splot} task within  {\fontfamily{lmtt}\selectfont IRAF}.

    The primary continuum source in the UV region is well known to be originated from the accretion disk {\citep[{e.g.,}][]{malkan82,wandel88,capellupo16}}. In the absence of extinction, the most widely-used  model for the continuum is a single power-law over a limited spectral range \citep[see e.g.,][]{sniegowskaetal20}. 
    We fitted a local continuum for two spectral ranges centred on the 1900 \AA\ blend and \oii, the most important emission lines relevant to this work (see Figure \ref{fig:spec_fits}). 
    
    Along the E1 main sequence, it is possible to model the \hb\ line profile with three components with blueshifted, unshifted and redshifted centroids (blue [BLUE], broad [BC] and very broad components [VBC], respectively; \citealt{marziani10}). Then, we can use this model for all strong broad lines by varying the relative intensity of the components. The model considering the BC and VBC separation applies to Pop. B sources and is consistent with stratification of the BLR \citep{sulenticetal00b,sneddengaskell07,wolfetal20}. The BC (hydrogen density \nh\ $\sim 10^{12}$ cm$^{-3}$, ionisation parameter $\log U \sim -2$\ and column density \nc\ $\gtrsim 10^{23} $ cm$^{-2}$) is present in almost all type-1 quasars and corresponds most likely to the virialized part of the BLR, while the VBC can be interpreted as the emitted gas in the innermost BLR \citep[e.g., ][]{sulenticetal00b,marzianietal03f,marziani10,wangli11,wolfetal20}. The BLUE component is apparent as a blueshifted excess superimposed to the blue wing of the BC \citep{leighlymoore04}. 
    
    The 1900\AA\ blend contains the same emission lines for both Pop. A and B: \al, \si, and \ciii\ which are the strongest features (see also Table 1 from \citealt{negrete12}). Figure \ref{fig:spec_fits} shows example fits of a Pop. A and B sources, where we used a Lorentzian function for Pop. A and a Gaussian function for Pop B. For Pop. B spectra, we included an additional Gaussian component for the VBC. A detailed description of the phenomenology of the line profiles can be found in \cite{sulentic00a,sulentic02,Marzianietal2022} and references therein.
    More in detail, we consider the following components described in Sections \ref{sssec:1900_fit} and \ref{sssec:OII_fit} for a complete model of each spectrum
 
\subsubsection{Region 1: 1750-2050\AA.} \label{sssec:1900_fit}
   
    \begin{enumerate}
    \item \textit{Continuum}: We adopt a single power-law to fit the region 1700-2050\AA, using a continuum window at 1700 \AA\ as seen in \cite{francis91}. 
    \item \textit{\feiii\ and \feii}: Emission of the \feiii\ multiplets can be strong in the vicinity of \ciii, as seen in the average quasar spectrum from \citet{vanden01}. They appear to be strong when \al\ is also strong \citep{HB86}. Strong \feiii\ and \al\ emission further strengthen the conclusion that the BLR densities, at least in Pop. A sources are very high (on the basis of photoionisation models,  $\sim 10^{11} - 10^{12}$ cm$^{-3}$, \citealt{korista97,kuraszkiewicz00}). We adopted for the \feiii\ template model the one obtained by \cite{vestergaard01}.
    The {\tt specfit} task scaled and broadened the template to reproduce the observed emission \citep{ByG}. In most cases, we fitted the multiplet \feUV\ {(seen in the blue-ward of the 1900\AA\ blend, \citealt{moore45})} as an isolated Gaussian in the rest-frame. {We adopted a Gaussian profile because the feature is a blend of several individual \feii\ lines belonging to the same multiplet}. If the \feii\ multiplet is prominent, an extra component is added: \feiii $\lambda1914$ to fit an excess seen near the red wing of \ciii\ associated with unresolved \feiii\ template emission in Pop. A quasars \citep{negrete12}. {This \feiii$\lambda$1914 emission is a single line, so we fitted it using a Lorentzian profile to be consistent with the profile of the BCs}. The spectrum of I Zw 1 shows this effect: both \cnl\ and \feiii $\lambda$1914 are needed to account for the double-peaked feature at 1910\AA\ that is too broad to be explained by a single line \citep{negrete12,negrete13}. This criterion rests on the assumption that \feiii $\lambda$1914 and the \feii\ UV multiplet \#191 are enhanced by \lya\ fluorescence \citep{sigutpradhan98,johansson00}.
    \item \textit{\ciii}: Strengths and FWHM were left free to vary in the \texttt{specfit} model with one restriction: FWHM(\ciii) $\leq$ FWHM(\aliii) or FWHM(\siiii) to avoid a divergence of the FWHM(\ciii) due to the \feiii\ emission on the red side of the blend. In the case where the model does not successfully follow this condition, the FWHM of both lines were forced to be the same to avoid a larger FWHM(\cnl). We also added a {narrow component} (NC) if needed with a fixed upper limit of FWHM $\sim$1000 \kms\ at the rest-frame wavelength as an initial condition. {The distribution of the peak emission around 1909 \AA\ shows a fraction of quasars with a shorter wavelength than the one expected for the laboratory wavelength of \ciii\ (Figure \ref{fig:c3NC}). This phenomenon could be due to two main reasons. First, we are looking at the prohibited line \ciiip\ (dotted line, Figure \ref{fig:c3NC}) also observed in the NLR. Second,  a blueshifted emission of \cnl\ NC due to an outflow in the NLR.} 
    
    \begin{figure}
        \centering
        \includegraphics[width=0.9\hsize]{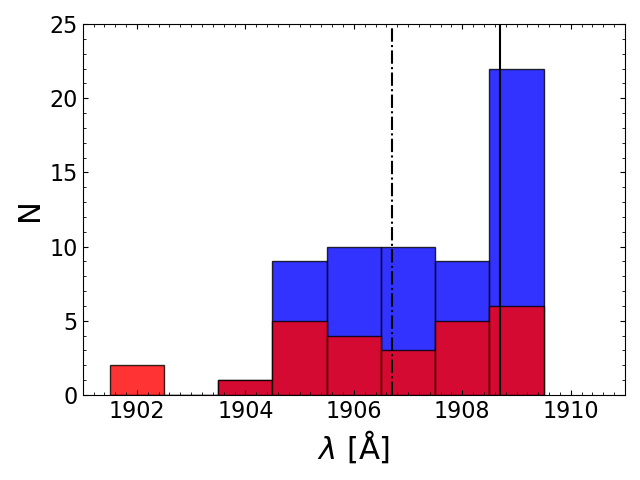}
        \caption{{Distribution of the peak blend wavelength around 1909\AA, for Pop. A (blue) and Pop. B (red). The vertical lines identify the rest-frame wavelength of  \ciiip\ (dotted) and \ciii\ (filled).}}        \label{fig:c3NC}
    \end{figure}

    For Pop. B objects we included a redshifted VBC whose strengths and FWHM were left free to vary with a FWHM lower limit $\approx$ 7000 \kms. In Figure \ref{fig:C31909} we illustrate the necessity of using an additional component in the \cnl\ profile. Looking at the residuals and the \chisq\ values, the best fit (according to the F distribution for the ratio of the $\chi^2$,  \citealt{bevington}) is the one with the VBC as seen for the SDSS spectrum J012726.39+154153.8 with a \chisq$_\mathrm{VBC} \approx$  0.1676 in contrast to \chisq $\approx$ 0.4451 obtained without VBC, respectively (see section \ref{ssec:popBC3_profile} for its interpretation). 
    
    \begin{figure}
        \centering
        \includegraphics[width=0.9\hsize]{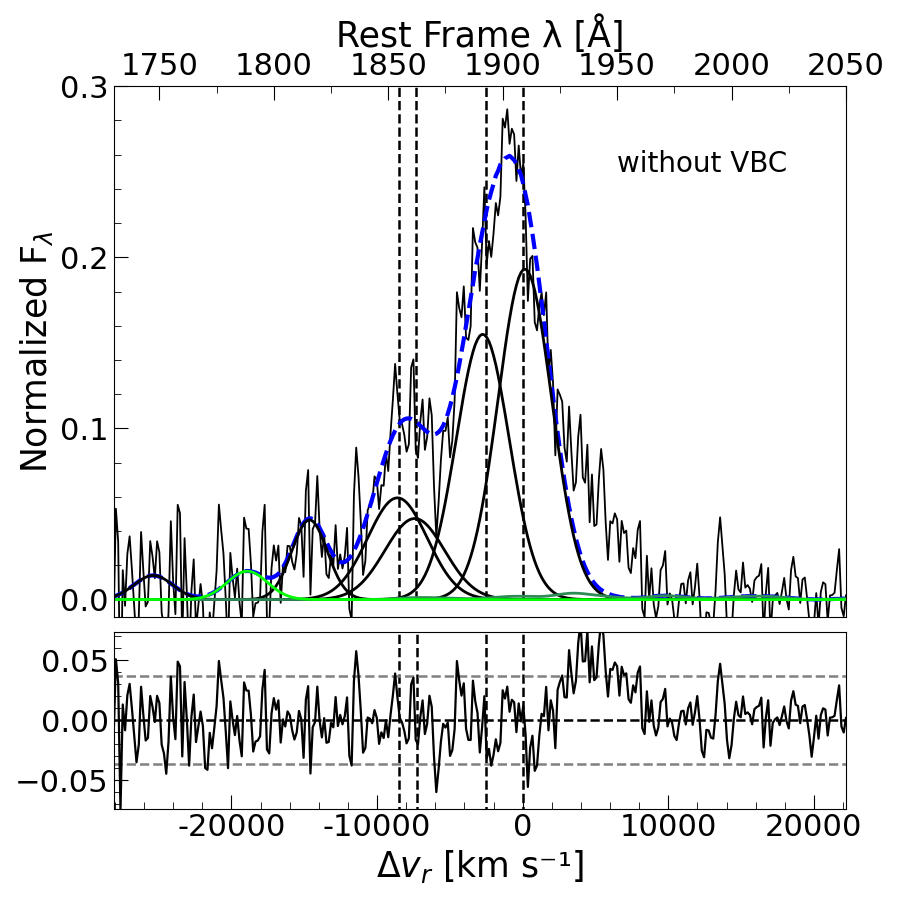}
        \includegraphics[width=0.9\hsize]{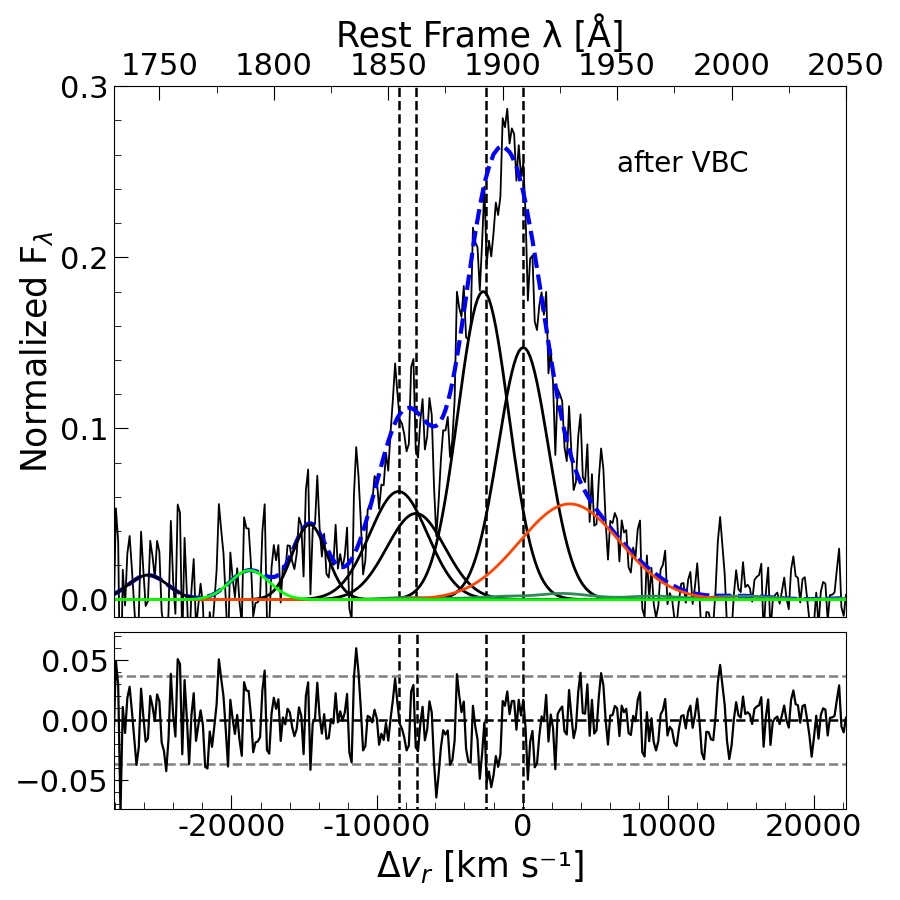}
        \caption{Analysis of the 1900\AA\ blend (continuum extracted for simplicity in both spectra) of SDSS J012726.39+154153.8 as described for Figure \ref{fig:spec_fits}. \cnl\ profile fitted with only a BC and BC+VBC, top and bottom, respectively. Abscissa is rest-frame wavelength in Å. Ordinate is the normalised specific flux obtained from \texttt{specfit}.}
        \label{fig:C31909}
    \end{figure}
    
    \item \textit{\siiii}: Strengths and shifts were free to vary, with one restriction: FWHM(\siiii) $\geq$ FWHM(\ciii). We had some difficulties due to the nature of the blend itself. The line tended to be blueshifted in order to “fill” its blue side, and therefore, \aliii\ also presented a blueshift as seen in \cite{mla18}. In the cases where the shift was completely unreal under a visual inspection, we fixed the central wavelength of \si\ to the rest-frame.
    \item \textit{\aliii}: {The doublet was resolved and the blue component shifts, FWHM, and intensity were allowed to vary, with the red one tied to the blue by identical FWHM and fixed wavelength ratio.  The ratio between the intensity of the red and blue component of the doublet was kept fixed 0.8 \citep{laor97}}. Rarely, a different ratio up to 0.98 was assumed according to be observed doublet profile. However doublet total strengths and shifts were left free to vary.
    Regarding \ciii, the condition  FWHM(\aliii) $\geq$ FWHM(\ciii) was imposed. {The FWHM(\cnl) limit comes from  the low value of the \ciii\ critical density. On the converse the \al \ line, emitted via a permitted transition, has no well-defined critical density \citep{baldwinetal95,korista97}.} 
    \item \textit{Other lines}: Two lines not as prominent as those described above in points 3-5 were detected in the blue side of the 1900\AA\ blend:  \niii \ and \siii. We assumed them to be at the rest-frame as an initial condition, although their shifts, strength and FWHM were left free to vary. 
    \end{enumerate}
    
    \subsubsection{Region 2: 3550-3950\AA.}\label{sssec:OII_fit}
    
     The \oii\ doublet emission line is one of the main emission features of this spectral range. {Hence}, the components are the same for all spectra: 
    \begin{enumerate}
    \item \textit{Continuum}: A strong pseudo-continuum associated with \feii\ emission is expected to be present in the spectral range around \oii\ 
    between 3500 \AA\ and 3850 \AA\ \citep{vanden01}. However, the limited range  3700-3770  \AA\ is smooth enough to permit the use of a power-law to model the sum of the AGN continuum and the FeII emission. 
    \item \textit{\oii}: Our spectra have unresolved or almost unresolved \oiiD\ lines because the spectral resolution at the observed wavelength around 8200 \AA\ is $\lambda/\delta \lambda \approx$ 2250, so the spectral purity is 3.64 \AA, which is larger than the doublet separation. Therefore, we used a single Gaussian fit \citep{bon20}.  
    \end{enumerate}
    
    \subsection{Spectral types along the E1}\label{ssec:spec_types}
    
    In order to classify the objects, we attempted to use as a first approximation the \cite{sulentic00b} spectral types. However, this classification is based on the FWHM \hb\ that increases systematically for higher $L$ objects, and we expect the same effect for the IILs.
    We identify 242 and 67 Pop. A and B objects, respectively. In a few cases, we had fits with both profiles, almost $\sim 5$ \% of the sample, but with different FWHM(\cnl). For these sources, with a value of FWHM near the 4000 \kms\ limit, it was necessary to choose the best fit according to the \chisq values using the F distribution \citep{bevington}.     
    {To separate highly accreting candidates of spectral type A3 and A4 (xA) from the  A1-A2 sources defined in Sec. \ref{sec:introduction}}, we used the UV diagnostics ratios of \citetalias{MS14}. Pop. A quasars located at the extreme of the MS, are considered to be sources radiating close to  the Eddington if they  satisfy the following criterion \citep[e.g.,][]{duetal16}:
    
    \begin{equation}
        R_{\mathrm{FeII}}=\frac{\mathrm{EW(FeII\lambda4750)}}{\mathrm{EW(H\beta)}} \geq 1.0
    \end{equation}

    An  equivalent condition has been proposed at intermediate to high redshift (z $\gtrsim$ 1) where the \hb\ line is no longer visible in the optical range, using the 1900\AA\ emission line blend of \aliii, \siiii\ and \ciii\ \citepalias{MS14}.  The blend involving these lines constrains the physical conditions in the broad-line emitting gas the same way as  extreme optical \feii\ emission. Measures of high S/N spectra of \citetalias{MS14} yield the selection criterion based on two related ratios:
    
    \begin{enumerate}
        \item \aliii / \siiii\ $\geq$ 0.5 and
        \item \ciii/\siiii \  $\leq$ 1.0
    \end{enumerate}
    
    The emitting region of the IILs corresponds to the densest emitting region likely associated with the production of LILs like the CaII IR triplet \citep{matsuoka08} and FeII \citep{baldwin04}. 

\begin{figure}
        \centering
        \includegraphics[width=\hsize]{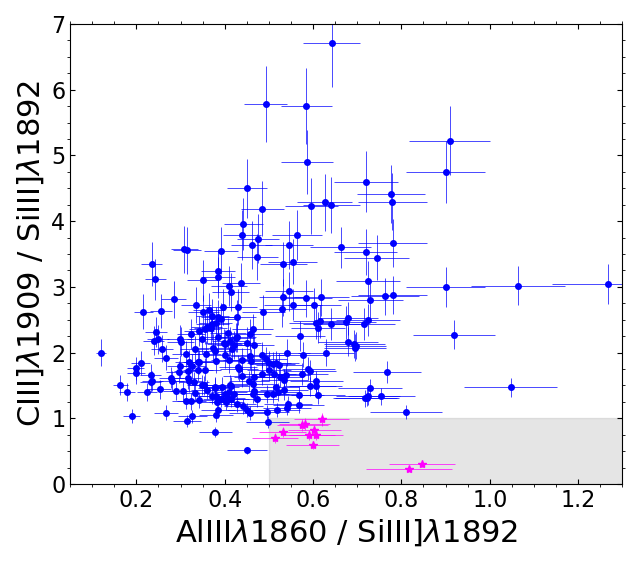}
        \caption{Distribution of Pop. A sources in the plane defined by the ratios \ciii/\siiii\ vs. \aliii/\siiii. The blue dots are quasars within populations A1-A2 (Pop. $\mathrm{\tilde{A}}$) and the xA sources are in magenta located in the lower-right grey area.}
        \label{fig:UV_MS14}
    \end{figure}

    We made a bin separation for the bins A1-A2 which we call Pop. $\mathrm{\tilde{A}}$\footnote{Note that in previous work Pop. A includes spectral types A1-A2 and A3-A4.} and the A3-A4 bins will be our Pop. xA candidates. In Figure \ref{fig:UV_MS14} we show the A1-A2 bins in blue, and in magenta we identify 11 xA quasars. 
    
\section{Results}\label{results}

    \begin{table*}[h]
    \centering
    \caption{Average and median values of the sample physical parameters by population.} 
    \label{tab:T1}
    \begin{tabular}{@{}lccccccl@{}}
    \toprule
    \toprule
     & \multicolumn{2}{c}{Pop $\mathrm{\tilde{A}}$ (231 quasars)} & \multicolumn{2}{c}{Pop xA (11 quasars)} & \multicolumn{2}{c}{Pop B (67 quasars)} \\  \cmidrule(l){2-3} \cmidrule(l){4-5} \cmidrule(l){6-7}
     & Average & Median & Average & Median & Average & Median & Notes\\ 
    \midrule
    FWHM(\cnl\ BC) & 3330$\pm$280 & 3370$\pm$280 & 3280$\pm$255 & 3420$\pm$260 & 4490$\pm$180 & 4380$\pm$180 & $a$ \\
    FWHM(\cnl\ VBC) & - & - & - & - & 8120$\pm$410 & 8120$\pm$410 & $a$ \\ 
    FWHM(\al) & 3560$\pm$270 & 3550$\pm$230 & 3560$\pm$230 & 3530$\pm$250 & 5270$\pm$240 & 5300$\pm$250 & $a$ \\    
    \midrule
    EW (\cnl) & -15.43$\pm$2.90 & -14.77$\pm$2.90 & -8.33$\pm$1.88 & -8.94$\pm$1.88 & -7.18$\pm$1.76 & -6.97$\pm$1.76 & $b$ \\
    EW (\al) & -3.55$\pm$0.94 & -3.27$\pm$0.94 & -7.08$\pm$0.79 & -7.05$\pm$0.79 & -5.02$\pm$0.97 & -4.92$\pm$0.97 & $b$ \\
    \midrule
    \al/\si & 0.48$\pm$0.11 & 0.45$\pm$0.11 & 0.63$\pm$0.02 & 0.60$\pm$0.02 & - & - & $c$ \\
    \cnl/\si & 2.17$\pm$0.58 & 1.88$\pm$0.58 & 0.72$\pm$0.10 & 0.75$\pm$0.10 & 0.94$\pm$0.22 & 0.86$\pm$0.22 & $c$ \\
    \cnl(BC+VBC)/\si & - & - & - & - & 1.68$\pm$0.32 & 1.53$\pm$0.32 & $d$ \\
    \midrule
    shift \cnl & 50$\pm$190 & 40$\pm$190 & 80$\pm$120 & 50$\pm$120 & 80$\pm$90 & 160$\pm$90 & $a$ \\
    shift \al & -30$\pm$120 & 10$\pm$120 & -200$\pm$232 & -240$\pm$230 & 30$\pm$120 & 10$\pm$120 & $a$ \\
    log \mbh(\cnl BC) & 8.82$\pm$0.14 & 8.81$\pm$0.14 & 8.79$\pm$0.13 & 8.82$\pm$0.13 & 9.15$\pm$0.12 & 9.13$\pm$0.12 & $e$ \\
    log \mbh(\al) & 8.80$\pm$0.10 & 8.78$\pm$0.10 & 8.79$\pm$0.09 & 8.83$\pm$0.09 & 9.14$\pm$0.11 & 9.12$\pm$0.11 & $e$ \\
    log L$_{1909}$  & 47.20 & 47.19 & 46.92 & 46.93 & 46.84 & 46.83 & $f$ \\
    log L$_{1860}$  & 46.56 & 46.56 & 46.87 & 46.79 & 46.73 & 46.70 & $f$ \\
    log L$_{Bol}$ & 46.87$\pm$2.65 & 46.86$\pm$2.65 & 46.85$\pm$1.84 & 46.87$\pm$1.84 & 46.84$\pm$2.59 & 46.79$\pm$2.59 & $g$ \\
    \redd (\cnl) & 0.95$\pm$0.20 & 0.85$\pm$0.20 & 0.98$\pm$0.24 & 0.87$\pm$0.24 & 0.39$\pm$0.07 & 0.39$\pm$0.07 & $h$  \\
    \redd (\al) & 0.96$\pm$0.19 & 0.86$\pm$0.19 & 0.91$\pm$0.13 & 0.85$\pm$0.13 & 0.41$\pm$0.07 & 0.39$\pm$0.07 & $h$  \\
    log L$_{vir}$ & - & - & 47.09 & 47.08 & - & - & $i$ \\
    \bottomrule
    \end{tabular}\\
    \tablefoot{
    \tablefoottext{a}{In units of \kms.}
    \tablefoottext{b}{Rest-frame equivalent widths reported with normalised spectra at 1700\AA\ are in units of \AA.}
    \tablefoottext{c}{UV diagnostic ratios from \citetalias{MS14} for Pop. A sources.} 
    \tablefoottext{d}{UV diagnostic ratio from \citetalias{MS14} for Pop. B sources.}
    \tablefoottext{e}{Log of \mbh\ are computed using the scale relations by \citetalias{Marzianietal2022} in units of M$_\odot$.}
    \tablefoottext{f}{Log of Line luminosity in units of \ergss; uncertainties are the 10$\%$ of the value.}
    \tablefoottext{g}{Log of bolometric luminosity in unit of \ergss computed using the continuum window at 1700\AA.}
    \tablefoottext{h}{\redd\ is the Eddington ratio.}
    \tablefoottext{i}{Log of virial luminosity in units of \ergss; uncertainties are the 10$\%$ of the value.}
    }
    \end{table*}

    Table \ref{tab:T3} of Appendix B lists the results of the line fitting procedures of Sec. \ref{sec:multifitting}, and the luminosity and \mbh\ computations of Section \ref{sec:discussion}. The Table also reports the redshift from the SDSS and our $z$ estimation using \o2, the continuum flux and the normalisation at 1700\AA, and the line profile classification (Lorentzian or Gaussian). From the {\tt specfit} analysis we report the intensity, FWHM, shift from the restframe, and EW for each emission line of the 1900\AA\ blend. For the \o2\ region, we report the intensity and FWHM. The last part of Table \ref{tab:T3} contains the UV diagnostic ratios, black hole mass, Eddington ratio, and virial luminosity (computed only for xA sources, see Sec. \ref{ssec:Al3_lvir}).  
    Table \ref{tab:T1} presents a summary of the physical parameter values where we report the median and average values by Population. The reported uncertainties are the semi-interquartile ranges (sIQR) of the parameter distributions. For luminosity estimates we adopted an uncertainty of 10\%.

     We organise the presentation of our results on line widths and shifts of the 1900\AA\ blend along the MS, separating Pops. $\mathrm{\tilde{A}}$ (A1-A2), xA and B. The MS is expected to trace changes in the dynamical and physical conditions inside the quasars \citep{marzianietal03f,lamuraetal09,popovicetal19}. {Line widths (e.g., FWHMs) of LILs and IILs measure the kinematics of the BLR. We assume that Doppler motions in a virialized region produce unshifted and symmetric line broadening. 
     Wavelength shifts of \aliii\ and \ciii\,  were measured with respect to the \oii\ rest frame. They may be due to Doppler effect because of radial gas motions plus obscuration along our line of sight. }
     The differences in line widths observed in the same spectrum might be due to emissions from regions of non-virialized motions (e.g., outflows), as usually seen in Pop. A and also in high luminosity Pop. B objects at high luminosity \citep{sulentic17}. {Line width differences in type 1 quasars are also associated with the stratification of the emitting region, where broader lines trace the kinematics of the regions closer to the SMBH  \citep[e.g.,][]{sulenticetal00b,petersonwandel00,sneddengaskell07,wolfetal20,Lietal2021}}. Last, FWHM differences may be due to different orientations of the accretion disk (expected to provide the reference plane of symmetry of the BLR) with respect to our line of sight.  
    
    \subsection{Systematic shifts}\label{ssec:syst_shifts}
    
    \begin{figure}
        \includegraphics[width=\hsize]{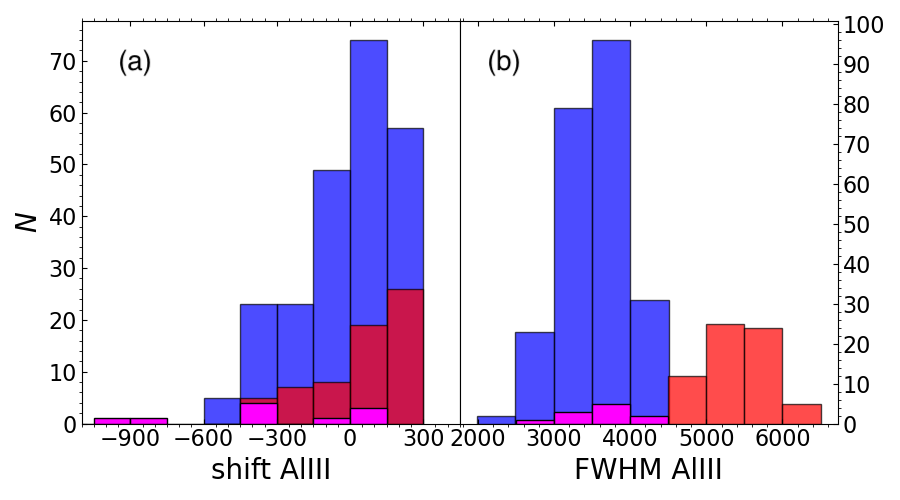}
        \includegraphics[width=\hsize]{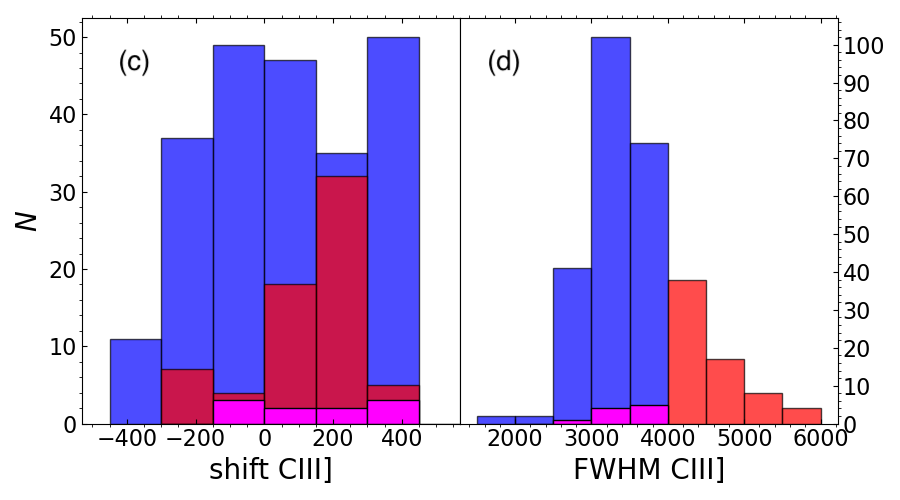}
        \caption{\textit{Upper}: Distribution of shift with respect to rest-frame and FWHM of \aliii. \textit{Bottom}: same but with \ciii. Shift and FWHM in units of \kms. Colour-code as Figure \ref{fig:Lbol_hist}.}  
        \label{fig:shift_Al3-C3}
    \end{figure}

    In the virialized region one can expect a modest shift ($\lesssim |\pm200|$ \kms) associated with the measurement of the uncertainties. We consider $\pm200$ \kms as an uncertainty limit, given the instrumental resolution of the SDSS spectra at their blue side, which is where the 1900\AA\ blend falls in the observed rest frame. Considering our complete sample, the median values of the \al\ and \cnl\ shifts in the histograms of Figure \ref{fig:shift_Al3-C3}a,c are 10$\pm$120 \kms\ and 40$\pm$190 \kms, respectively (see also Table \ref{tab:T1}). In almost 90\% of Pop. $\mathrm{\tilde{A}}$ and B \al\ profiles we find that the shifts are lower than the uncertainty limit. However, Figure \ref{fig:shift_Al3-C3}a  shows an asymmetric distribution of \aliii\ shifts with an extended tail of blueshifts reaching several hundred \kms. Blueshifts larger than 200 \kms\ imply that we are most likely looking at a mixture of two non-resolved components in the line profiles: a virialized plus an outflow component. 
    Even though a blueshifted component in \al\ may not be as intense as the blue component of \civ\ it is essential to be aware of its presence: significant shifts would introduce a bias in the estimation of the rest-frame, as the \al\  blue component would broadens and shift the full profile. 
    The \ciii\ line shows a more uniform distribution shifts in Pop. $\mathrm{\tilde{A}}$ (Fig. \ref{fig:shift_Al3-C3}c), with a slight net shift to the red $\sim 100$ \kms, smaller than the typical uncertainty in the individual shift measurements. 
    
    \begin{figure*}
        \centering
        \includegraphics[width=0.42\textwidth]{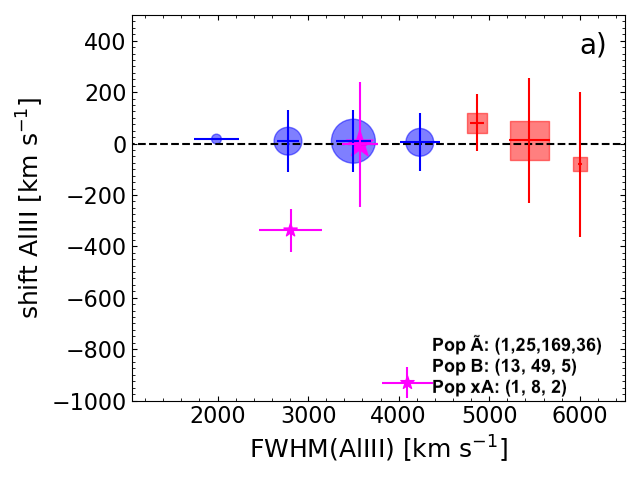}
        \includegraphics[width=0.42\textwidth]{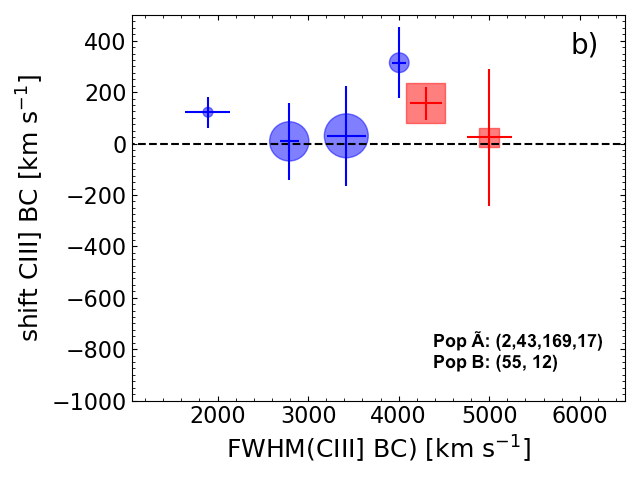}\\
        \includegraphics[width=0.42\textwidth]{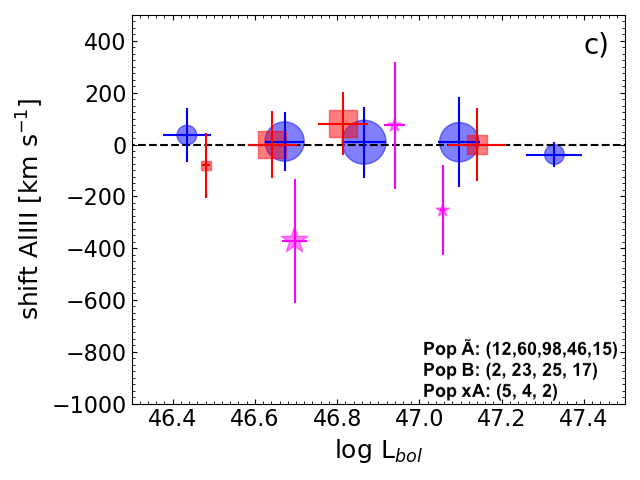}
        \includegraphics[width=0.42\textwidth]{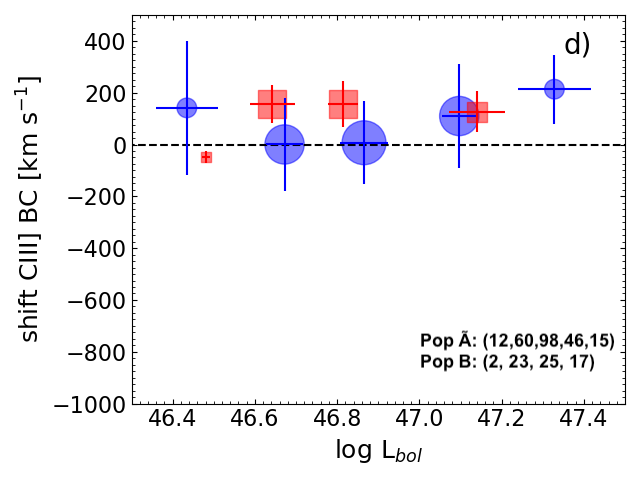}\\
        \includegraphics[width=0.42\textwidth]{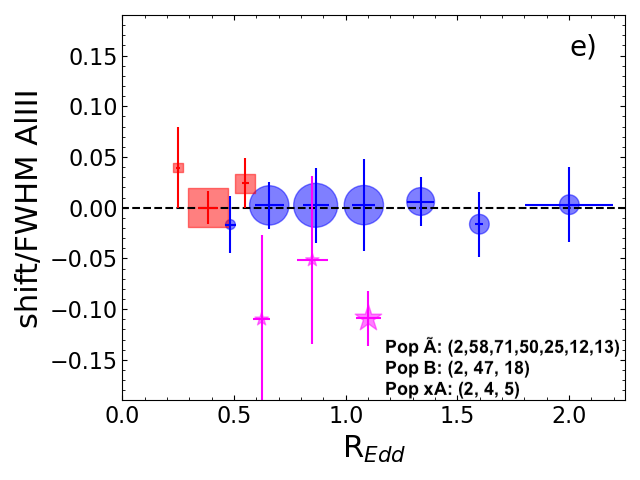}
        \includegraphics[width=0.42\textwidth]{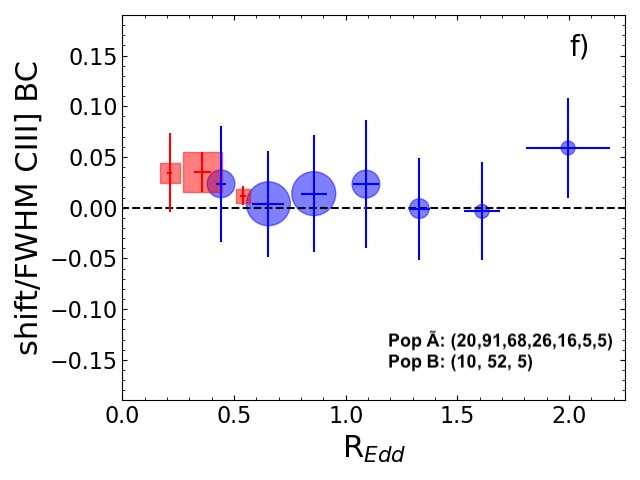}\\
        \includegraphics[width=0.42\textwidth]{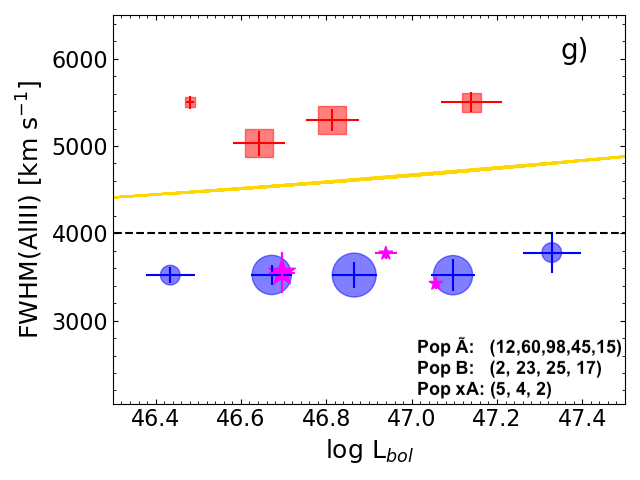}
        \includegraphics[width=0.42\textwidth]{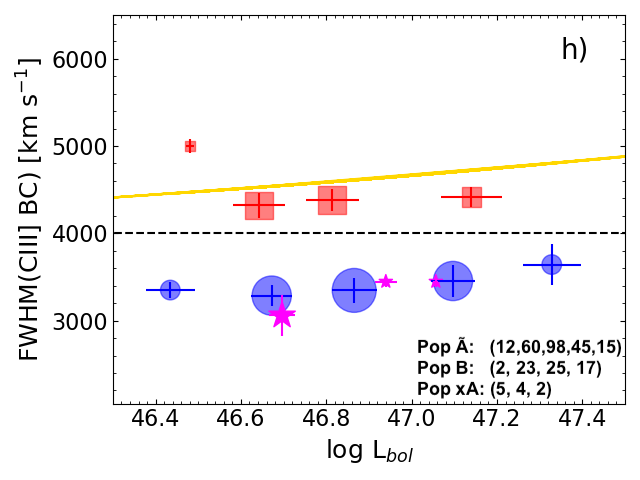}
        \caption{Behaviour of \aliii\ (left) and \ciii\ (right) by population. 
        \textbf{a,b}: shift with respect to rest-frame vs. FWHM. 
        \textbf{c,d}: shift vs. log L$_{bol}$. 
        \textbf{e,f}: ratio of shift over FWHM vs. \redd. 
        \textbf{g,h}: FWHM vs $\log$ \lbol. 
        Lines are the luminosity-dependent limit between Pop. A and B. of \citeauthor{sulentic17} (\citeyear{sulentic17}, gold) and the empirical separation of \citeauthor{sulentic00b}  (\citeyear{sulentic00b}, dashed). 
        Colour-coding: Pop. $\mathrm{\tilde{A}}$ (blue circles), Pop. xA (magenta stars) and Pop. B (red squares). Reported values are sub-sample medians, error bars are  sIQRs.  Marker sizes are as indicated in the legend of each plot. }
        \label{fig:shift_fw}
    \end{figure*}

    \subsubsection{Population $\mathrm{\tilde{A}}$}\label{sssec:popA}
    
    Figure \ref{fig:shift_Al3-C3}a shows that only 39 out of our 231 Pop. $\mathrm{\tilde{A}}$ objects have \al\ blueshifts larger than the uncertainty limit (33 objects have shifts $<$ -300 \kms). This trend can also be seen in Figure \ref{fig:shift_fw}a were we plot the \al\ shift as a function of its FWHM in bins\footnote{Throughout Figure \ref{fig:shift_fw}, we report the median values for each bin.} of $\Delta$FWHM(\al)=1000 \kms. The plot shows that, on average, Pop. $\mathrm{\tilde{A}}$ sources (blue circles) do not present systematic shifts in \al\ that significantly affect the FWHM measurements. This behaviour confirms the reliability of the rest-frame of the \aliii\ for sources within the A1-A2 populations. 
    In the relations of the \al\ shift with the bolometric luminosity or Eddington ratio (Figure \ref{fig:shift_fw}c, e) we also do not find displacements larger than the uncertainty limit. Data were divided in sub-samples of $\Delta$log\lbol=0.2 dex and $\Delta$\redd=0.5. 
    
    Regarding the behaviour of \cnl\ in Pop. $\mathrm{\tilde{A}}$ sources, in Figure \ref{fig:shift_Al3-C3}b we find 11 objects ($\sim 4\%$) that show blueshifts larger than -300 \kms. As observed for \al\ relations, in \cnl, we do not see clear tendencies of \lbol\ and \redd(\cnl) with the shift (Fig. \ref{fig:shift_fw}d,f), although $\sim$17\% Pop. A sources shows a displacement as large as $\sim +300$ \kms (52 objects, Figure \ref{fig:shift_Al3-C3}c). This displacement toward the red is most likely due to the effect of the strong \feiii\ emission heavily blended with \cnl.
    
    \subsubsection{Extreme Population A}\label{sssec:popxA}
    
    The spectral fitting of our 11 xA objects are shown in Figure \ref{fig:xA_spectra} of Appendix \ref{app_A}. Figures \ref{fig:shift_fw}(a,c), present our xA sub-sample in magenta points which show \aliii\ blueshifts reaching several hundred \kms. However, those shifts are much lower than those found in \civ\ (e.g., \citealt{sulentic07}; Section \ref{ssec:civ_al3}).
    Nine out of 11 xA sources of Figure \ref{fig:shift_fw}c show a blueshift in \al, with a median shift of $\sim -$340 \kms\ and a maximum of $\sim -1000$ \kms. The Figure reveals that there is no dependence on luminosity for the \al\ shift. Figure \ref{fig:shift_fw}e shows that not only a blueshift is detected for xA sources but that the blueshift is also significant, $\sim$10\%\ of the FWHM.
    In the other hand, we note that 50 Pop. A objects ($\sim$17\% of the sample) have \redd\ higher than the one of xA sources. As described in Sec. \ref{ssec:spec_types} and discussed in Section \ref{ssec:mass_Redd}, we expect a higher \redd\ for xA objects).
    
    \ciii\ shifts in Pop. xA sources seem to be slightly redshifted (the median shift is 50$\pm$120 \kms), so we do not take them into account for the analysis. A redshift of $\sim$300 \kms\ was find in three xA spectra were \cnl\ is weak and is also affected by the \feiii\ emission at $\approx$ 1915 - 1920 \AA.

    \subsubsection{Population B}\label{sssec:popB}

    Our Pop. B sample is represented by red squares in Fig. \ref{fig:shift_fw}. In general, the Pop. B \al\ profile show small displacements from the rest frame wavelength within the uncertainty limit. Figure \ref{fig:shift_fw}a shows that the shift distribution is symmetric around 0. Figures \ref{fig:shift_fw}c,e are consistent with this trend: symmetric displacements around 0 shift, and no dependency on the bolometric luminosity and Eddington ratio obtained with \al. 
    
    In Fig. \ref{fig:shift_fw}b we observe a peak in the \cnl\ shift with a median value of 160$\pm$90 \kms. As seen in Fig. \ref{fig:C31909}, this small redshift could indicate that sources above the 4000\kms\ limit tends to ``cover'' the VBC spectral range.
    This behaviour has also been observed for the LIL \hb\ line \citep{zamfir10}. From Figure \ref{fig:shift_fw}d,f we observe a consistent behaviour for both \cnl\ and \al:  there is no significant dependency on \lbol\ and \redd. 
    
    \subsection{Line widths}\label{ssec:lwidths}
    
    \subsubsection{Population $\mathrm{\tilde{A}}$}\label{sssec:lw_popA}
    
    \begin{figure}
        \includegraphics[width=\hsize]{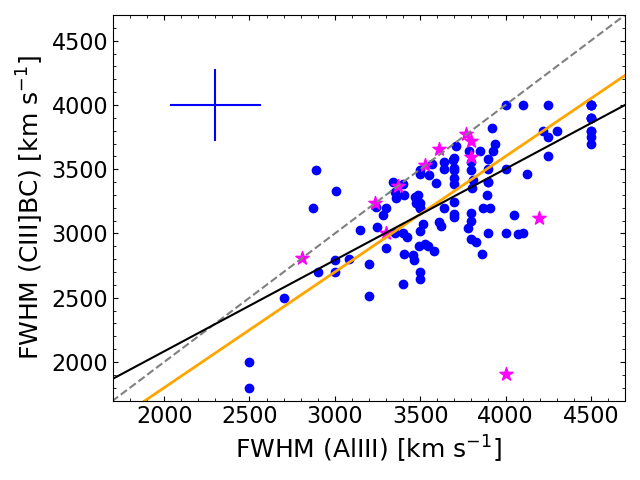}
        \caption{FWHM(\aliii) vs. FWHM(\ciii) for Pop. $\mathrm{\tilde{A}}$* [FWHM(\cnl) $\neq$ FWHM(\al)] sources. The dots are colour-coded as Figures above by population. The dashed line traces the 1:1 relation; the black line is the best fit obtained using the least-square method and the orange one is the line of FWHM(\cnl) = 0.9 FWHM(\al).  {The cross   shows   sIQRs of the uncertainties.}}
        \label{fig:FWHM_Al3C3}
    \end{figure}
    
    From our spectral fitting,  111 Pop A objects (excluding the 11 xA sources) have FWHM of \cnl\ and \al\ not forced to be equal. We call them Pop. $\mathrm{\tilde{A}}$*. In Figure \ref{fig:FWHM_Al3C3} we show  Pop. $\mathrm{\tilde{A}}$* quasars in blue and xA quasars as magenta points. The grey line indicates the 1:1 relation, the black line is the best fit for the Pop. $\mathrm{\tilde{A}}$* sources. Using the least-square method it yields the equation: FWHM(\cnl) $\approx$ (663 $\pm$ 348) + (0.709 $\pm$ 0.061) FWHM(\al). The \al\ FWHM median value is 3550$\pm$230\kms. The orange line is set at FWHM(\cnl) = 0.9 FWHM(\al), according to the findings of \citet[][hereafter \citetalias{Marzianietal2022}]{Marzianietal2022}. The value that relates the FWHM of \cnl\ and \al\ should be in the range $0.8 - 1.1$ for Pop A1-A2 sources. {Indeed, as seen in Fig. \ref{fig:FWHM_Al3C3} we have three objects with FWHM(\cnl\ BC) $\sim$ 1.1 FWHM(\aliii)}. This behaviour indicates that A1-A2 lines are narrower than \hb\ by $\approx$ 10\%.

    \subsubsection{Extreme Population A}
    
    The physical reason to use the ratio of the shift from the central wavelength over the FWHM in Figure \ref{fig:shift_fw}e,f resides in the effect of outflow motions on the BC. An outflow component leads to an increase in the profile width \citep[as seen for \civ\ and \mgii, e.g.,][]{marzianietal16a,marziani13,denney12,mcLure2002}. The decomposition of the xA \aliii\ profiles shown in Appendix \ref{app_A} validates this assumption. In our xA sub sample, SDSS J003546.29-034118.2 is the source with the highest blueshift, -1010$\pm$80\kms, that is as well the quasar with the largest FWHM(\al)=4200$\pm$340\kms. Another example is SDSS J152314.49+375928.9, with a shift of -850$\pm$30\kms, and a FWHM(\al)=4000$\pm$400\kms. The other 9 xA \al \ profiles show  shifts between  -500 and 270\kms with FWHM(\al) between 2500 and 4000\kms. 
    In the \al\ case (Fig. \ref{fig:shift_fw}e), the largest ratio of shift/FWHM is $\lesssim$ -0.15 {which implies that the broadening effect might significantly affect the  \al\ line width}.
     
    \subsubsection{Population B}
    
    The highest value of \al\ FWHM observed in Pop. B objects is 6500 \kms. Only one object shows a FWHM of 4000 \kms, indicating that the \al\ line is broader than in Pop. A spectra but not as wide as the \cnl\ VBC. The median values are FWHM (\cnl\ BC)=5300$\pm$250\kms and FWHM(\cnl\ VBC)=8124$\pm$410\kms. 
    The inclusion of a VBC of \cnl\ in the red side of the Pop. B spectra is evident as seen in Figure \ref{fig:C31909}. In Pop. B objects we do not expect strong contribution of \feiii, so the residual seen in the fit with no VCB (Fig. \ref{fig:C31909}, upper panel) should be also part of \cnl. 
    
    The case of a VBC in \al\ and \si\ is not so evident. It could be that for \al\ and \si\ it is also necessary to add a VBC due to the large FWHM observed (up to $\sim$6500 \kms\ for \al\ and 5800\kms\ for \si). In the case of \al, the blending is not extremely severe, so we are able to efficiently deblend the BC (as seen in Figure \ref{fig:spec_fits}).
    The blend profile suggests that if \al\ VBC is present, it might be very weak, probably unresolved and not dominant in the emission line profile. A VBC might not be detected and lost in the spectral noise. 
   
    In Section \ref{ssec:popBC3_profile} we derive constraints on the   \al\ and \si\ VBC.

    \section{Discussion}\label{sec:discussion}
    
     \begin{figure*}
        \centering
        \includegraphics[width=\hsize]{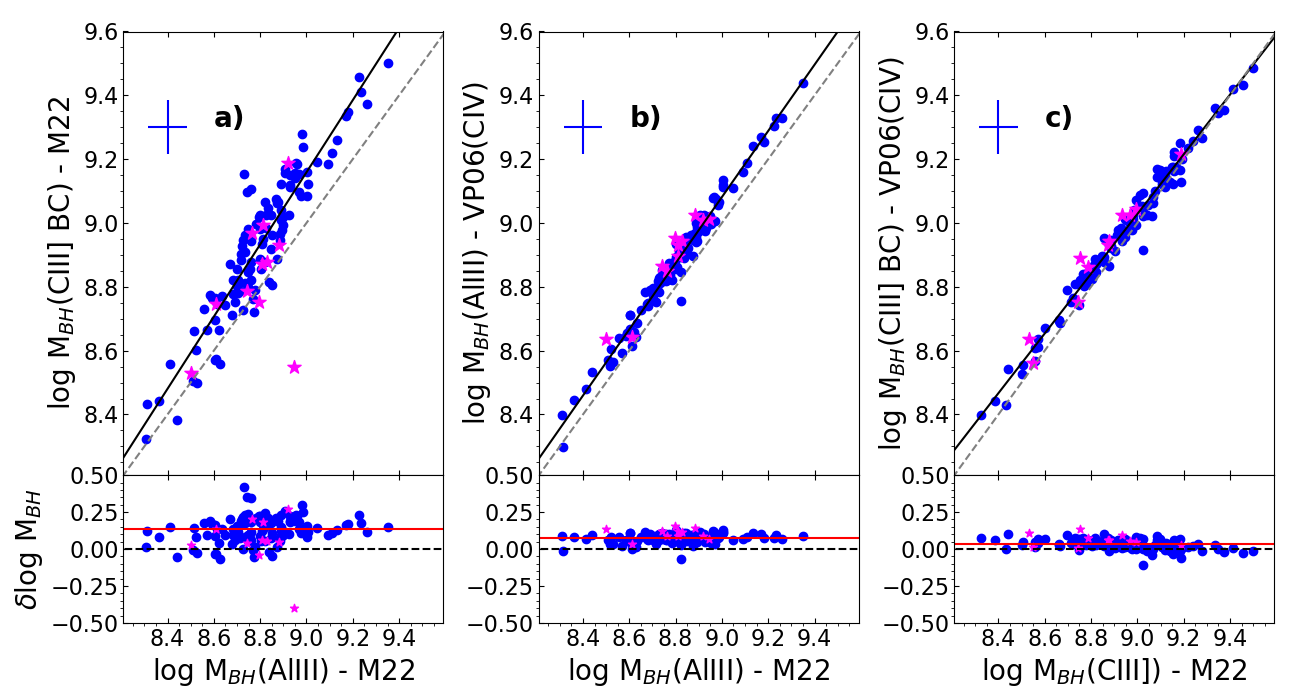} 
        \caption{\textit{Upper panels}: Log space of \mbh(\aliii) vs. \mbh(\ciii) computed with the scale relations of \citetalias{Marzianietal2022} and \citetalias{vp06} for the Pop. $\mathrm{\tilde{A}}$* sources [FWHM(\cnl) $\neq$ FWHM(\al)]: a) log \mbh(\al) vs. log \mbh(\cnl) both using \citetalias{Marzianietal2022} SR; b) log \mbh(\al) using Eq. \ref{SR1} by \citetalias{Marzianietal2022} vs. Eq. 7 by \citetalias{vp06}; c) log \mbh(\cnl) using Eq. \ref{SR2} by \citetalias{Marzianietal2022} vs. Eq. 7 by \citetalias{vp06}. \textit{Lower panels}: Residuals of each \mbh\ computation, $\delta$log\mbh. Colour-code as Figure \ref{fig:FWHM_Al3C3}. Dashed line traces the 1:1 relation; the filled line is the best fit obtained using the least-square method and the red line is the median value of $\delta$log \mbh. The uncertainties of the Pop. $\mathrm{\tilde{A}}$* sample are marked with a cross and calculated as the  sIQR. }
        \label{Al3-C3_SR}
    \end{figure*}
    
    The \mbh\  computations are closely related to the FWHM of prominent broad components and the underlying continuum. Therefore, the decomposition of the line profile becomes important to isolate the virialized component from other components, either coming from an outflow (blueshifted) or possibly coming from an inflow region (redshifted, \citealt{wang17}). We explore different \mbh\ estimators 
    (Sections \ref{ssec:mass_Redd} and \ref{ssec:Al3_lvir}) not as affected by shifts as \civ\ (Sect. \ref{ssec:civ_al3}). For Pop. B objects, we analyse the possibility of a  VBC in \al\ and \si\ (Section \ref{ssec:popBC3_profile}).

    \subsection{Virial mass and Eddington ratio estimates  with \aliii}\label{ssec:mass_Redd}

    Using the FWHM of two prominent lines of the 1900\AA\ blend, \al\ and \cnl, we compute the virial \mbh\ with two methods: (1) \citetalias{Marzianietal2022} (equations \ref{SR1}-\ref{SR2}) derived from the comparison of the FWHM of \hb\ with \al\ and \cnl; (2) \citeauthor{vp06} (\citeyear{vp06}, hereafter \citetalias{vp06}, Equations 5 and 7) that are based on the \hb\ and \civ\ line widths.  
    
    The \citetalias{Marzianietal2022} scaling laws take the form: 
     
    \begin{multline}
        \mathrm{log} M_{BH} (\mathrm{AlIII}) \approx  (0.580^{+0.035}_{-0.040})\mathrm{log} L_{1700,44} + \\ 
        2 \mathrm{log (\xi_{AlIII} FWHM(AlIII))} + (0.51^{+0.05}_{-0.05})
    \label{SR1}
    \end{multline}

    \begin{multline}
        \mathrm{log} M_{BH} (\mathrm{CIII]}) \approx  (0.645^{+0.045}_{-0.045})\mathrm{log} L_{1700,44} + \\
        2 \mathrm{log (\xi_{CIII]} FWHM(CIII]))} + (0.355^{+0.075}_{-0.045}).
    \label{SR2}        
    \end{multline}

    Here $\xi$ is a correction needed in the FWHM  (Section \ref{sssec:lw_popA}), for $\xi_{AIII}\approx$ 1 and $\xi_{CIII]}\approx$ 1.25.

    \begin{figure}
        \centering
        \includegraphics[width=\hsize]{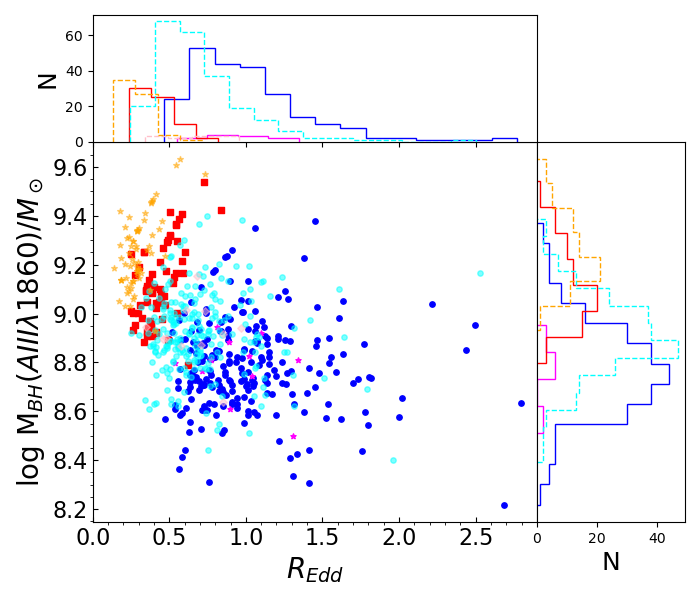}
        \caption{Distribution of log \mbh(\aliii) vs. \redd. Colour-code is as follows: the scale relations of \citetalias{Marzianietal2022} in blue, red and magenta and for the one of {\citetalias{vp06} (Equation 5)} in light blue, orange and light magenta for Pops. $\mathrm{\tilde{A}}$, B and xA, respectively.}
        \label{Al3_Mbh}
    \end{figure}
    
    \citetalias{vp06} use the optical continuum luminosity $L_{\lambda}$(5100\AA) along with FWHM(\hbbc) and the UV continuum $L_{\lambda}$(1350\AA) along with FWHM(\civbc). Considering the different continuum windows of these equations, we found it necessary to extrapolate the continuum obtained from {\tt specfit}\ to a wavelength as close as possible to 5100 \AA\ or 1350 \AA.
    {We applied  the ``surrogate''  lines \aliii\ and \ciii\ in both equations of \citetalias{vp06}. This means that we can directly compare the  \mbh(\al) and  \mbh(\cnl) from the scaling relations of \citetalias{Marzianietal2022} to the ones of \citetalias{vp06} using FWHM \aliii\ or \ciii\ in place of FWHM \civ\ or \hb.}
    
    The comparison of \mbh\ from equations \ref{SR1} and \ref{SR2} is presented in Figure \ref{Al3-C3_SR} along with the residuals of each set with $\delta$log\mbh = log\mbh(\al) - log\mbh(\cnl). In the Figure, the grey line indicates the 1:1 relation, and the black line is the best fit for the Pop. $\mathrm{\tilde{A}}$* sources using the least-square method. 
    For the \citetalias{Marzianietal2022} results (Figure \ref{Al3-C3_SR}a) the  equation is $\log$ \mbh(\cnl) $\approx (-1.017 \pm 0.095) + (1.131 \pm 0.085) \log$ \mbh(\al); the rms of the linear fit is 0.043 and the deviation from the 1:1 relation is 0.14. In the other two panels of Figure \ref{Al3-C3_SR}, we used Eq. 7 of \citetalias{vp06} by replacing the FWHM of \civonly\ with the FWHM of \al\ and \cnl.  Figure \ref{Al3-C3_SR}b shows the relation for \mbh(\al) using Eq. \ref{SR1} vs. \citetalias{vp06} Eq. 7. There is very good agreement between the estimations:  $\log$ \mbh(\al)$_{M22}$ $\approx (-0.249 \pm 0.045) + (1.037 \pm 0.065) \log$ \mbh(\al)$_{VP06}$ with a Pearson correlation of 0.98; the standard error (STD err) of the linear fit is 0.015 and the deviation from the 1:1 relation is 0.08.
    Figure \ref{Al3-C3_SR}c displays the relation of \mbh(\cnl) using Eq. \ref{SR2} vs. \citetalias{vp06} Eq. 7. In this case, we also observe a good agreement between each estimation. The resultant equation is log \mbh(\cnl)$_{M22}$ $\approx (0.609 \pm 0.085) + (0.935 \pm 0.065)$ log \mbh(\cnl)$_{VP06}$ with a Pearson correlation of 0.99; the STD err of the lineal fit is 0.012 and the deviation from the 1:1 relation is 0.04. The scatter of Figure \ref{Al3-C3_SR}b,c is smaller than in  Figure \ref{Al3-C3_SR}a due to fact that the equations of \citetalias{Marzianietal2022} are based on \citetalias{vp06}.
    
    As for the case of  Eq. 5 of \citetalias{vp06} by replacing the FWHM of \hb\ for the FWHM of \al\ and \cnl, we observed a discrepancy between estimations, a much larger scatter and systematic changes associated probably to the extrapolation of the continuum from $\approx$ 4000\AA\ to 5100\AA. We see a similar situation for \al\ and \cnl\ for both comparisons, both of them with a Pearson correlation of $\sim$0.8 and a resultant equation of log \mbh(\al,\cnl)$_{M22} \approx$0.6 log \mbh(\al,\cnl)$_{VP06}$. These findings proves that \al\ and \cnl\ are equivalent as virial broadening estimators for quasars (with a Pearson correlation coefficient of 0.93 for \citetalias{Marzianietal2022}) at intermediate $z$ from observations obtained from large surveys such as the SDSS. 
     
    In HILs such as \civ, the contamination of an outflow introduces a bias in the black hole mass estimations (see Sec. \ref{ssec:civ_al3}), because of over-broadening of the lines. The dynamic associated with a virialized system is different from the outflow that emerges from a system dominated by radiation pressure. A similar effect could be seen in the IIL \al\ for xA objects, but the contribution of the outflow is much lower than in the case of \civ.  This accounts for the good agreement found between the scaling laws for \civ\ by \citetalias{vp06} and the one of \citetalias{Marzianietal2022}.  The \civ\ scaling law of  \citetalias{vp06} was built around the assumption that the \civ\ FWHM was as good as \hb\ for virial mass estimation.  Using the \ciii\ (or \aliii), we use a line that is truly equivalent to \hb\ \citepalias{Marzianietal2022} and is not strongly affected by any non-virial component. 
    
    Several authors proposed that the Eddington ratio is driving the E1 MS \citep{marziani01,marziani03,shenho14,sun15}. Trends in \redd\ are also reflected in the X-ray properties \citep{boller96,wbb96,laor97}, \civ\ line profiles \citep{wills99,sulentic00b,sulentic07}, and in virial BH mass estimates using the width of the broad emission lines \citep{laor00,boroson02,dong11}. 
    The distribution of \mbh(\al) vs \redd\ is shown in Figure \ref{Al3_Mbh}. {Eddington luminosities} have been calculated based on masses obtained from the FWHM(\aliii) following the relation: $L_\mathrm{Edd}\approx 1.5\times10^{38} (M_{BH}/M_{\odot})$ [erg s$^{-1}$] \citep[e.g.,][]{NM10,netzer15}. The bolometric correction for $L_\lambda$ (1700\AA), 6.3   was taken from \citetalias{MS14}, and from \cite{richards06} the ones for 1350\AA\ (5.75) and for 5100\AA\ (10.3). 
    
    Fig. \ref{Al3_Mbh} shows that Pops. B and $\mathrm{\tilde{A}}$ appear to be segregated mainly on the basis of \redd: few Pop. B sources are in excess of \redd\ $\approx$ 0.5. The wide majority of Pop. $\mathrm{\tilde{A}}$ is constrained in the range $0.4 \lesssim$ \redd $\lesssim 1.2$. A minority of data points scatter in the range $1.2 \lesssim$ \redd $\lesssim 3$. If orientation plays a role, and if pole-on orientation leads to narrower lines \citep{mcLure2002,collin06,decarlietal11,mejia-restrepoetal17,mejia-restrepoetal18a}, the \mbh\ might be severely underestimate and \redd\ overestimated. A similar effect has been already seen in the \mbh\ vs. luminosity diagram \citep[e.g.,][]{marzianietal06}.     
    
    \begin{figure*}
        \centering
        \includegraphics[width=0.4\textwidth]{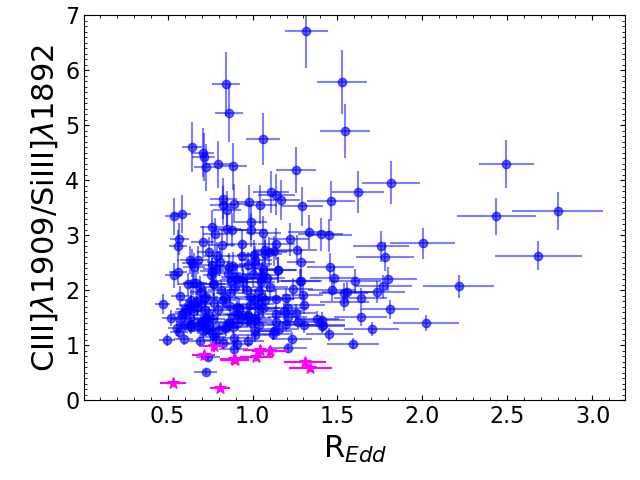}
        \includegraphics[width=0.4\textwidth]{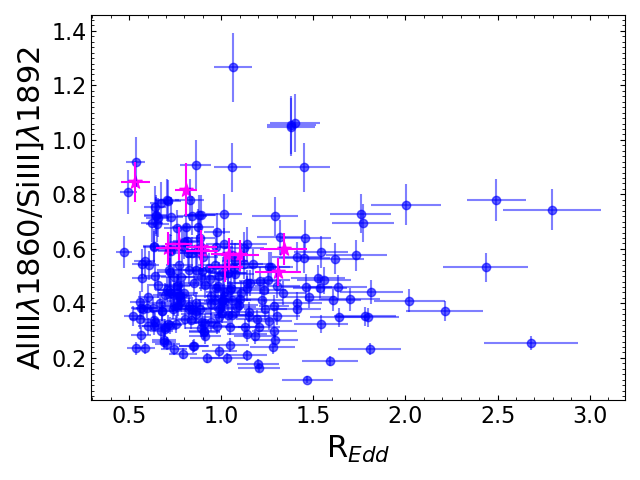}
        \caption{Distribution of the UV diagnostic ratios of \citetalias{MS14}: \aliii/\siiii\ (left) and \ciii/\siiii\ (right) with respect \redd\ (\al). for Pop. $\mathrm{\tilde{A}}$ and xA sources only}
        \label{fig:UVratios_Redd}
    \end{figure*}

    We also analysed the parameter space of the UV diagnostic ratios $vs.$ \redd\ plane for the Pop. $\mathrm{\tilde{A}}$ sources, however, we found no strong correlations (Figure \ref{fig:UVratios_Redd}). Table \ref{tab:T1} reports the median values of the line ratios. The condition \ciii/\siiii $\lesssim$1 seems to be sufficient to identify xA quasars. Yet, xA quasars are not associated with the highest \redd\ (see also Figure \ref{Al3_Mbh}).  This might be a consequence of an overbroadening due to an outflow component, increasing \mbh\ for xA sources and therefore lowering \redd. The median excess in the virial luminosity (rose bars in Fig. \ref{dlogLvir}) suggests a median \mbh\ overestimate $\delta \log $\mbh $\approx$ 0.2. Pop. $\mathrm{\tilde{A}}$  sources with \redd $\gtrsim 2$ might be oriented preferentially pole-on, leading to a strong underestimate of the black hole mass and hence to an overestimate of the Eddington ratio {(as observed in Figure \ref{Al3_Mbh}, the \redd\ is up to $\sim$2.5)}.  
    
    \subsection{Virial luminosities and outflow broadening}\label{ssec:Al3_lvir}

    The physical parameters of xA quasars are correspondingly extreme, with maximum radiative output per unit of mass close to their Eddington limit. This condition is predicted by accretion disk theory at high accretion rates: radiative efficiency low and Eddington ratio saturating towards a limiting value (\citealt{mineshigeetal00,sadowski11,sadowski14}, and references therein).  We also expect that the intensity ratios of the intermediate and low ionisation lines in xA quasars  remain almost the same: only the line width increases with luminosity \citep{negrete12,negrete13}. The spectral invariance with luminosity implies that the radius of the emitting regions should rigorously scale as $L^{1/2}$; if not, the ionisation parameter $U$ should change with luminosity \citep{m20_iau}. Putting together these considerations: (1) $L/L_{Edd} = const.$, (2) $r \propto L^{1/2}$, together with (3) the definition of Eq. \ref{eq:mbh} $M_{BH} \propto rFWHM^2$, we obtain a relation linking luminosity and line width, known as the ``virial luminosity equation'' \citepalias{MS14}: 
    
    \begin{equation}
        L_{Vir} = L_{0} \cdot (\mathrm{FWHM})^4_{1000}\ \mathrm{erg\ s}^{-1}
    \label{Lvir}
    \end{equation}
    
    \noindent where $L_0$ = 7.88 $\times$ 10$^{44}$ and the FWHM is of the virialized BC in units of 1000 \kms\ (see eq. 6 in \citetalias{MS14} for the complete derivation). The FWHM of \al\ is the adopted virial broadening estimator for our work. We calculated \lvir\ for the 11 xA sources and the average and median values are reported in Table \ref{tab:T1}. However, there are two effects that can significantly affect the luminosity estimations: an outflow that broadens the virial component, and an orientation effect that narrows it.
    
    \begin{figure}
        \centering
        \includegraphics[width=0.35\textwidth]{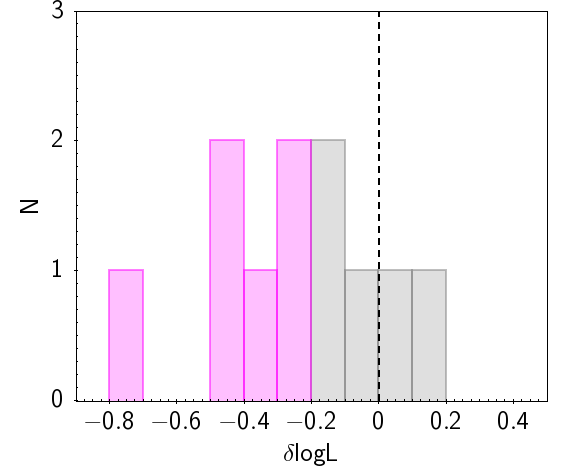}
        \caption{Distribution of the luminosity difference between concordance cosmology determinations and virial ones, $\delta \log L$. Colour-code: Grey bars are the 5 objects that lie under our uncertainties of $\delta \log L$ and the 6 objects in magenta are the ones that shows an outflow effect ($\delta \log L \lesssim -0.2$).}
        \label{dlogLvir}
    \end{figure}
    
    For xA sources, the \al\ shift/FWHM ratio can be up to $\sim -0.1$ (Fig. \ref{fig:shift_fw},e). The dominant effect on our sources may be due to an outflow since the virial luminosities are larger than the cosmological ones. We have 11 extreme sources, of which six objects have $\delta \log  L = \log $ \lbol\ $ - \log$ \lvir $\approx - 0.2$ (magenta bars in Figure \ref{dlogLvir}). SDSS J003546.29, J152314.49 and J023055.54 showed a difference between the cosmological and virial luminosities under -0.2 and are sources with an \al\ blueshift. SDSS J003546.29 having a $\delta$log$L$=-0.75 and shift \al $\approx -1000$ \kms.
    
    \subsection{\civ\ and \aliii\ inter-comparison}\label{ssec:civ_al3}

    \begin{figure*}
        \centering
        \includegraphics[width=0.475\textwidth]{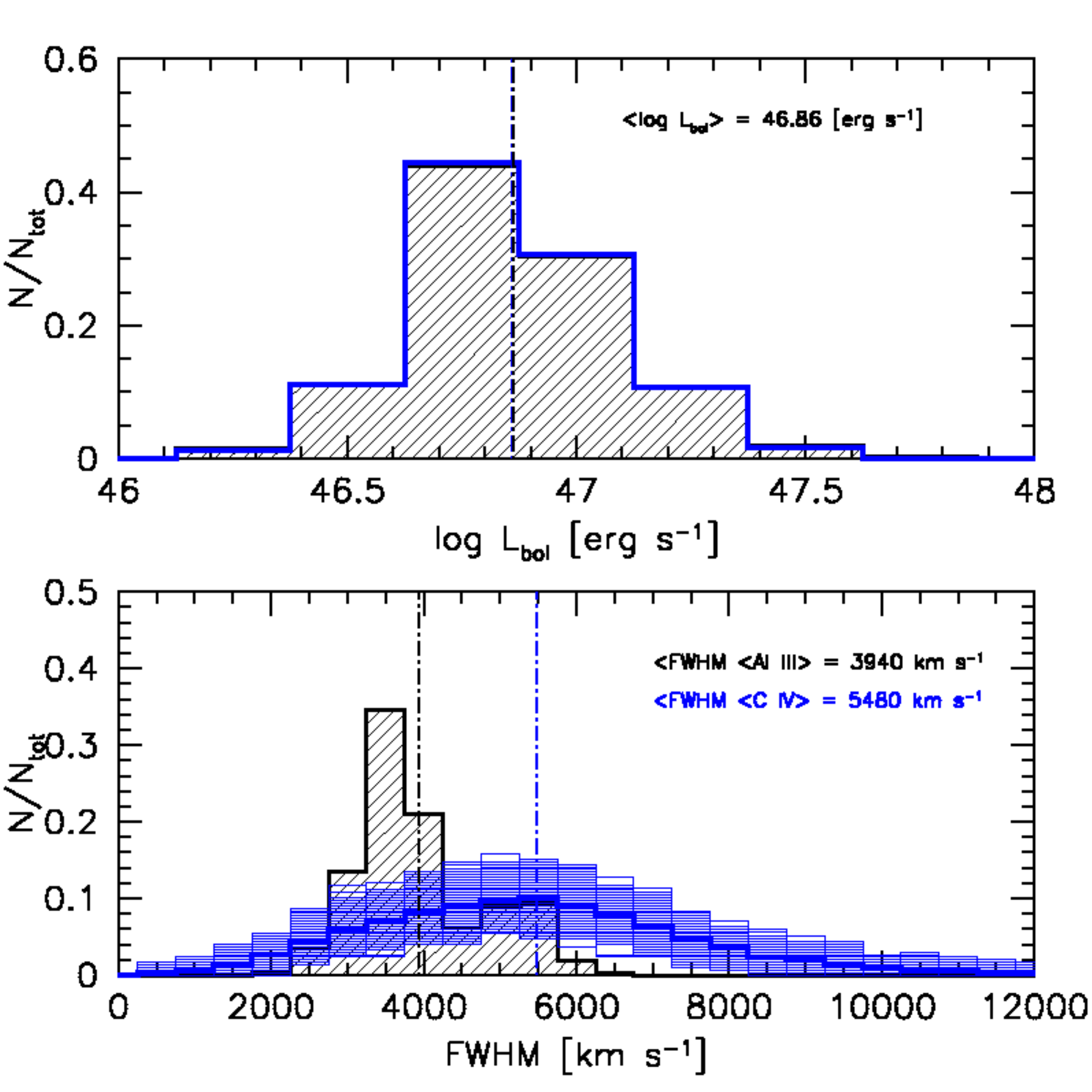}
        \includegraphics[width=0.475\textwidth]{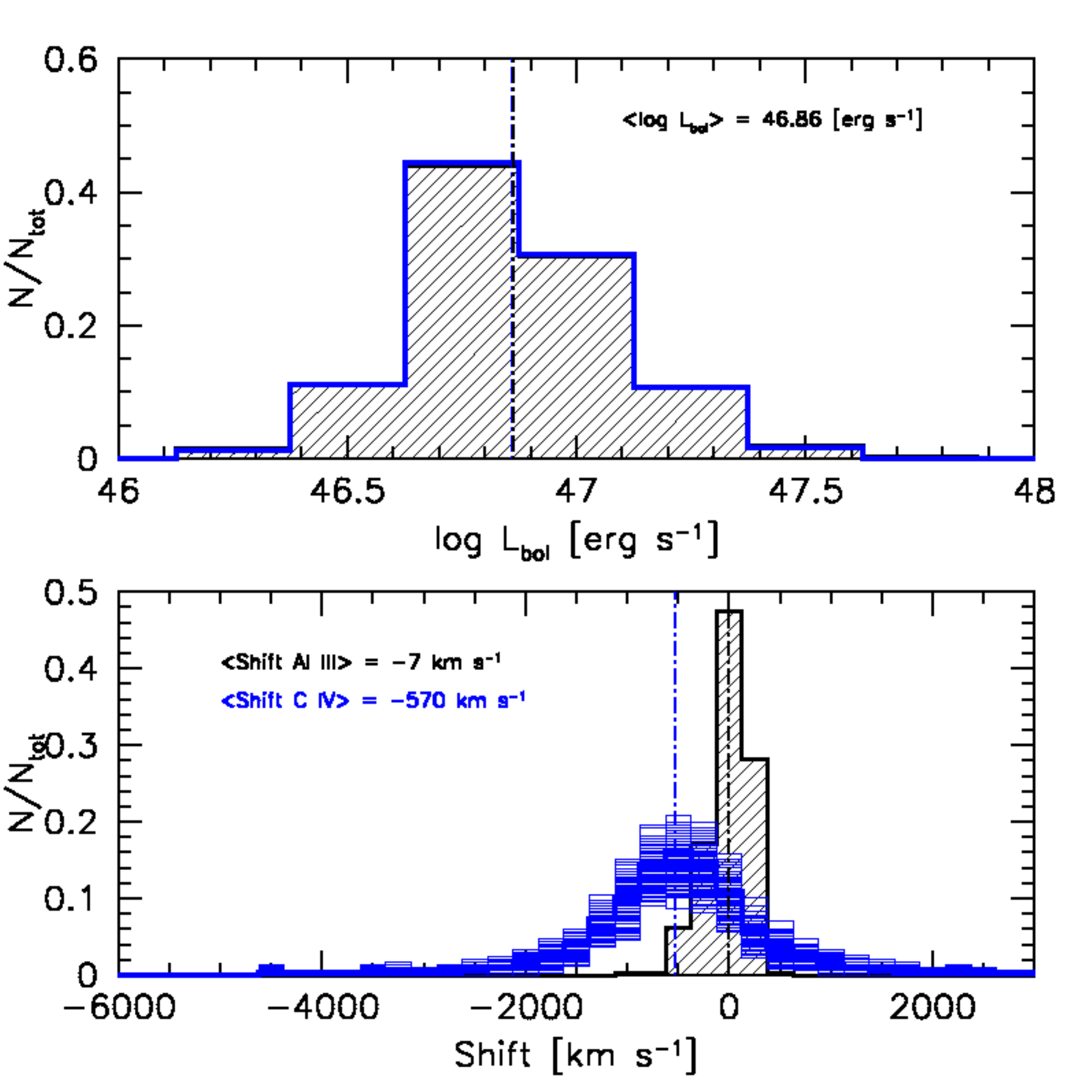}
        \caption{\textit{Left}: inter-comparison between the \aliii\ of the present sample (hatched histograms) and the {\civ\ FWHM} distribution of the sample of \citet{shen11}, for matching luminosity distributions (top panel).  The bottom panel thin blue lines show the binned distributions of  bootstrap replications of the \citet{shen11} data; the thick blue line is their average. \textit{Right}: same for peak shift. 
        \label{fig:al3c4}}
    \end{figure*}

     The blueshift of \civ\ is usually evidence of strong outflows (e.g. \citealt{richards11}) that, most likely, results from the presence of a radiation line-driven accretion-disc wind (\citealt{gallagher15} and references therein). Therefore, a prominent blue component over the line profile is expected, especially at high/intermediate redshift  \citep{mla18}. However, when compared to samples with lower redshift, Pop. A sources at intermediate redshifts tend to show broader and more blueshifted components of \civonly. They are indicative of wind activities surrounding the central region \citep{deconto21}. Therefore, black hole masses based on the FWHM(\civonly) emission line can be overestimated by a factor of 4-5 at large blueshifts and are biased due to this non-virial component \citep{coatman16,denney12}.
     
     A sub-sample from \cite{shen11} was extracted  to compare the \civ\ profile with our \al\ sample. The criteria used were: luminosity distribution consistent with the one of the  \aliii\ sample. The results are shown in Fig. \ref{fig:al3c4}, for the FWHM and peak shift of the line.  
     The blue lines show the distribution of bootstrap replications of the bolometric luminosity distribution, for $300$\ objects pooled out of $\sim 50,000$\ sources from the \citet{shen11} catalogue. The luminosity distributions of the bootstrapped samples  overlay the one of the present samples since the source from \citet{shen11} where pooled preserving the relative prevalence of the \aliii\ luminosity been (shaded histogram in Fig. \ref{fig:al3c4}). For both \civ\ FWHM and shift, the distribution imply extremely high probabilities that they are not consistent. In particular, the FWHM \civ\ appears to be systematically broader than the one of \al\ by $\approx 1500$ \kms. While the \al\ blueshifts are modest (within $|\delta v_\mathrm{r}| \lesssim 500$ \kms, and the distribution appears centred at rest frame and only slightly skewed to the blue, the \civ\ line presents  a systematic blueshift by  $\approx -600 $ \kms. 
     
     The \civ\ shift has been analysed with respect to \hb\ \citep[e.g.,][]{leighlymoore04,marziani10,sulentic17,vietrietal18}, and interpreted as a strong wind contribution affecting the \civ\ profile in the form of an excess blueshifted emission. The same difference has been revealed in a detailed same-source, inter-line comparison between \al\ and \civ\ in $\approx 20$ xA sources \citep{martinez-aldamaetal18}.  Fig. \ref{fig:al3c4} provides a statistical confirmation that the \civ\  blueshifted emission is broadening and shifting considerably the \civ\ profile with respect to the one of \aliii. 
     
     \subsection{A model for the Population B \ciii\ profile}\label{ssec:popBC3_profile}
   
     \begin{figure*}
        \centering
        \includegraphics[width=0.45\textwidth]{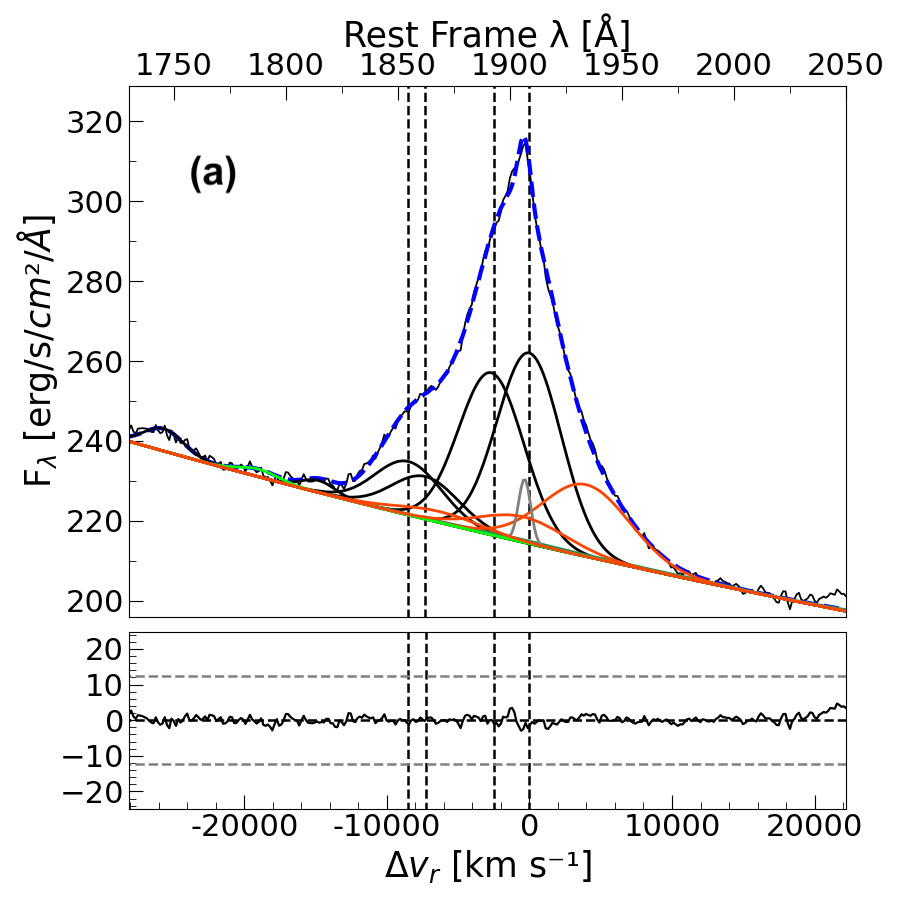}
        \includegraphics[width=0.48\textwidth]{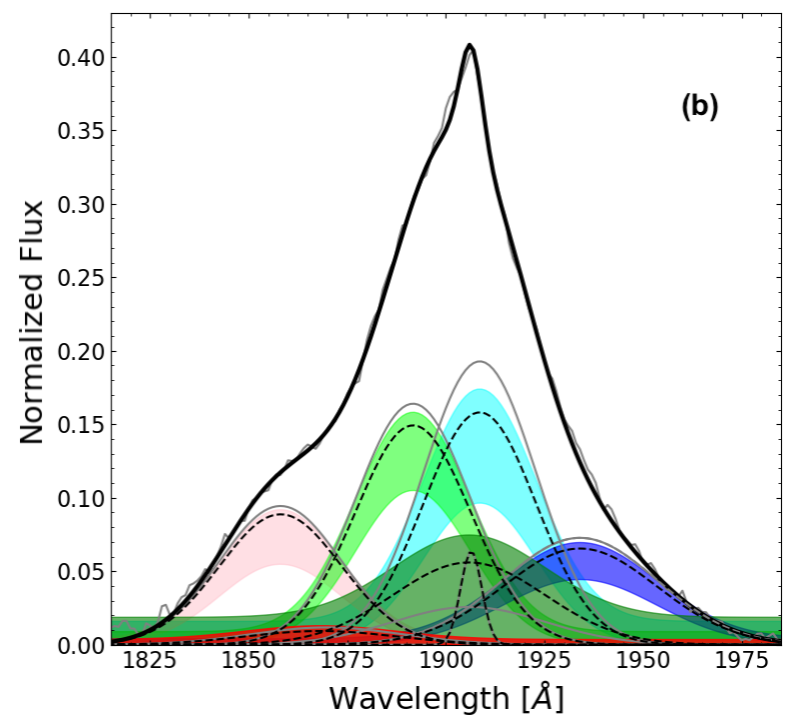}
        \includegraphics[width=0.5\textwidth]{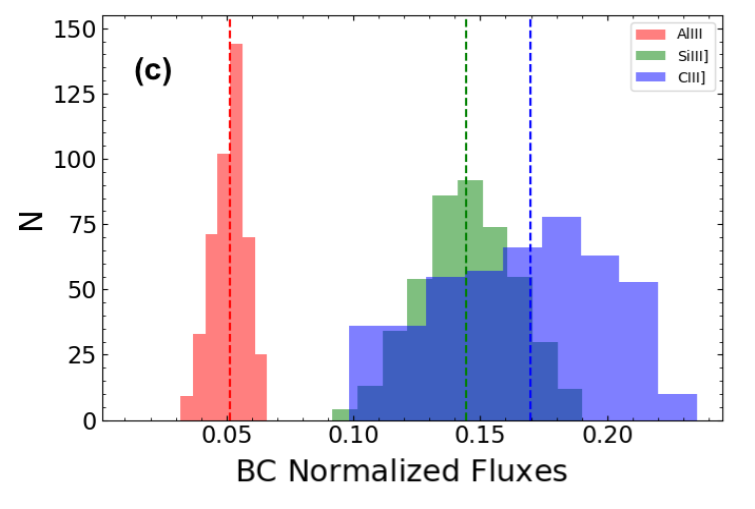}
        \includegraphics[width=0.4\textwidth]{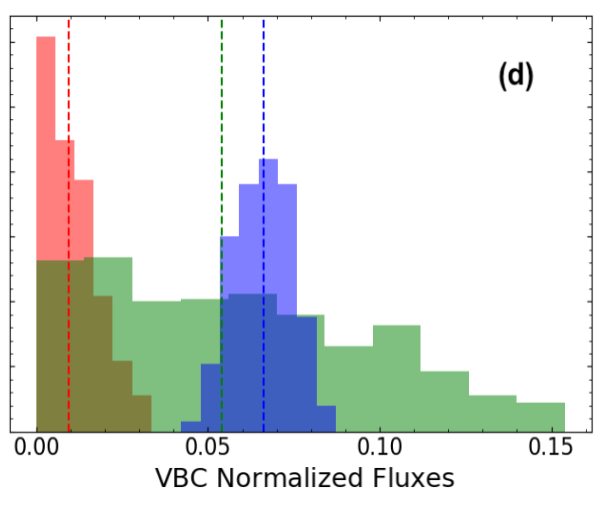}
        \caption{ \textit{Upper left}: \texttt{specfit} model (M3) of the composite spectra of Pop. B sources with VBC (in red) added to the profile of \al, \si\ and \cnl. Abscissa scales are rest-frame wavelength in Å. Ordinate scale is the specific flux in units of 10$^{-17}$\ergss\ cm$^{-2}$ \AA$^{-1}$. \textit{Upper right}: Synthetic model (M4) with BC+VBC in all three lines with the median value (dashed line) and their uncertainties regions of each profile in light and dark tone: red, green and blue for \al, \si\ and \cnl\ BC - VBC respectively. Grey lines are the M3 flux values as the initial condition. \textit{Lower panels:} Gaussian distributions of the BC - VBC fluxes obtained with a million random iterations of M3 with values that satisfies the condition of \chisq(M4)/\chisq$_{min}$(M4) < F(2$\sigma$).} 
        \label{fig:popB_VBC}
    \end{figure*}
   
    Is our model of the 1900 blend adequate? This work has convincingly shown the need to include a VBC to account for the \ciii\ profile. Even if the \siiii\ is heavily blended, the fits detect an emission peak between \al\ and \ciii, implying that the \siiii\ core component is always prominent. There is no doubt about the existence of a core component (i.e., the BC) for \al. However \al\ is the weakest line in the blend, and some VBC emission could be lost in noise. 
    
    We can analyse the expectation of VBC emission considering that (1) the velocity field of the emitting regions is predominantly virial in Pop. B sources, as established by early reverberation mapping studies \citep{petersonwandel99,petersonwandel00},  and that (2) the VBC is a heuristic representation of the innermost part of the BLR. 
    {This VBC could be physically associated to inflowing gas \citep{wang17,giustini19} or due to an effect of the gravitational redshift \citep{netzer77,zheng90,corbin95,Liuetal2017,mediavilla18}. This component has been observed in sources with masses in the range 10$^8$--10$^{10}$ M$_{\odot}$ \citep{bon15}, comparable to the Pop. B mass values of the present sample, with  mean $\log$\ \mbh= 9.1 [M$_{\odot}$].  }
   
   In this way, three empirical approaches are in order: (1) a fit with only the BCs (M1); (2) a fit in which one VBC is assumed for \ciii\ only (as done for all Pop. B sources, M2); (3) a fit in which 3BCs and 3VBCs are introduced, with restriction on consistent shifts and widths for the BCs and VBCs (M3) as seen in Figure \ref{fig:popB_VBC}a. This last option implies 8 free parameters (peak shifts and wavelengths are locked). It is probably the most appropriate in physical terms, but is very difficult to implement for individual sources. 
    The three fits were carried out on an average composite for all Pop. B sources, and the resulting ratios are reported for the three models in Table \ref{tab:T2}. 
    The basic inference is that the BC ratios \aliii/\siiii\ and \siiii/\ciii\  remain consistent if different models are assumed. A second result is that the VBC/BC ratio is $<$1 and $\ll$1 for \siiii\ and \aliii, respectively. 
    
    The physical implications for the line emitting regions have been analysed using CLOUDY 17.02 \citep{ferlandetal17} arrays of photoionisation simulations computed for an unrelated work \citep{sniegowska21}. Briefly, they assumed a standard AGN continuum implemented in CLOUDY, solar metallicity, canonical value of the Hydrogen column density (10$^{23}$ cm$^{-2}$), no micro-turbulence. They were computed for an  array  of ionisation parameter and density covering the ranges (in log) $-4.5 - 1$, and $7 - 14$\ [cm$^{-3}$], respectively.  
    
    \begin{figure}
        \centering
        \includegraphics[width=0.8\hsize]{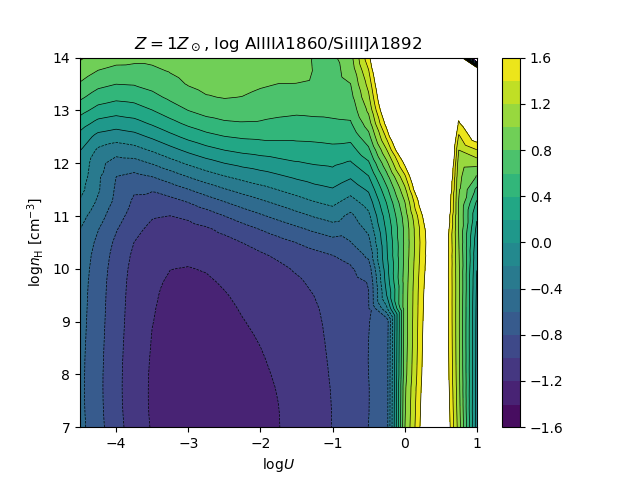}
        \includegraphics[width=0.8\hsize]{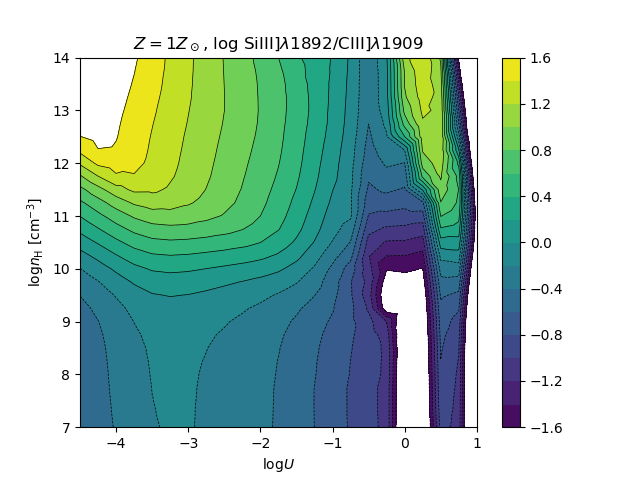}
        \includegraphics[width=0.8\hsize]{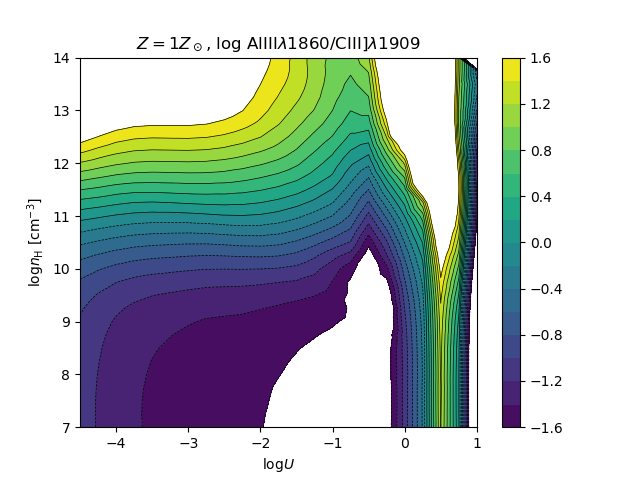}\\
        \caption{From top to bottom, maps of intensity ratios  as a function of ionisation parameter and Hydrogen number density: $\log$ \aliii/\siiii, $\log$ \siiii/\ciii\ and $\log$ \aliii/\ciii.  \label{fig:nu}}
    \end{figure}
    
    Fig. \ref{fig:nu} shows the behaviour of the ratios \aliii/\siiii, \siiii/\ciii\ and \aliii/\ciii\ as  a function of ionisation parameter and hydrogen density. The typical ratios measured on the Pop. B sample and on the composite spectrum indicate that the BC is emitted in a region of moderate density and high ionisation ($U \sim 10^{-1}$, \nh $\sim 10^{11} $ cm$^{-3}$). Similar values are found for the VBC. 
    
    To push the analysis one step forward we consider the ratios estimated for Model 4 (a synthetic model with VBC and BC for the three lines, Table \ref{tab:T2}) as seen in Figure \ref{fig:popB_VBC}b. This model was made using M3 flux values as an initial condition, then let the model adapt to the better statistical values (along very well defined physical ranges) varying the fit with a million random iterations. {
    The significance of \chisq\ variations is described by F statistics appropriate for ratios of \chisq\ values \citep{bevington}, F = $\chi^2_{\nu}/\chi^2_{\nu,min}$, with degrees of freedom $\nu\approx$ 165. F $\approx$ 1.30 provides 2$\sigma$ conﬁdence ranges on the parameters. The final fluxes obtained for the M4 BC and VBC fluxes are the ones that satisfies F within a 2$\sigma$ conﬁdence level (F(2$\sigma$), Figure \ref{fig:popB_VBC}b)}. In the Figure \ref{fig:popB_VBC}b we can see the distributions in light-dark red, green and blue for \al, \si\ and \cnl\ BC - VBC, respectively. The dotted line of each distribution are the median values used in the UV ratios in Table \ref{tab:T2}.
    {Our synthetic models that satisfied the condition of the F(2$\sigma$) showed Gaussian distributions for the BC and VBC fluxes centred in one very well defined value}, except for \si\ VBC (Figure \ref{fig:popB_VBC}c,d). The median values are marked for each distribution and corresponds to the dotted black lines of Figure \ref{fig:popB_VBC}b.
    
    \begin{table*}[h]
        \centering
        \caption{ UV diagnostic ratios of \citetalias{MS14} values obtained on an average composite for all Pop. B sources with different considerations for the \al, \cnl\ and \si\ profile lines.} \label{tab:T2}
        \begin{tabular}{@{}ccccc@{}}
        \toprule
        \toprule
                                   & M1 & M2 & M3 & M4 \\ 
                                   & (a) & (b) & (c) & (d) \\ \midrule
        \al\ BC/ \si BC & 0.62$\pm$ 0.19 & 0.65$\pm$0.39 & 0.61$\pm$0.06 & 0.37$\pm$0.73\\
        \cnl BC/ \si BC & 1.34$\pm$0.25 & 1.28$\pm$0.32 & 1.16$\pm$0.07 & 1.39$\pm$0.79  \\
        \al\ BC/ \cnl\ BC & 0.46$\pm$0.14 & 0.51$\pm$0.28 & 0.53$\pm$0.08 & 0.27$\pm$0.78 \\
        \al\ VBC / \si VBC  & -  & -  & 0.42$\pm$0.06 & 0.19$\pm$3.95 \\
        \cnl VBC / \si VBC  & -  & -  & 2.87$\pm$0.07 & 1.65$\pm$2.23 \\
        \al\ VBC+BC / \si BC+VBC  & - & -  & 0.58$\pm$0.12 & 0.32$\pm$1.10  \\
        \cnl VBC+BC / \si BC+VBC & -  & -  & 1.45$\pm$0.14 & 1.46$\pm$0.90 \\
        \al\ VBC / \al\ BC & - & - & 0.15$\pm$0.03 & 0.17$\pm$2.45 \\
        \si\ VBC / \si\ BC & - & - & 0.21$\pm$0.01 & 0.33$\pm$2.23 \\
        \cnl\ VBC / \cnl\ BC & - & - & 0.51$\pm$0.01 & 0.39$\pm$0.79 \\
        \chisq  & 0.11994 & 0.01146 & 0.01084 & 1.05202 \\
        FWHM \cnl\ BC & 5356 &5075 & 5299 & 5299 \\
        FWHM \si\ BC & 5704 & 5416 & 5444 & 5444\\
        FWHM \al\ BC & 5708 & 5578 & 5674 & 5674 \\
        FWHM VBC & - & 7024 & 7128 & 7128 
        \\ \bottomrule
        \end{tabular}
        \tablefoot{
        \tablefoottext{a}{Model 1: Fit with only BCs as shown in Figure \ref{fig:C31909} (top panel).}
        \tablefoottext{b}{Model 2: Fit with only a VBC in \cnl\ as described in section \ref{sec:multifitting} (Figure \ref{fig:spec_fits}, bottom panel).}
        \tablefoottext{c}{Model 3: Fit with BC + VBCs in all three lines as Figure \ref{fig:popB_VBC}a.}
        \tablefoottext{c}{Model 4: Synthetic model fit with BC+VBC in all three lines as Figure \ref{fig:popB_VBC}b. }
        }
    \end{table*}

\begin{figure}
        \centering
        \includegraphics[width=1\hsize]{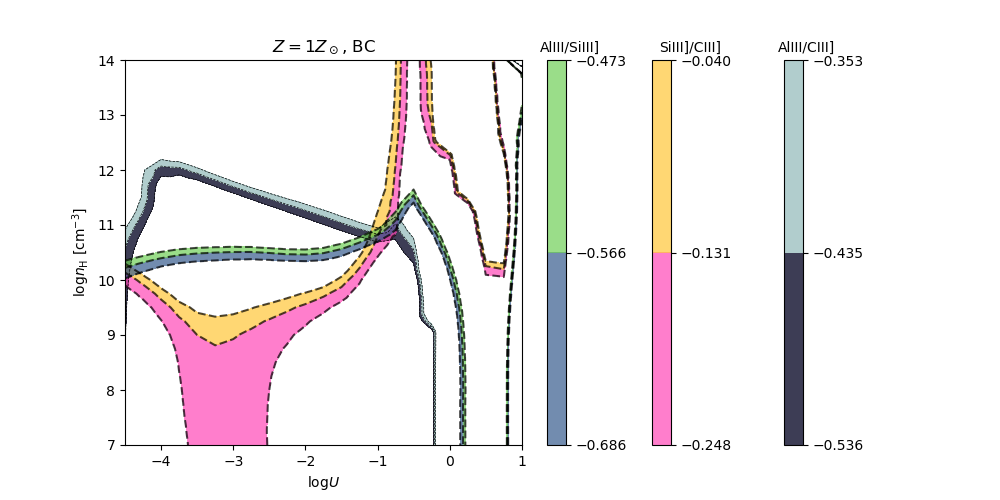}\\
        \includegraphics[width=1\hsize]{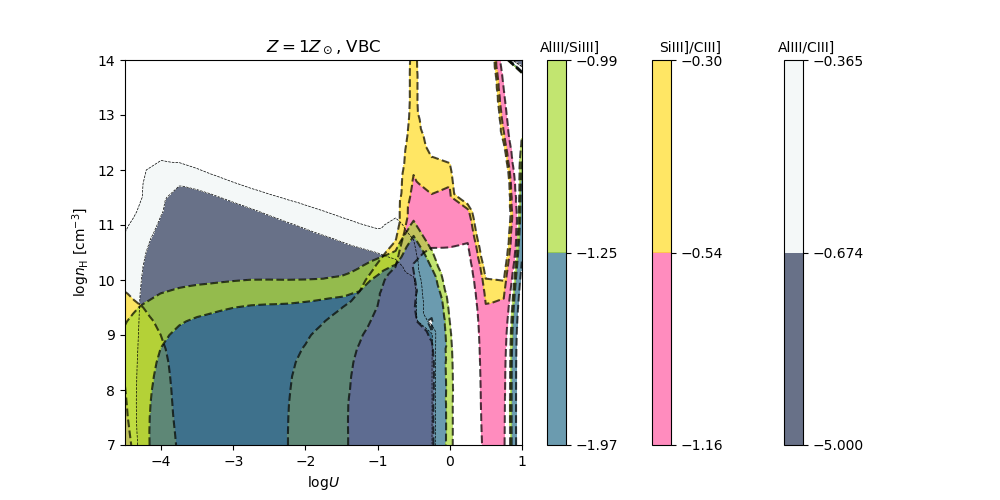}\\
        \includegraphics[width=1\hsize]{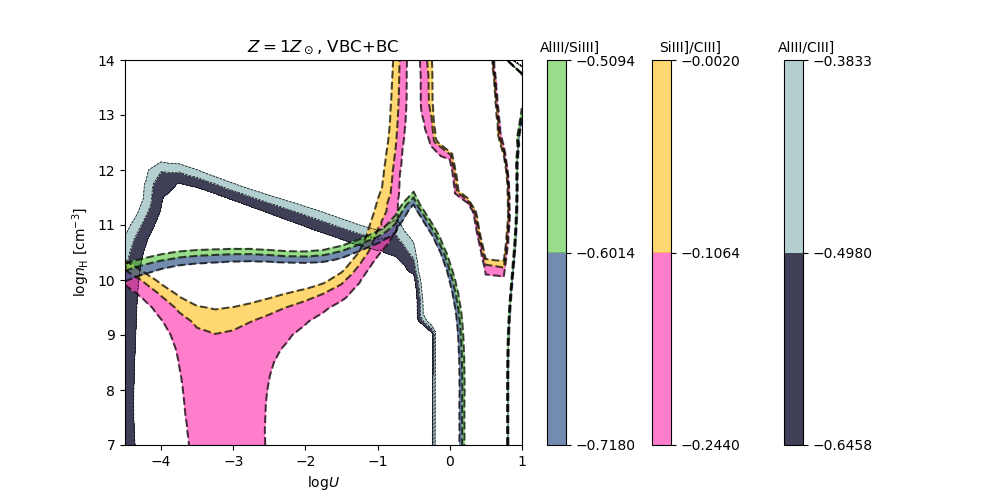}\\
                \caption{Isophotes tracing the loci of the parameter plane ($\log U$, $\log $ \nh) consistent with the observed intensity ratios \aliii/\siiii, \siiii/\ciii\ and \aliii/\ciii\ (shown in log scale in the Figure). The crossing region defines the ($U$, \nh) parameter range  consistent with the values of the three ratios. \textit{Up}: BC only, \textit{middle}: VBC, \textit{bottom}: BC+VBC.  }
        \label{fig:crossing}
    \end{figure}

    Fig. \ref{fig:crossing} shows the regions in the parameter plane that are consistent with the  ratios built from the three lines in the blend. For the BC there is a very well defined ($U$, \nh) region where the three ratios cross; it means that in this region the values of  ($U$, \nh) are able to reproduce the observed ratios: $\log U \sim -1.00^{+0.12}_{-0.28}$, $\log$ \nh $\sim 10.78^{+0.28}_{-0.08}$ cm$^{-3}$. Similar values are derived if the BC and VBC are added together: $\log U \sim -1.03^{+0.31}_{-0.19}$, $\log$ \nh $\sim 10.72^{+0.19}_{-0.15}$ cm$^{-3}$, where the uncertainty range has been set from the $\pm 1 \sigma$ uncertainties for the individual line ratios. These values indicate moderate density and fairly high ionisation as expected for Population B sources \citep{negrete13,negrete14}. We warn that our single zone model is certainly not adequate to represent the complexity of the emitting region. In the case of Pop. B, there is a most likely a range of densities, column densities, and ionisation parameters that makes the locally-optimised cloud model \citep{baldwinetal95,koristaetal97} the most appropriate.  
    
    The case of the VBC  deserves special attention. We are dealing with emission that is well-constrained only in the case of \ciii, and that is presumably much weaker than the corresponding BC emission for \aliii\ and \siiii. Using  the ratios of the best fit to the synthetic spectrum we obtain 
    $\log U \sim -0.72^{+0.27}_{-1.53}$, $\log$ \nh $\sim 10.27^{+0.19}_{-\ldots}$\ [cm$^{-3}$]. Due to the very low \siiii/\ciii\ and \aliii/\siiii\ intensity ratios derived from the fit (actually consistent with 0 within the uncertainties), the lower limit of both $U$\ and \nh\ are practically unconstrained. 
    
    Further clues can be obtained considering that the line width should be inversely proportional to the square root of the radius of the emitting region: FWHM $\propto 1/r^{\frac{1}{2}}$, as per Eq. \ref{eq:mbh}. The BC over VBC FWHM ratio is $\approx 0.8$, implying that the ratio of the radii should be $\approx 0.64$. For constant \nh, this would imply an increase in $\delta \log U \sim +0.38$. The diagnostics from the blend for the VBC are very poor (a more refined analysis should involve measurements of at least \civ\ and \heiiuv\ which are not covered in the spectra of our sample). However, there is slight increase in $U$ moving from the BLR to the VBLR\ that does not suggest any gross inconsistency with the virial approach. 
    
    Fig. \ref{fig:nu} indicates that 
    that, for a likely value of \nh $\sim 10^{11}$ cm$^{-3}$, moving from $\log U \sim -1$\ toward higher $U$\ values, we may expect a lowering of the \siiii/\ciii\ ratio to level that may make the VBC of \siiii\ difficult to detect. At the same time, the \aliii/\siiii\ ratio might increase sharply for $\log U \gtrsim -0.5$, reaching $\approx 1  $\ for $\log U \gtrsim -0.0$. At that value of the ionisation parameter, the \aliii\ VBC should be stronger than the one of  \ciii.  The composite profile (Fig. \ref{fig:popB_VBC}) disfavours the possibility that the \aliii\ VBC and in turn the  $U$\ could be that high: in Pop. B, the intensity of \aliii\ is   lower than the one of \siiii\ \citep{bachev04,kuraszkiewiczetal04,liraetal17,liraetal18}. ionisation parameter  $\log U \sim -0.25$\ might be a possibility entailing \aliii/\ciii$\approx 0.3$, \aliii/\siiii $\approx$2, and \siiii/\ciii $\approx$0.15 (assuming \nh $= 10^{11}$ cm$^{-3}$).
    
    In summary, these consideration justify the neglect of a \siiii\ VBC. The possibility of an \aliii\ VBC dominating the \aliii\ cannot be excluded if the ionisation parameter is high. Against this prospect goes the empirical fact that the FWHM of \aliii\ and \siiii\ BC are consistent. 
    
\section{Summary and conclusions}\label{conclusions}

    The present investigation has shown that the intermediate ionisation lines are little affected by outflows, and that the  \al\ and \cnl\ are equivalent (with some caveats) as virial broadening estimators for quasars, providing a suitable tool for \mbh\ estimates at intermediate $z$ from observations obtained from a big survey such as the SDSS. More in detail, the results of the present investigation can be summarised as follows:

    \begin{itemize}
    \item We carried out a redshift correction of the sample spectra using the narrow LIL \o2\ rest-frame wavelength. The rest-frame of the 1900\AA\ blend corrected in redshift for the \o2\ line proved the effectiveness of \al\ and \cnl\ as rest-frame estimators.
    \item We subdivided the sample into Population A and B. We took into account the luminosity-dependent relation of \cite{sulentic17}. Within Pop. A, extreme Population A have been considered separately. 
    \item Pop. A quasars constitute 78\% of the sample with 11 sources classified as extreme accretors, and 22\%\ as Pop. B quasars out of a sample of 309 objects. We observed a bias against  high Eddington ratio sources due to the absence of \o2\ in the spectra, and  a Malmquist-type bias at low Eddington ratio.
    \item Applying the \texttt{specfit} routine of {\fontfamily{lmtt}\selectfont IRAF}, we were able to fit the most prominent emission lines of the 1900\AA\ blend simultaneously, proving that we can measure widths of \al\ and \cnl\ (and \si) even if they are blended. 
    \item In terms of tendencies observed for each population: Pop. $\mathrm{\tilde{A}}$ has shown no shifts in the median sub-samples from the \al\ profile, Pop. B shows symmetric shifts around 0; only Pop. xA show a median blueshift of -300 \kms\ indicating a mixture of two unresolved components: a virialized plus an outflow component. The xA sub-sample showed an \al\ shift/FWHM ratio $\sim$ 10 to 15\%, indicating that the displacement significantly affects the line width.
    \item The virial black hole mass estimations of our sample using the FWHM (\al) are consistent with the ones obtained with FWHM(\cnl), using the \citetalias{vp06} and \citetalias{Marzianietal2022} scale relations.
    \item Our xA sample (11 quasars) showed a broad consistency between the cosmological and virial luminosity computed with the equation derived by \citetalias{MS14}; however, an excess in the virial luminosity with respect to the concordance one  indicates that in this case the \al\ width is affected by a non-virial broadening.
    \item The comparison of the \al\  and a large numbers of \civonly\ bootstrapped  samples extracted from \cite{shen11} matching the luminosity distribution showed that the 
    \al\ and \civonly\ FWHM and shift distributions  differ fundamentally, in the sense that the \civ\ FWHM and shift distributions are much broader than those of \al. Shift amplitudes of \civ\ are a factor $\sim$ 10 larger than the \al\ ones.  
    \item  Our single zone model proved that there is a very well defined region in the log plane $n_{H},U$ for the BC and BC+VBC models for the composite Pop. B spectra. As for the case of only VBC, $n_{H},U$ are not fully constrained. Nonetheless, the appearance o -f the blend and the intensity ratios of the components are consistent with the predominance of a virial velocity field, with a stratification of emission properties.
    \end{itemize}
    
     {In conclusion, we can use the IILs \al\ and \cnl\ as a reliable surrogate mass estimator  for Pops. $\mathrm{\tilde{A}}$ and B objects. Highly accreting quasars show  smaller blueshifts (on average $\ll$ 1000\kms) compared to the ones observed in \civonly, and the   method discussed in this paper may provide slight \mbh\ over-estimates by a factor $\lesssim 2$ {as described in Secs. \ref{ssec:mass_Redd} and \ref{ssec:Al3_lvir}}.} 

\begin{acknowledgements}
D. Dultzin and C. A. Negrete acknowledge support form grant IN111422 PAPIIT UNAM. C. A. Negrete acknowledge support form CONACyT project Paradigmas y Controversias de la Ciencia 2022-320020. The work of T. M. Buendia-Rios has been sponsored by CONACYT-Mexico through the Ph.D. scholarship No. 760641. Funding for the SDSS and SDSS-II has been provided by the Alfred P. Sloan Foundation, the Participating Institutions, the National Science Foundation, the U.S. Department of Energy, the National Aeronautics and Space Administration, the Japanese Monbukagakusho, the Max Planck Society, and the Higher Education Funding Council for England. The SDSS website is http://www.sdss.org. The SDSS is managed by the Astrophysical Research Consortium for the Participating Institutions listed at the http://www.sdss.org.
\end{acknowledgements}
    
\bibliographystyle{aa}
\bibliography{biblio.bib}

\begin{thebibliography}{131}
\expandafter\ifx\csname natexlab\endcsname\relax\def\natexlab#1{#1}\fi

\bibitem[{{Bachev} {et~al.}(2004){Bachev}, {Marziani}, {Sulentic}, {Zamanov},
  {Calvani}, \& {Dultzin-Hacyan}}]{bachev04}
{Bachev}, R., {Marziani}, P., {Sulentic}, J.~W., {et~al.} 2004, \apj, 617, 171

\bibitem[{{Baldwin} {et~al.}(1995){Baldwin}, {Ferland}, {Korista}, \&
  {Verner}}]{baldwinetal95}
{Baldwin}, J., {Ferland}, G., {Korista}, K., \& {Verner}, D. 1995, ApJL, 455,
  L119+

\bibitem[{{Baldwin} {et~al.}(2004){Baldwin}, {Ferland}, {Korista}, {Hamann}, \&
  {LaCluyz{\'e}}}]{baldwin04}
{Baldwin}, J.~A., {Ferland}, G.~J., {Korista}, K.~T., {Hamann}, F., \&
  {LaCluyz{\'e}}, A. 2004, \apj, 615, 610

\bibitem[{{Bensch} {et~al.}(2015){Bensch}, {del Olmo}, {Sulentic}, {Perea}, \&
  {Marziani}}]{bensch15}
{Bensch}, K., {del Olmo}, A., {Sulentic}, J., {Perea}, J., \& {Marziani}, P.
  2015, Journal of Astrophysics and Astronomy, 36, 467

\bibitem[{{Bentz} {et~al.}(2009){Bentz}, {Walsh}, {Barth}, {Baliber},
  {Bennert}, {Canalizo}, {Filippenko}, {Ganeshalingam}, {Gates}, {Greene},
  {Hidas}, {Hiner}, {Lee}, {Li}, {Malkan}, {Minezaki}, {Sakata}, {Serduke},
  {Silverman}, {Steele}, {Stern}, {Street}, {Thornton}, {Treu}, {Wang}, {Woo},
  \& {Yoshii}}]{bentz09}
{Bentz}, M.~C., {Walsh}, J.~L., {Barth}, A.~J., {et~al.} 2009, \apj, 705, 199

\bibitem[{{Bevington} \& {Robinson}(2003)}]{bevington}
{Bevington}, P.~R. \& {Robinson}, D.~K. 2003, {Data reduction and error
  analysis for the physical sciences}

\bibitem[{{Boller} {et~al.}(1996){Boller}, {Brandt}, \& {Fink}}]{boller96}
{Boller}, T., {Brandt}, W.~N., \& {Fink}, H. 1996, \aap, 305, 53

\bibitem[{{Bon} {et~al.}(2015){Bon}, {Bon}, {Marziani}, \&
  {Jovanovi{\'c}}}]{bon15}
{Bon}, N., {Bon}, E., {Marziani}, P., \& {Jovanovi{\'c}}, P. 2015, \apss, 360,
  7

\bibitem[{{Bon} {et~al.}(2020){Bon}, {Marziani}, {Bon}, {Negrete}, {Dultzin},
  {del Olmo}, {D'Onofrio}, \& {Mart{\'\i}nez-Aldama}}]{bon20}
{Bon}, N., {Marziani}, P., {Bon}, E., {et~al.} 2020, \aap, 635, A151

\bibitem[{{Boroson}(2002)}]{boroson02}
{Boroson}, T.~A. 2002, \apj, 565, 78

\bibitem[{{Boroson} \& {Green}(1992)}]{ByG}
{Boroson}, T.~A. \& {Green}, R.~F. 1992, \apjs, 80, 109

\bibitem[{{Capellupo} {et~al.}(2016){Capellupo}, {Netzer}, {Lira},
  {Trakhtenbrot}, \& {Mej{\'\i}a-Restrepo}}]{capellupo16}
{Capellupo}, D.~M., {Netzer}, H., {Lira}, P., {Trakhtenbrot}, B., \&
  {Mej{\'\i}a-Restrepo}, J. 2016, \mnras, 460, 212

\bibitem[{{Coatman} {et~al.}(2016){Coatman}, {Hewett}, {Banerji}, \&
  {Richards}}]{coatman16}
{Coatman}, L., {Hewett}, P.~C., {Banerji}, M., \& {Richards}, G.~T. 2016,
  \mnras, 461, 647

\bibitem[{{Coatman} {et~al.}(2017){Coatman}, {Hewett}, {Banerji}, {Richards},
  {Hennawi}, \& {Prochaska}}]{coatman17}
{Coatman}, L., {Hewett}, P.~C., {Banerji}, M., {et~al.} 2017, \mnras, 465, 2120

\bibitem[{{Collin} {et~al.}(2006){Collin}, {Kawaguchi}, {Peterson}, \&
  {Vestergaard}}]{collin06}
{Collin}, S., {Kawaguchi}, T., {Peterson}, B.~M., \& {Vestergaard}, M. 2006,
  \aap, 456, 75

\bibitem[{{Corbin}(1995)}]{corbin95}
{Corbin}, M.~R. 1995, \apj, 447, 496

\bibitem[{{Dalla Bont{\`a}} {et~al.}(2020){Dalla Bont{\`a}}, {Peterson},
  {Bentz}, {Brandt}, {Ciroi}, {De Rosa}, {Fonseca Alvarez}, {Grier}, {Hall},
  {Hern{\'a}ndez Santisteban}, {Ho}, {Homayouni}, {Horne}, {Kochanek}, {Li},
  {Morelli}, {Pizzella}, {Pogge}, {Schneider}, {Shen}, {Trump}, \&
  {Vestergaard}}]{dallabontaetal20}
{Dalla Bont{\`a}}, E., {Peterson}, B.~M., {Bentz}, M.~C., {et~al.} 2020, \apj,
  903, 112

\bibitem[{{Decarli} {et~al.}(2011){Decarli}, {Dotti}, \&
  {Treves}}]{decarlietal11}
{Decarli}, R., {Dotti}, M., \& {Treves}, A. 2011, \mnras, 413, 39

\bibitem[{{Deconto-Machado} {et~al.}(2021){Deconto-Machado}, {del Olmo},
  {Marziani}, {Perea}, \& {Stirpe}}]{deconto21}
{Deconto-Machado}, A., {del Olmo}, A., {Marziani}, P., {Perea}, J., \&
  {Stirpe}, G. 2021, Astronomische Nachrichten
  [\eprint{https://onlinelibrary.wiley.com/doi/pdf/10.1002/asna.20210084}]

\bibitem[{{Denney}(2012)}]{denney12}
{Denney}, K.~D. 2012, \apj, 759, 44

\bibitem[{{Denney} {et~al.}(2010){Denney}, {Peterson}, {Pogge}, {Adair},
  {Atlee}, {Au-Yong}, {Bentz}, {Bird}, {Brokofsky}, {Chisholm}, {Comins},
  {Dietrich}, {Doroshenko}, {Eastman}, {Efimov}, {Ewald}, {Ferbey}, {Gaskell},
  {Hedrick}, {Jackson}, {Klimanov}, {Klimek}, {Kruse}, {Lad{\'e}route}, {Lamb},
  {Leighly}, {Minezaki}, {Nazarov}, {Onken}, {Petersen}, {Peterson},
  {Poindexter}, {Sakata}, {Schlesinger}, {Sergeev}, {Skolski}, {Stieglitz},
  {Tobin}, {Unterborn}, {Vestergaard}, {Watkins}, {Watson}, \&
  {Yoshii}}]{denney10}
{Denney}, K.~D., {Peterson}, B.~M., {Pogge}, R.~W., {et~al.} 2010, \apj, 721,
  715

\bibitem[{{Di Matteo} {et~al.}(2003){Di Matteo}, {Croft}, {Springel}, \&
  {Hernquist}}]{dimatteo03}
{Di Matteo}, T., {Croft}, R. A.~C., {Springel}, V., \& {Hernquist}, L. 2003,
  \apj, 593, 56

\bibitem[{{Dong} {et~al.}(2011){Dong}, {Wang}, {Ho}, {Wang}, {Fan}, {Wang},
  {Zhou}, \& {Yuan}}]{dong11}
{Dong}, X.-B., {Wang}, J.-G., {Ho}, L.~C., {et~al.} 2011, \apj, 736, 86

\bibitem[{{Du} {et~al.}(2016){Du}, {Wang}, {Hu}, {Ho}, {Li}, \&
  {Bai}}]{duetal16}
{Du}, P., {Wang}, J.-M., {Hu}, C., {et~al.} 2016, \apjl, 818, L14

\bibitem[{{Feng} {et~al.}(2021){Feng}, {Hu}, {Li}, {Liu}, {Bai}, {Xing},
  {Wang}, {Yang}, {Xiao}, \& {Lu}}]{feng21}
{Feng}, H.-C., {Hu}, C., {Li}, S.-S., {et~al.} 2021, \apj, 909, 18

\bibitem[{{Ferland} {et~al.}(2017){Ferland}, {Chatzikos}, {Guzm{\'a}n},
  {Lykins}, {van Hoof}, {Williams}, {Abel}, {Badnell}, {Keenan}, {Porter}, \&
  {Stancil}}]{ferlandetal17}
{Ferland}, G.~J., {Chatzikos}, M., {Guzm{\'a}n}, F., {et~al.} 2017, \rmxaa, 53,
  385

\bibitem[{{Ferland} {et~al.}(2009){Ferland}, {Hu}, {Wang}, {Baldwin}, {Porter},
  {van Hoof}, \& {Williams}}]{ferlandetal09}
{Ferland}, G.~J., {Hu}, C., {Wang}, J., {et~al.} 2009, \apjl, 707, L82

\bibitem[{Fraix-Burnet {et~al.}(2017)Fraix-Burnet, Marziani, D'Onofrio, \&
  Dultzin}]{fraix-burnetetal17}
Fraix-Burnet, D., Marziani, P., D'Onofrio, M., \& Dultzin, D. 2017, Frontiers
  in Astronomy and Space Sciences, 4, 1

\bibitem[{{Francis} {et~al.}(1991){Francis}, {Hewett}, {Foltz}, {Chaffee},
  {Weymann}, \& {Morris}}]{francis91}
{Francis}, P.~J., {Hewett}, P.~C., {Foltz}, C.~B., {et~al.} 1991, \apj, 373,
  465

\bibitem[{{Gallagher} {et~al.}(2015){Gallagher}, {Everett}, {Abado}, \&
  {Keating}}]{gallagher15}
{Gallagher}, S.~C., {Everett}, J.~E., {Abado}, M.~M., \& {Keating}, S.~K. 2015,
  \mnras, 451, 2991

\bibitem[{{Gaskell}(1982)}]{gaskell82}
{Gaskell}, C.~M. 1982, \apj, 263, 79

\bibitem[{{Giustini} \& {Proga}(2019)}]{giustini19}
{Giustini}, M. \& {Proga}, D. 2019, \aap, 630, A94

\bibitem[{{Grupe} {et~al.}(2004){Grupe}, {Wills}, {Leighly}, \&
  {Meusinger}}]{grupe04}
{Grupe}, D., {Wills}, B.~J., {Leighly}, K.~M., \& {Meusinger}, H. 2004, \aj,
  127, 156

\bibitem[{{Hartig} \& {Baldwin}(1986)}]{HB86}
{Hartig}, G.~F. \& {Baldwin}, J.~A. 1986, \apj, 302, 64

\bibitem[{{Hewett} \& {Wild}(2010)}]{hewett10}
{Hewett}, P.~C. \& {Wild}, V. 2010, \mnras, 405, 2302

\bibitem[{{Johansson} {et~al.}(2000){Johansson}, {Zethson}, {Hartman},
  {Ekberg}, {Ishibashi}, {Davidson}, \& {Gull}}]{johansson00}
{Johansson}, S., {Zethson}, T., {Hartman}, H., {et~al.} 2000, \aap, 361, 977

\bibitem[{{Komossa} \& {Xu}(2007)}]{komossa07}
{Komossa}, S. \& {Xu}, D. 2007, \apjl, 667, L33

\bibitem[{{Korista} {et~al.}(1997{\natexlab{a}}){Korista}, {Baldwin},
  {Ferland}, \& {Verner}}]{korista97}
{Korista}, K., {Baldwin}, J., {Ferland}, G., \& {Verner}, D.
  1997{\natexlab{a}}, \apjs, 108, 401

\bibitem[{{Korista} {et~al.}(1997{\natexlab{b}}){Korista}, {Baldwin},
  {Ferland}, \& {Verner}}]{koristaetal97}
{Korista}, K., {Baldwin}, J., {Ferland}, G., \& {Verner}, D.
  1997{\natexlab{b}}, ApJS, 108, 401

\bibitem[{{Kriss}(1994)}]{kriss94}
{Kriss}, G. 1994, in Astronomical Society of the Pacific Conference Series,
  Vol.~61, Astronomical Data Analysis Software and Systems III, ed. D.~R.
  {Crabtree}, R.~J. {Hanisch}, \& J.~{Barnes}, 437

\bibitem[{{Kuraszkiewicz} {et~al.}(2000){Kuraszkiewicz}, {Wilkes}, {Czerny}, \&
  {Mathur}}]{kuraszkiewicz00}
{Kuraszkiewicz}, J., {Wilkes}, B.~J., {Czerny}, B., \& {Mathur}, S. 2000, \apj,
  542, 692

\bibitem[{{Kuraszkiewicz} {et~al.}(2004){Kuraszkiewicz}, {Green}, {Crenshaw},
  {Dunn}, {Forster}, {Vestergaard}, \& {Aldcroft}}]{kuraszkiewiczetal04}
{Kuraszkiewicz}, J.~K., {Green}, P.~J., {Crenshaw}, D.~M., {et~al.} 2004,
  \apjs, 150, 165

\bibitem[{{La Mura} {et~al.}(2009){La Mura}, {Di Mille}, {Ciroi},
  {Popovi{\'c}}, \& {Rafanelli}}]{lamuraetal09}
{La Mura}, G., {Di Mille}, F., {Ciroi}, S., {Popovi{\'c}}, L.~{\v C}., \&
  {Rafanelli}, P. 2009, \apj, 693, 1437

\bibitem[{{Laor}(2000)}]{laor00}
{Laor}, A. 2000, \apjl, 543, L111

\bibitem[{{Laor} {et~al.}(1997){Laor}, {Fiore}, {Elvis}, {Wilkes}, \&
  {McDowell}}]{laor97}
{Laor}, A., {Fiore}, F., {Elvis}, M., {Wilkes}, B.~J., \& {McDowell}, J.~C.
  1997, \apj, 477, 93

\bibitem[{{Leighly} \& {Moore}(2004)}]{leighlymoore04}
{Leighly}, K.~M. \& {Moore}, J.~R. 2004, \apj, 611, 107

\bibitem[{{Li} {et~al.}(2021){Li}, {Yang}, {Yang}, {Chen}, {Songsheng}, {Liu},
  {Du}, {Luo}, {Yu}, {Hu}, {Jiang}, {Bao}, {Guo}, {Zhang}, {Li}, {Xiao}, {Lu},
  {Ho}, {Bai}, {Bian}, {Aceituno}, {Minezaki}, {Horne}, {Kokubo}, \&
  {Wang}}]{Lietal2021}
{Li}, S.-S., {Yang}, S., {Yang}, Z.-X., {et~al.} 2021, \apj, 920, 9

\bibitem[{{Lira} {et~al.}(2017){Lira}, {Botti}, {Kaspi}, \&
  {Netzer}}]{liraetal17}
{Lira}, P., {Botti}, I., {Kaspi}, S., \& {Netzer}, H. 2017, Frontiers in
  Astronomy and Space Sciences, 4, 71

\bibitem[{{Lira} {et~al.}(2018){Lira}, {Kaspi}, {Netzer}, {Botti}, {Morrell},
  {Mej{\'{\i}}a-Restrepo}, {S{\'a}nchez-S{\'a}ez}, {Mart{\'{\i}}nez-Palomera},
  \& {L{\'o}pez}}]{liraetal18}
{Lira}, P., {Kaspi}, S., {Netzer}, H., {et~al.} 2018, \apj, 865, 56

\bibitem[{{Liu} {et~al.}(2017){Liu}, {Feng}, \& {Bai}}]{Liuetal2017}
{Liu}, H.~T., {Feng}, H.~C., \& {Bai}, J.~M. 2017, \mnras, 466, 3323

\bibitem[{{Lyke} {et~al.}(2020){Lyke}, {Higley}, {McLane}, {Schurhammer},
  {Myers}, {Ross}, {Dawson}, {Chabanier}, {Martini}, {Busca}, {Mas des
  Bourboux}, {Salvato}, {Streblyanska}, {Zarrouk}, {Burtin}, {Anderson},
  {Bautista}, {Bizyaev}, {Brandt}, {Brinkmann}, {Brownstein}, {Comparat},
  {Green}, {de la Macorra}, {Mu{\~n}oz Guti{\'e}rrez}, {Hou}, {Newman},
  {Palanque-Delabrouille}, {P{\^a}ris}, {Percival}, {Petitjean}, {Rich},
  {Rossi}, {Schneider}, {Smith}, {Vivek}, \& {Weaver}}]{Lykeetal2020}
{Lyke}, B.~W., {Higley}, A.~N., {McLane}, J.~N., {et~al.} 2020, \apjs, 250, 8

\bibitem[{{Malkan} \& {Sargent}(1982)}]{malkan82}
{Malkan}, M.~A. \& {Sargent}, W.~L.~W. 1982, \apj, 254, 22

\bibitem[{{Marconi} {et~al.}(2009){Marconi}, {Axon}, {Maiolino}, {Nagao},
  {Pietrini}, {Risaliti}, {Robinson}, \& {Torricelli}}]{marconietal09}
{Marconi}, A., {Axon}, D.~J., {Maiolino}, R., {et~al.} 2009, \apjl, 698, L103

\bibitem[{{Mart{\'\i}nez-Aldama} {et~al.}(2018){Mart{\'\i}nez-Aldama}, {del
  Olmo}, {Marziani}, {Sulentic}, {Negrete}, {Dultzin}, {D'Onofrio}, \&
  {Perea}}]{mla18}
{Mart{\'\i}nez-Aldama}, M.~L., {del Olmo}, A., {Marziani}, P., {et~al.} 2018,
  \aap, 618, A179

\bibitem[{Mart{\'\i}nez-Aldama {et~al.}(2018)Mart{\'\i}nez-Aldama, Del~Olmo,
  Marziani, Sulentic, Negrete, Dultzin, Perea, \&
  D'Onofrio}]{martinez-aldamaetal18}
Mart{\'\i}nez-Aldama, M.~L., Del~Olmo, A., Marziani, P., {et~al.} 2018,
  Frontiers in Astronomy and Space Sciences, 4, 65

\bibitem[{{Marziani} {et~al.}(2022){Marziani}, {del Olmo}, {Negrete},
  {Dultzin}, {Piconcelli}, {Vietri}, {Martinez-Aldama}, {D'Onofrio}, {Bon},
  {Bon}, {Deconto Machado}, {Stirpe}, \& {Buendia Rios}}]{Marzianietal2022}
{Marziani}, P., {del Olmo}, A., {Negrete}, C.~A., {et~al.} 2022, arXiv
  e-prints, arXiv:2205.07034

\bibitem[{{Marziani} {et~al.}(2020){Marziani}, {del Olmo}, {Perea},
  {D'Onofrio}, \& {Panda}}]{m20}
{Marziani}, P., {del Olmo}, A., {Perea}, J., {D'Onofrio}, M., \& {Panda}, S.
  2020, Atoms, 8, 94

\bibitem[{{Marziani} {et~al.}(2021){Marziani}, {Dultzin}, {del Olmo},
  {D'Onofrio}, {de Diego}, {Stirpe}, {Bon}, {Bon}, {Czerny}, {Perea}, {Panda},
  {Loli Martinez-Aldama}, \& {Negrete}}]{m20_iau}
{Marziani}, P., {Dultzin}, D., {del Olmo}, A., {et~al.} 2021, in Nuclear
  Activity in Galaxies Across Cosmic Time, ed. M.~{Povi{\'c}}, P.~{Marziani},
  J.~{Masegosa}, H.~{Netzer}, S.~H. {Negu}, \& S.~B. {Tessema}, Vol. 356,
  66--71

\bibitem[{{Marziani} {et~al.}(2018){Marziani}, {Dultzin}, {Sulentic}, {Del
  Olmo}, {Negrete}, {Mart{\'\i}nez-Aldama}, {D'Onofrio}, {Bon}, {Bon}, \&
  {Stirpe}}]{M18}
{Marziani}, P., {Dultzin}, D., {Sulentic}, J.~W., {et~al.} 2018, Frontiers in
  Astronomy and Space Sciences, 5, 6

\bibitem[{{Marziani} {et~al.}(2006){Marziani}, {Dultzin-Hacyan}, \&
  {Sulentic}}]{marzianietal06}
{Marziani}, P., {Dultzin-Hacyan}, D., \& {Sulentic}, J.~W. 2006, in New
  Developments in Black Hole Research, ed. P.~V. {Kreitler} (Nova Press, New
  York), 123

\bibitem[{{Marziani} {et~al.}(2016){Marziani}, {Mart{\'{\i}}nez Carballo},
  {Sulentic}, {Del Olmo}, {Stirpe}, \& {Dultzin}}]{marzianietal16a}
{Marziani}, P., {Mart{\'{\i}}nez Carballo}, M.~A., {Sulentic}, J.~W., {et~al.}
  2016, \apss, 361, 29

\bibitem[{{Marziani} {et~al.}(2017){Marziani}, {Olmo}, {Mart{\'\i}nez-Aldama},
  {Dultzin}, {Negrete}, {Bon}, {Bon}, \& {D'Onofrio}}]{marziani17}
{Marziani}, P., {Olmo}, A., {Mart{\'\i}nez-Aldama}, M., {et~al.} 2017, Atoms,
  5, 33

\bibitem[{{Marziani} \& {Sulentic}(2012)}]{ms12}
{Marziani}, P. \& {Sulentic}, J.~W. 2012, The Astronomical Review, 7, 33

\bibitem[{{Marziani} \& {Sulentic}(2014)}]{MS14}
{Marziani}, P. \& {Sulentic}, J.~W. 2014, \mnras, 442, 1211

\bibitem[{{Marziani} {et~al.}(1996){Marziani}, {Sulentic}, {Dultzin-Hacyan},
  {Calvani}, \& {Moles}}]{marziani96}
{Marziani}, P., {Sulentic}, J.~W., {Dultzin-Hacyan}, D., {Calvani}, M., \&
  {Moles}, M. 1996, \apjs, 104, 37

\bibitem[{{Marziani} {et~al.}(2010){Marziani}, {Sulentic}, {Negrete},
  {Dultzin}, {Zamfir}, \& {Bachev}}]{marziani10}
{Marziani}, P., {Sulentic}, J.~W., {Negrete}, C.~A., {et~al.} 2010, \mnras,
  409, 1033

\bibitem[{{Marziani} {et~al.}(2013){Marziani}, {Sulentic}, {Plauchu-Frayn}, \&
  {del Olmo}}]{marziani13}
{Marziani}, P., {Sulentic}, J.~W., {Plauchu-Frayn}, I., \& {del Olmo}, A. 2013,
  \apj, 764, 150

\bibitem[{{Marziani} {et~al.}(2003{\natexlab{a}}){Marziani}, {Sulentic},
  {Zamanov}, \& {Calvani}}]{marzianietal03f}
{Marziani}, P., {Sulentic}, J.~W., {Zamanov}, R., \& {Calvani}, M.
  2003{\natexlab{a}}, Memorie della Societa Astronomica Italiana, 74, 490

\bibitem[{{Marziani} {et~al.}(2001){Marziani}, {Sulentic}, {Zwitter},
  {Dultzin-Hacyan}, \& {Calvani}}]{marziani01}
{Marziani}, P., {Sulentic}, J.~W., {Zwitter}, T., {Dultzin-Hacyan}, D., \&
  {Calvani}, M. 2001, \apj, 558, 553

\bibitem[{{Marziani} {et~al.}(2003{\natexlab{b}}){Marziani}, {Zamanov},
  {Sulentic}, \& {Calvani}}]{marziani03}
{Marziani}, P., {Zamanov}, R.~K., {Sulentic}, J.~W., \& {Calvani}, M.
  2003{\natexlab{b}}, \mnras, 345, 1133

\bibitem[{{Matsuoka} {et~al.}(2008){Matsuoka}, {Kawara}, \&
  {Oyabu}}]{matsuoka08}
{Matsuoka}, Y., {Kawara}, K., \& {Oyabu}, S. 2008, \apj, 673, 62

\bibitem[{{McLure} \& {Jarvis}(2002)}]{mcLure2002}
{McLure}, R.~J. \& {Jarvis}, M.~J. 2002, \mnras, 337, 109

\bibitem[{{Mediavilla} {et~al.}(2018){Mediavilla}, {Jim{\'e}nez-Vicente},
  {Fian}, {Mu{\~n}oz}, {Falco}, {Motta}, \& {Guerras}}]{mediavilla18}
{Mediavilla}, E., {Jim{\'e}nez-Vicente}, J., {Fian}, C., {et~al.} 2018, \apj,
  862, 104

\bibitem[{{Mej{\'{\i}}a-Restrepo} {et~al.}(2017){Mej{\'{\i}}a-Restrepo},
  {Lira}, {Netzer}, {Trakhtenbrot}, \& {Capellupo}}]{mejia-restrepoetal17}
{Mej{\'{\i}}a-Restrepo}, J.~E., {Lira}, P., {Netzer}, H., {Trakhtenbrot}, B.,
  \& {Capellupo}, D. 2017, Frontiers in Astronomy and Space Sciences, 4, 70

\bibitem[{{Mej{\'{\i}}a-Restrepo} {et~al.}(2018){Mej{\'{\i}}a-Restrepo},
  {Lira}, {Netzer}, {Trakhtenbrot}, \& {Capellupo}}]{mejia-restrepoetal18a}
{Mej{\'{\i}}a-Restrepo}, J.~E., {Lira}, P., {Netzer}, H., {Trakhtenbrot}, B.,
  \& {Capellupo}, D.~M. 2018, Nature Astronomy, 2, 63

\bibitem[{{Mineshige} {et~al.}(2000){Mineshige}, {Kawaguchi}, {Takeuchi}, \&
  {Hayashida}}]{mineshigeetal00}
{Mineshige}, S., {Kawaguchi}, T., {Takeuchi}, M., \& {Hayashida}, K. 2000,
  \pasj, 52, 499

\bibitem[{{Moore}(1945)}]{moore45}
{Moore}, C.~E. 1945, Contributions from the Princeton University Observatory,
  20, 1

\bibitem[{{Negrete} {et~al.}(2012){Negrete}, {Dultzin}, {Marziani}, \&
  {Sulentic}}]{negrete12}
{Negrete}, C.~A., {Dultzin}, D., {Marziani}, P., \& {Sulentic}, J.~W. 2012,
  \apj, 757, 62

\bibitem[{{Negrete} {et~al.}(2013){Negrete}, {Dultzin}, {Marziani}, \&
  {Sulentic}}]{negrete13}
{Negrete}, C.~A., {Dultzin}, D., {Marziani}, P., \& {Sulentic}, J.~W. 2013,
  \apj, 771, 31

\bibitem[{{Negrete} {et~al.}(2014){Negrete}, {Dultzin}, {Marziani}, \&
  {Sulentic}}]{negrete14}
{Negrete}, C.~A., {Dultzin}, D., {Marziani}, P., \& {Sulentic}, J.~W. 2014,
  \apj, 794, 95

\bibitem[{{Netzer}(1977)}]{netzer77}
{Netzer}, H. 1977, \mnras, 181, 89

\bibitem[{{Netzer}(2015)}]{netzer15}
{Netzer}, H. 2015, \araa, 53, 365

\bibitem[{{Netzer} \& {Marziani}(2010)}]{NM10}
{Netzer}, H. \& {Marziani}, P. 2010, \apj, 724, 318

\bibitem[{Netzer \& Peterson(1997)}]{NP97}
Netzer, H. \& Peterson, B.~M. 1997, Astronomical Time Series, 85

\bibitem[{{Panda} {et~al.}(2019{\natexlab{a}}){Panda}, {Mart{\'\i}nez-Aldama},
  \& {Zaja{\v{c}}ek}}]{panda19}
{Panda}, S., {Mart{\'\i}nez-Aldama}, M.~L., \& {Zaja{\v{c}}ek}, M.
  2019{\natexlab{a}}, Frontiers in Astronomy and Space Sciences, 6, 75

\bibitem[{{Panda} {et~al.}(2019{\natexlab{b}}){Panda}, {Marziani}, \&
  {Czerny}}]{P19}
{Panda}, S., {Marziani}, P., \& {Czerny}, B. 2019{\natexlab{b}}, \apj, 882, 79

\bibitem[{{Peterson} {et~al.}(1993){Peterson}, {Ali}, {Horne}, {Bertram},
  {Lame}, {Pogge}, \& {Wagner}}]{petersonetal93}
{Peterson}, B.~M., {Ali}, B., {Horne}, K., {et~al.} 1993, \apj, 402, 469

\bibitem[{{Peterson} \& {Wandel}(1999)}]{petersonwandel99}
{Peterson}, B.~M. \& {Wandel}, A. 1999, \apjl, 521, L95

\bibitem[{{Peterson} \& {Wandel}(2000)}]{petersonwandel00}
{Peterson}, B.~M. \& {Wandel}, A. 2000, \apjl, 540, L13

\bibitem[{{Petrucci} {et~al.}(2020){Petrucci}, {Gronkiewicz}, {Rozanska},
  {Belmont}, {Bianchi}, {Czerny}, {Matt}, {Malzac}, {Middei}, {De Rosa},
  {Ursini}, \& {Cappi}}]{petrucci20}
{Petrucci}, P.~O., {Gronkiewicz}, D., {Rozanska}, A., {et~al.} 2020, \aap, 634,
  A85

\bibitem[{{Popovi{\'c}} {et~al.}(2019){Popovi{\'c}},
  {Kova{\v{c}}evi{\'c}-Doj{\v{c}}inovi{\'c}}, \& {Mar{\v{c}}eta-Mand
  i{\'c}}}]{popovicetal19}
{Popovi{\'c}}, L.~{\v{C}}., {Kova{\v{c}}evi{\'c}-Doj{\v{c}}inovi{\'c}}, J., \&
  {Mar{\v{c}}eta-Mand i{\'c}}, S. 2019, \mnras, 484, 3180

\bibitem[{{Richards} {et~al.}(2011){Richards}, {Kruczek}, {Gallagher}, {Hall},
  {Hewett}, {Leighly}, {Deo}, {Kratzer}, \& {Shen}}]{richards11}
{Richards}, G.~T., {Kruczek}, N.~E., {Gallagher}, S.~C., {et~al.} 2011, \aj,
  141, 167

\bibitem[{{Richards} {et~al.}(2006){Richards}, {Strauss}, {Fan}, {Hall},
  {Jester}, {Schneider}, {Vanden Berk}, {Stoughton}, {Anderson}, {Brunner},
  {Gray}, {Gunn}, {Ivezi{\'c}}, {Kirkland}, {Knapp}, {Loveday}, {Meiksin},
  {Pope}, {Szalay}, {Thakar}, {Yanny}, {York}, {Barentine}, {Brewington},
  {Brinkmann}, {Fukugita}, {Harvanek}, {Kent}, {Kleinman}, {Krzesi{\'n}ski},
  {Long}, {Lupton}, {Nash}, {Neilsen}, {Nitta}, {Schlegel}, \&
  {Snedden}}]{richards06}
{Richards}, G.~T., {Strauss}, M.~A., {Fan}, X., {et~al.} 2006, \aj, 131, 2766

\bibitem[{{Sadowski}(2011)}]{sadowski11}
{Sadowski}, A. 2011, arXiv e-prints, arXiv:1108.0396

\bibitem[{{Salviander} {et~al.}(2007){Salviander}, {Shields}, {Gebhardt}, \&
  {Bonning}}]{salviander07}
{Salviander}, S., {Shields}, G.~A., {Gebhardt}, K., \& {Bonning}, E.~W. 2007,
  \apj, 662, 131

\bibitem[{{Schlafly} \& {Finkbeiner}(2011)}]{SanF11}
{Schlafly}, E.~F. \& {Finkbeiner}, D.~P. 2011, \apj, 737, 103

\bibitem[{Shen(2013)}]{shen13}
Shen, Y. 2013, arXiv preprint arXiv:1302.2643

\bibitem[{{Shen} \& {Ho}(2014)}]{shenho14}
{Shen}, Y. \& {Ho}, L.~C. 2014, \nat, 513, 210

\bibitem[{{Shen} {et~al.}(2011){Shen}, {Richards}, {Strauss}, {Hall},
  {Schneider}, {Snedden}, {Bizyaev}, {Brewington}, {Malanushenko},
  {Malanushenko}, {Oravetz}, {Pan}, \& {Simmons}}]{shen11}
{Shen}, Y., {Richards}, G.~T., {Strauss}, M.~A., {et~al.} 2011, \apjs, 194, 45

\bibitem[{{Sigut} \& {Pradhan}(1998)}]{sigutpradhan98}
{Sigut}, T.~A.~A. \& {Pradhan}, A.~K. 1998, \apjl, 499, L139

\bibitem[{{Singh} {et~al.}(1985){Singh}, {Garmire}, \& {Nousek}}]{singh85}
{Singh}, K.~P., {Garmire}, G.~P., \& {Nousek}, J. 1985, \apj, 297, 633

\bibitem[{{S{\k{a}}dowski} {et~al.}(2014){S{\k{a}}dowski}, {Narayan},
  {McKinney}, \& {Tchekhovskoy}}]{sadowski14}
{S{\k{a}}dowski}, A., {Narayan}, R., {McKinney}, J.~C., \& {Tchekhovskoy}, A.
  2014, \mnras, 439, 503

\bibitem[{{Small} \& {Blandford}(1992)}]{small92}
{Small}, T.~A. \& {Blandford}, R.~D. 1992, \mnras, 259, 725

\bibitem[{{Snedden} \& {Gaskell}(2007)}]{sneddengaskell07}
{Snedden}, S.~A. \& {Gaskell}, C.~M. 2007, ApJ, 669, 126

\bibitem[{{{\'S}niegowska} {et~al.}(2020){{\'S}niegowska}, {Koz{\l}owski},
  {Czerny}, {Panda}, \& {Hryniewicz}}]{sniegowskaetal20}
{{\'S}niegowska}, M., {Koz{\l}owski}, S., {Czerny}, B., {Panda}, S., \&
  {Hryniewicz}, K. 2020, \apj, 900, 64

\bibitem[{{{\'S}niegowska} {et~al.}(2021){{\'S}niegowska}, {Marziani},
  {Czerny}, {Panda}, {Mart{\'\i}nez-Aldama}, {del Olmo}, \&
  {D'Onofrio}}]{sniegowska21}
{{\'S}niegowska}, M., {Marziani}, P., {Czerny}, B., {et~al.} 2021, \apj, 910,
  115

\bibitem[{{Sulentic} {et~al.}(2007){Sulentic}, {Bachev}, {Marziani}, {Negrete},
  \& {Dultzin}}]{sulentic07}
{Sulentic}, J.~W., {Bachev}, R., {Marziani}, P., {Negrete}, C.~A., \&
  {Dultzin}, D. 2007, \apj, 666, 757

\bibitem[{{Sulentic} {et~al.}(2017){Sulentic}, {del Olmo}, {Marziani},
  {Mart{\'\i}nez-Carballo}, {D'Onofrio}, {Dultzin}, {Perea},
  {Mart{\'\i}nez-Aldama}, {Negrete}, {Stirpe}, \& {Zamfir}}]{sulentic17}
{Sulentic}, J.~W., {del Olmo}, A., {Marziani}, P., {et~al.} 2017, \aap, 608,
  A122

\bibitem[{{Sulentic} {et~al.}(2014){Sulentic}, {Marziani}, {del Olmo},
  {Dultzin}, {Perea}, \& {Negrete}}]{sulentic14}
{Sulentic}, J.~W., {Marziani}, P., {del Olmo}, A., {et~al.} 2014, \aap, 570,
  A96

\bibitem[{{Sulentic} {et~al.}(2000{\natexlab{a}}){Sulentic}, {Marziani}, \&
  {Dultzin-Hacyan}}]{sulentic00a}
{Sulentic}, J.~W., {Marziani}, P., \& {Dultzin-Hacyan}, D. 2000{\natexlab{a}},
  \araa, 38, 521

\bibitem[{{Sulentic} {et~al.}(2002){Sulentic}, {Marziani}, {Zamanov}, {Bachev},
  {Calvani}, \& {Dultzin-Hacyan}}]{sulentic02}
{Sulentic}, J.~W., {Marziani}, P., {Zamanov}, R., {et~al.} 2002, \apjl, 566,
  L71

\bibitem[{{Sulentic} {et~al.}(2000{\natexlab{b}}){Sulentic}, {Marziani},
  {Zwitter}, {Dultzin-Hacyan}, \& {Calvani}}]{sulenticetal00b}
{Sulentic}, J.~W., {Marziani}, P., {Zwitter}, T., {Dultzin-Hacyan}, D., \&
  {Calvani}, M. 2000{\natexlab{b}}, ApJL, 545, L15

\bibitem[{Sulentic {et~al.}(2007)Sulentic, Zamfir, Marziani, \&
  Dultzin}]{sulentic08}
Sulentic, J.~W., Zamfir, S., Marziani, P., \& Dultzin, D. 2007, arXiv preprint
  arXiv:0709.2499

\bibitem[{{Sulentic} {et~al.}(2000{\natexlab{c}}){Sulentic}, {Zwitter},
  {Marziani}, \& {Dultzin-Hacyan}}]{sulentic00b}
{Sulentic}, J.~W., {Zwitter}, T., {Marziani}, P., \& {Dultzin-Hacyan}, D.
  2000{\natexlab{c}}, \apjl, 536, L5

\bibitem[{{Sulentic} {et~al.}(2000{\natexlab{d}}){Sulentic}, {Zwitter},
  {Marziani}, \& {Dultzin-Hacyan}}]{sulenticetal00c}
{Sulentic}, J.~W., {Zwitter}, T., {Marziani}, P., \& {Dultzin-Hacyan}, D.
  2000{\natexlab{d}}, ApJL, 536, L5

\bibitem[{Sun \& Shen(2015)}]{sun15}
Sun, J. \& Shen, Y. 2015, The Astrophysical Journal Letters, 804, L15

\bibitem[{{Tody}(1986)}]{tody86}
{Tody}, D. 1986, in Society of Photo-Optical Instrumentation Engineers (SPIE)
  Conference Series, Vol. 627, Instrumentation in astronomy VI, ed. D.~L.
  {Crawford}, 733

\bibitem[{{Vanden Berk} {et~al.}(2001){Vanden Berk}, {Richards}, {Bauer},
  {Strauss}, {Schneider}, {Heckman}, {York}, {Hall}, {Fan}, {Knapp},
  {Anderson}, {Annis}, {Bahcall}, {Bernardi}, {Briggs}, {Brinkmann}, {Brunner},
  {Burles}, {Carey}, {Castander}, {Connolly}, {Crocker}, {Csabai}, {Doi},
  {Finkbeiner}, {Friedman}, {Frieman}, {Fukugita}, {Gunn}, {Hennessy},
  {Ivezi{\'c}}, {Kent}, {Kunszt}, {Lamb}, {Leger}, {Long}, {Loveday}, {Lupton},
  {Meiksin}, {Merelli}, {Munn}, {Newberg}, {Newcomb}, {Nichol}, {Owen}, {Pier},
  {Pope}, {Rockosi}, {Schlegel}, {Siegmund}, {Smee}, {Snir}, {Stoughton},
  {Stubbs}, {SubbaRao}, {Szalay}, {Szokoly}, {Tremonti}, {Uomoto}, {Waddell},
  {Yanny}, \& {Zheng}}]{vanden01}
{Vanden Berk}, D.~E., {Richards}, G.~T., {Bauer}, A., {et~al.} 2001, \aj, 122,
  549

\bibitem[{{Vestergaard} \& {Peterson}(2006)}]{vp06}
{Vestergaard}, M. \& {Peterson}, B.~M. 2006, \apj, 641, 689

\bibitem[{{Vestergaard} \& {Wilkes}(2001)}]{vestergaard01}
{Vestergaard}, M. \& {Wilkes}, B.~J. 2001, \apjs, 134, 1

\bibitem[{{Vietri} {et~al.}(2018){Vietri}, {Piconcelli}, {Bischetti}, {Duras},
  {Martocchia}, {Bongiorno}, {Marconi}, {Zappacosta}, {Bisogni}, {Bruni},
  {Brusa}, {Comastri}, {Cresci}, {Feruglio}, {Giallongo}, {La Franca},
  {Mainieri}, {Mannucci}, {Ricci}, {Sani}, {Testa}, {Tombesi}, {Vignali}, \&
  {Fiore}}]{vietrietal18}
{Vietri}, G., {Piconcelli}, E., {Bischetti}, M., {et~al.} 2018, \aap, 617, A81

\bibitem[{{Walter} \& {Fink}(1993)}]{WF93}
{Walter}, R. \& {Fink}, H.~H. 1993, \aap, 274, 105

\bibitem[{{Wandel} \& {Petrosian}(1988)}]{wandel88}
{Wandel}, A. \& {Petrosian}, V. 1988, \apjl, 329, L11

\bibitem[{{Wang} \& {Li}(2011)}]{wangli11}
{Wang}, J. \& {Li}, Y. 2011, \apjl, 742, L12

\bibitem[{{Wang} {et~al.}(2017){Wang}, {Du}, {Brotherton}, {Hu}, {Songsheng},
  {Li}, {Shi}, \& {Zhang}}]{wang17}
{Wang}, J.-M., {Du}, P., {Brotherton}, M.~S., {et~al.} 2017, Nature Astronomy,
  1, 775

\bibitem[{{Wang} {et~al.}(1996){Wang}, {Brinkmann}, \& {Bergeron}}]{wbb96}
{Wang}, T., {Brinkmann}, W., \& {Bergeron}, J. 1996, \aap, 309, 81

\bibitem[{{Wills} {et~al.}(1999){Wills}, {Laor}, {Brotherton}, {Wills},
  {Wilkes}, {Ferland}, \& {Shang}}]{wills99}
{Wills}, B.~J., {Laor}, A., {Brotherton}, M.~S., {et~al.} 1999, \apjl, 515, L53

\bibitem[{{Wolf} {et~al.}(2020){Wolf}, {Salvato}, {Coffey}, {Merloni},
  {Buchner}, {Arcodia}, {Baron}, {Carrera}, {Comparat}, {Schneider}, \&
  {Nandra}}]{wolfetal20}
{Wolf}, J., {Salvato}, M., {Coffey}, D., {et~al.} 2020, \mnras, 492, 3580

\bibitem[{{Zamanov} {et~al.}(2002){Zamanov}, {Marziani}, {Sulentic}, {Calvani},
  {Dultzin-Hacyan}, \& {Bachev}}]{zamanov02}
{Zamanov}, R., {Marziani}, P., {Sulentic}, J.~W., {et~al.} 2002, \apjl, 576, L9

\bibitem[{{Zamfir} {et~al.}(2010){Zamfir}, {Sulentic}, {Marziani}, \&
  {Dultzin}}]{zamfir10}
{Zamfir}, S., {Sulentic}, J.~W., {Marziani}, P., \& {Dultzin}, D. 2010, \mnras,
  403, 1759

\bibitem[{{Zheng} \& {Sulentic}(1990)}]{zheng90}
{Zheng}, W. \& {Sulentic}, J.~W. 1990, \apj, 350, 512

\end{thebibliography}

\onecolumn
\begin{appendix}

\section{Extreme accretor sources multi-component fits}\label{app_A}
    
    The results of line profile fitting for the 1900\AA\ blend are shown in the following Figure \ref{fig:xA_spectra}. Notes for some of the objects within the xA sub-sample:
    
    \begin{itemize}
        \item J152314.49+375928.9: One of the highest blueshift sources: shift(\al)= -849.96$\pm$25 \kms, with a FWHM(\al)=4000$\pm$400 \kms. This source is also the one with an additional component in \feiii$\lambda$1914 due to the high amount of iron emission observed in the red side of \cnl.
        \item J100827.67+210931.1: Brightest quasar with a log \lbol=47.06, with a FWHM(\al)=3612$\pm$362 \kms\ it give us also one of the most massive xA quasar with a log \mbh(\al)=8.92.
        \item J235157.59+003610.6: This object was affected by the host galaxy emission, it was fitted with positive continuum and a few absorption lines in the red side of \siiii\ and \ciii\ are observed.
        \item J095531.45+174340.3: One of the brightest sources with a log \lbol=47.04, with a relatively high shift(\al)=-427$\pm$43 \kms. This object has the highest Eddington ratio in the sub-sample, \redd=1.34.
        \item J092612.68+202326.6: This object presented an \al\ redshift of 262$\pm$27\kms with a FWHM(\al) = 3771$\pm$370.
        \item J003546.29-034118.2: Highest blueshifted source, shift(\al) = -1011.69$\pm$81\kms,  FWHM(\al) = 4194.93$\pm$342\kms. With a $\delta$log\lvir = -0.75. 
    \end{itemize}
    
    \begin{figure*}
    \centering
        \includegraphics[width=0.3\textwidth]{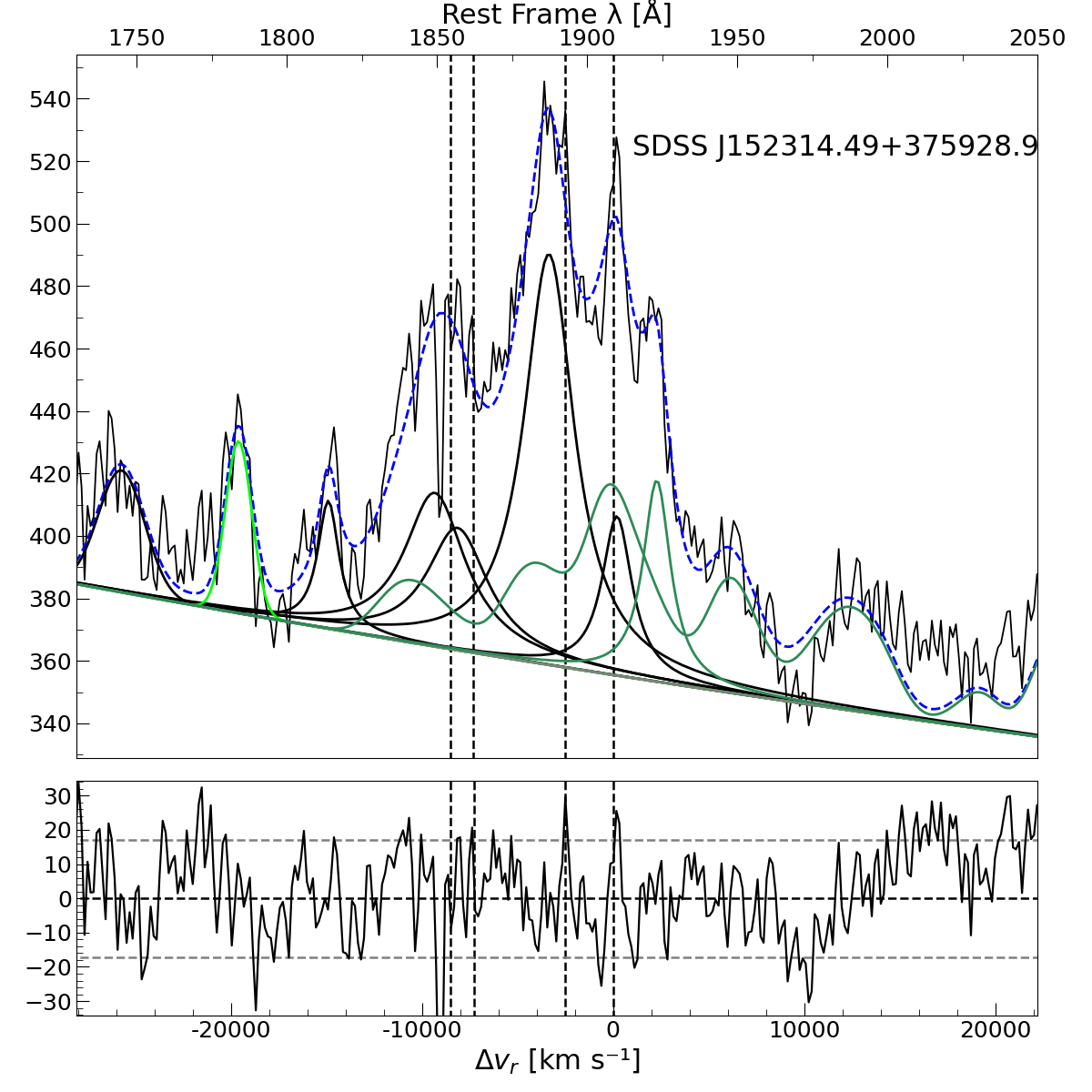}
        \includegraphics[width=0.3\textwidth]{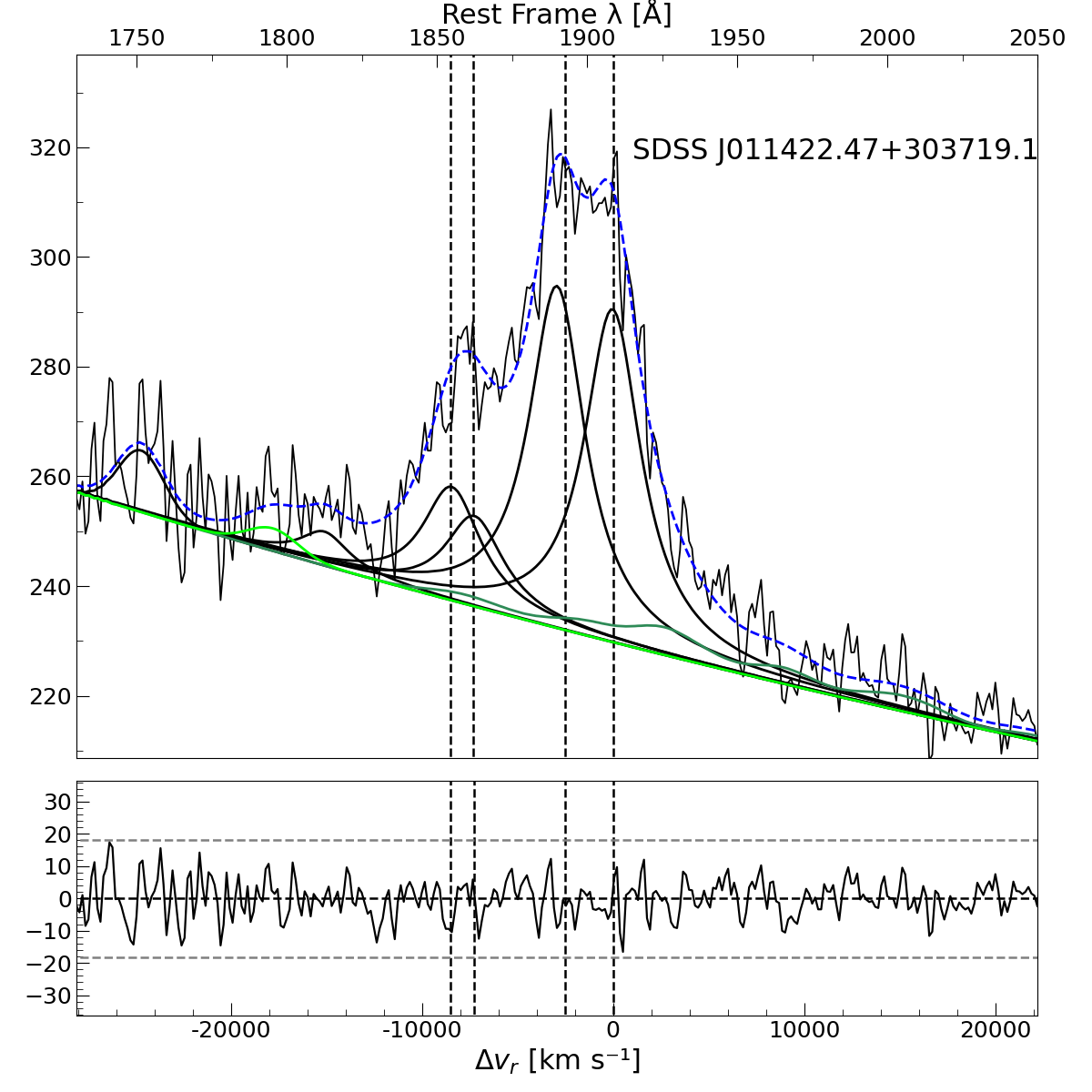}
        \includegraphics[width=0.3\textwidth]{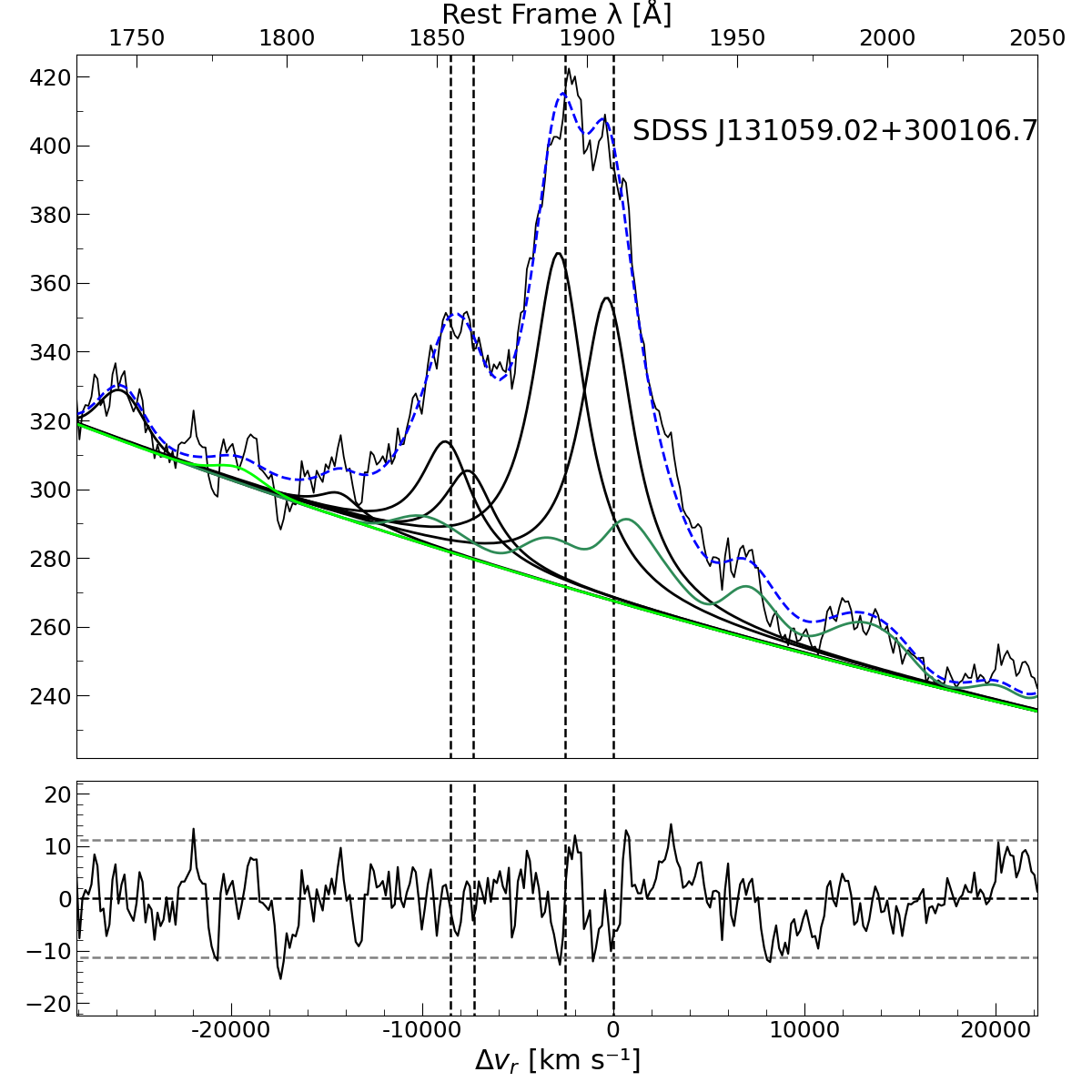}\\
        \includegraphics[width=0.3\textwidth]{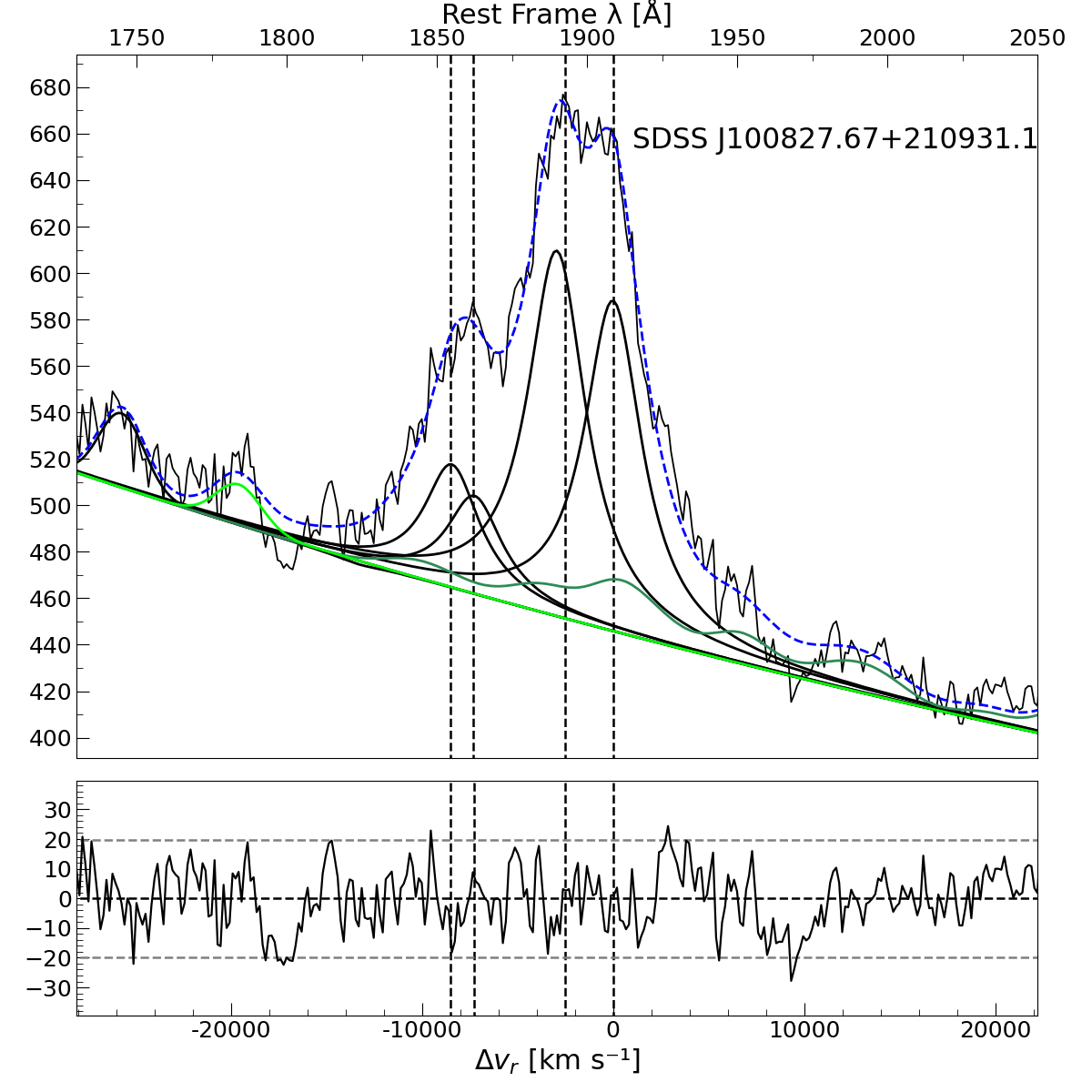}
        \includegraphics[width=0.3\textwidth]{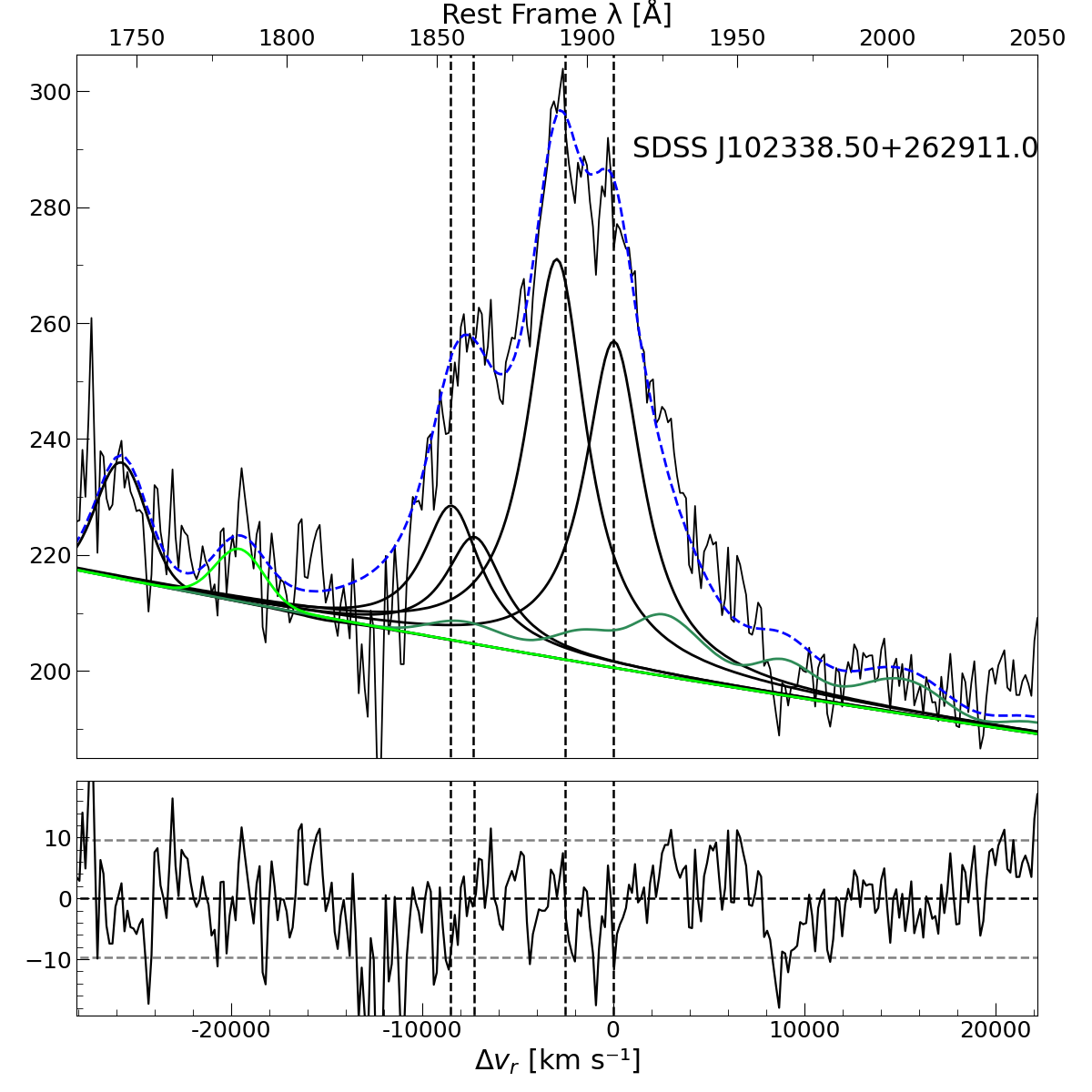}
        \includegraphics[width=0.3\textwidth]{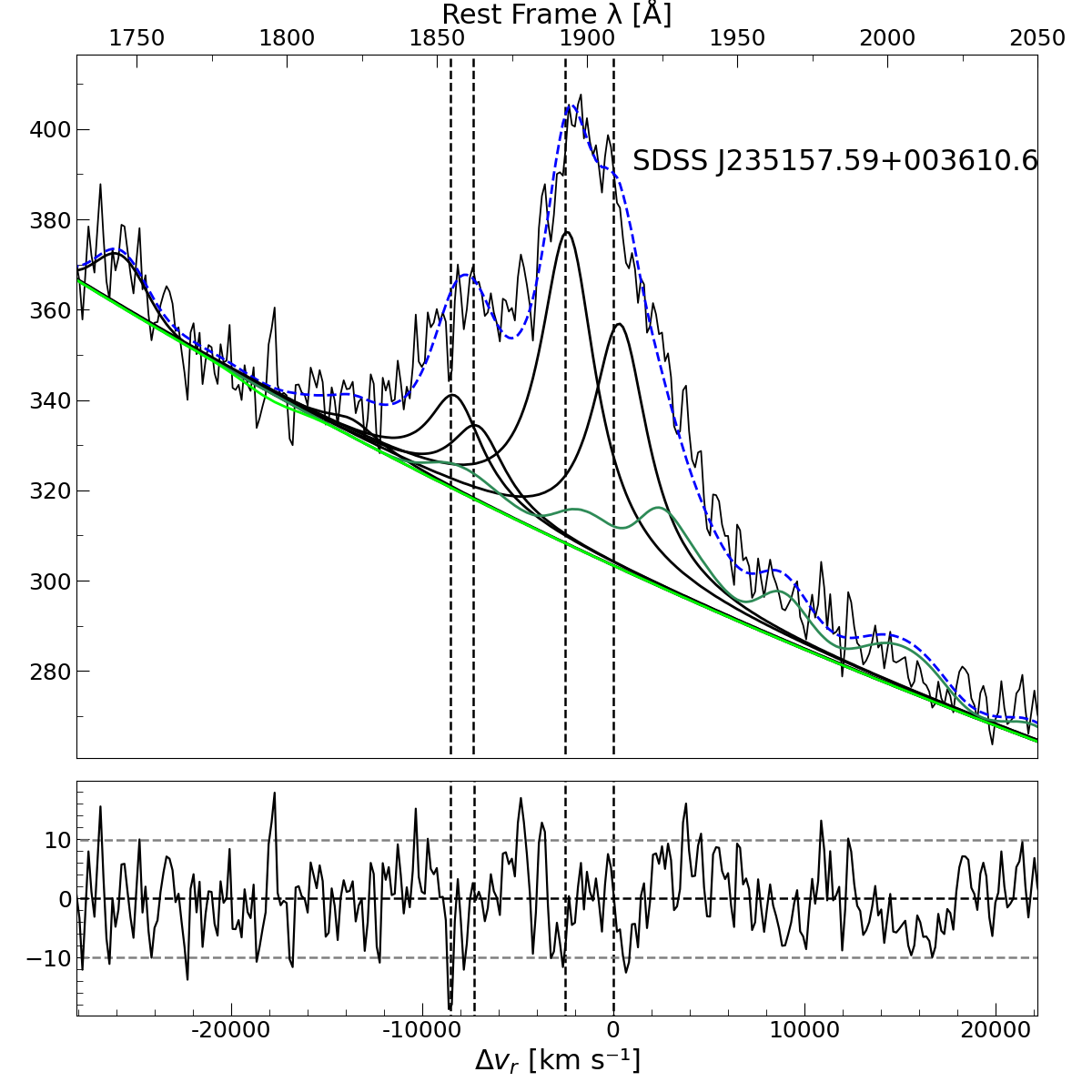}\\
        \includegraphics[width=0.3\textwidth]{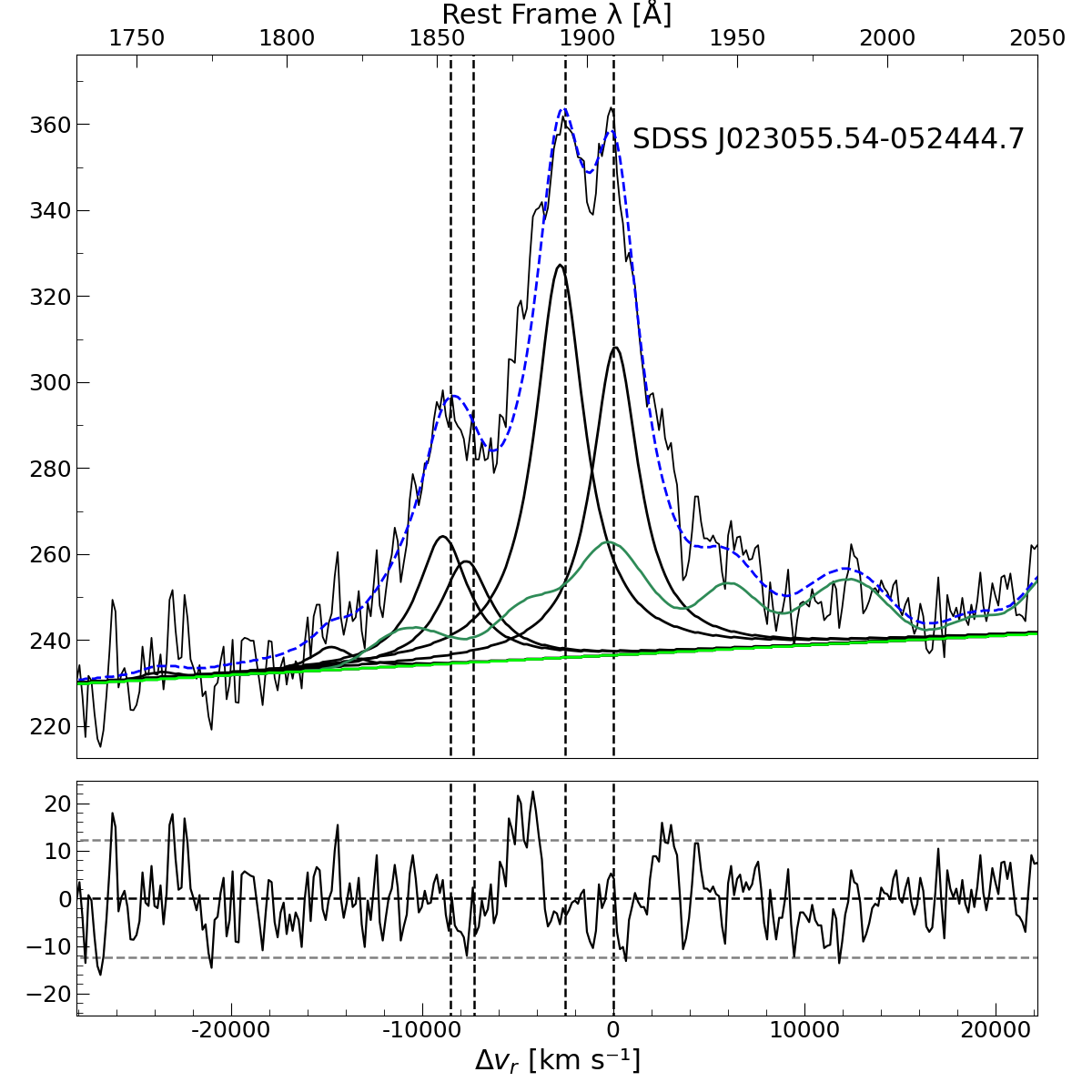}
        \includegraphics[width=0.3\textwidth]{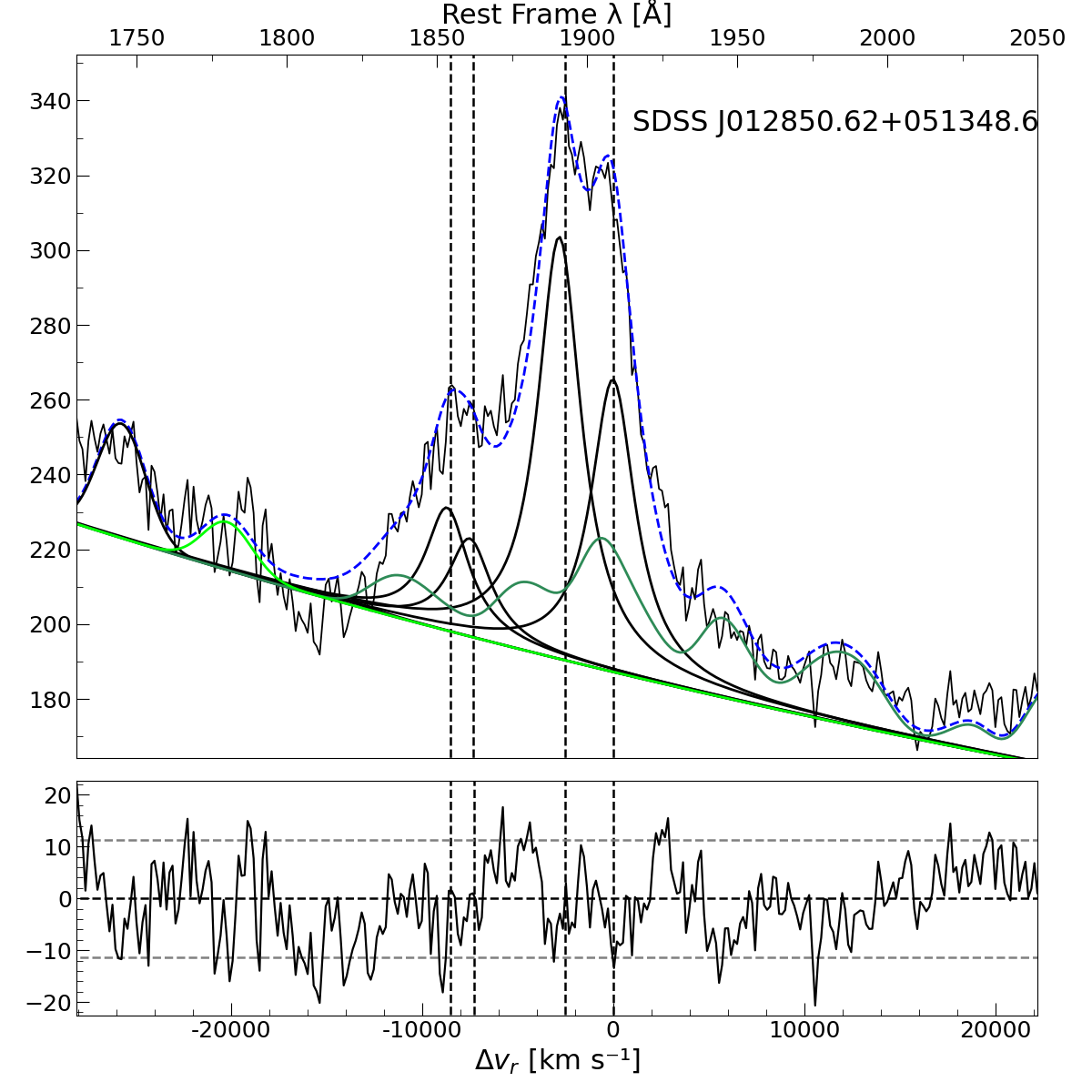}
        \includegraphics[width=0.3\textwidth]{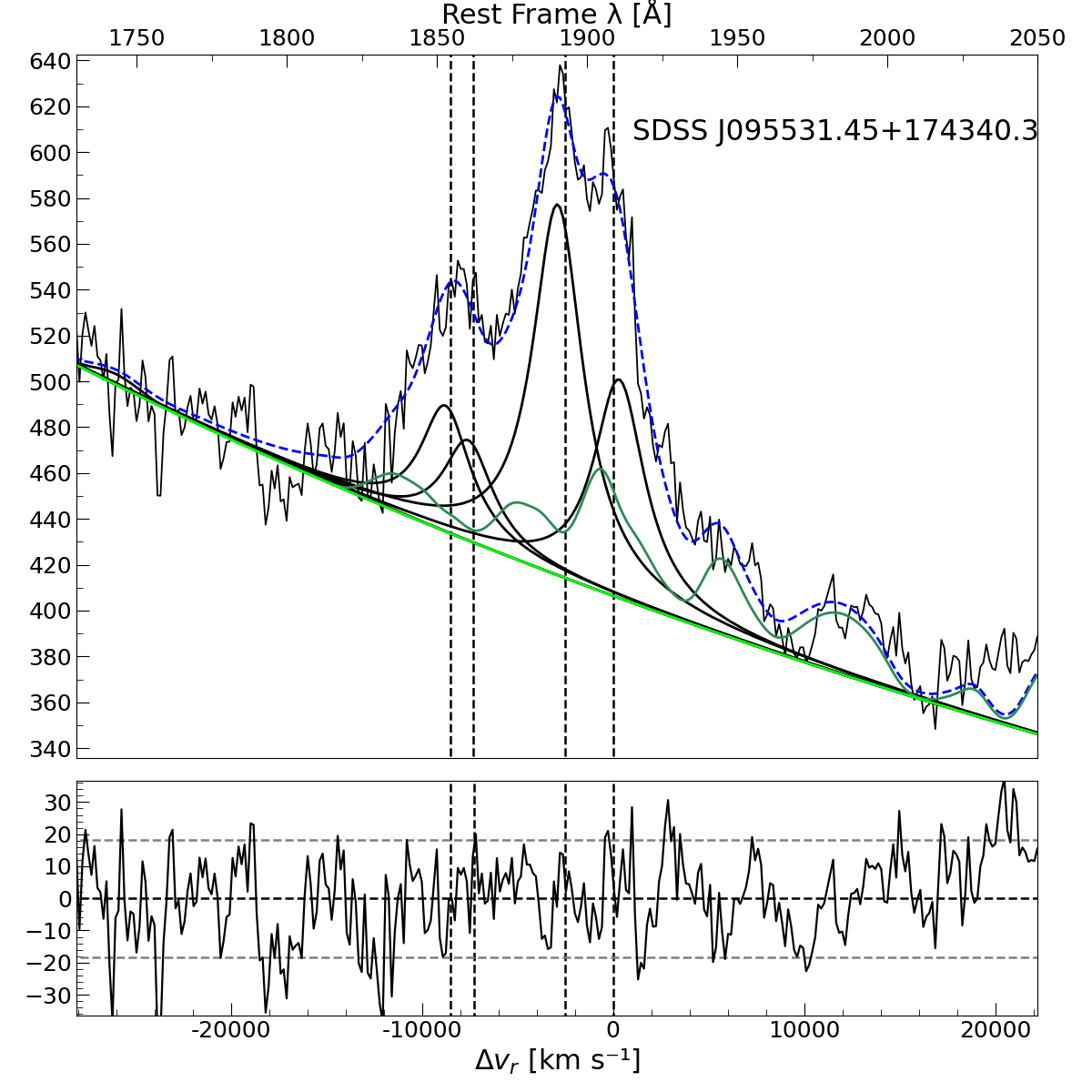}\\
        \includegraphics[width=0.3\textwidth]{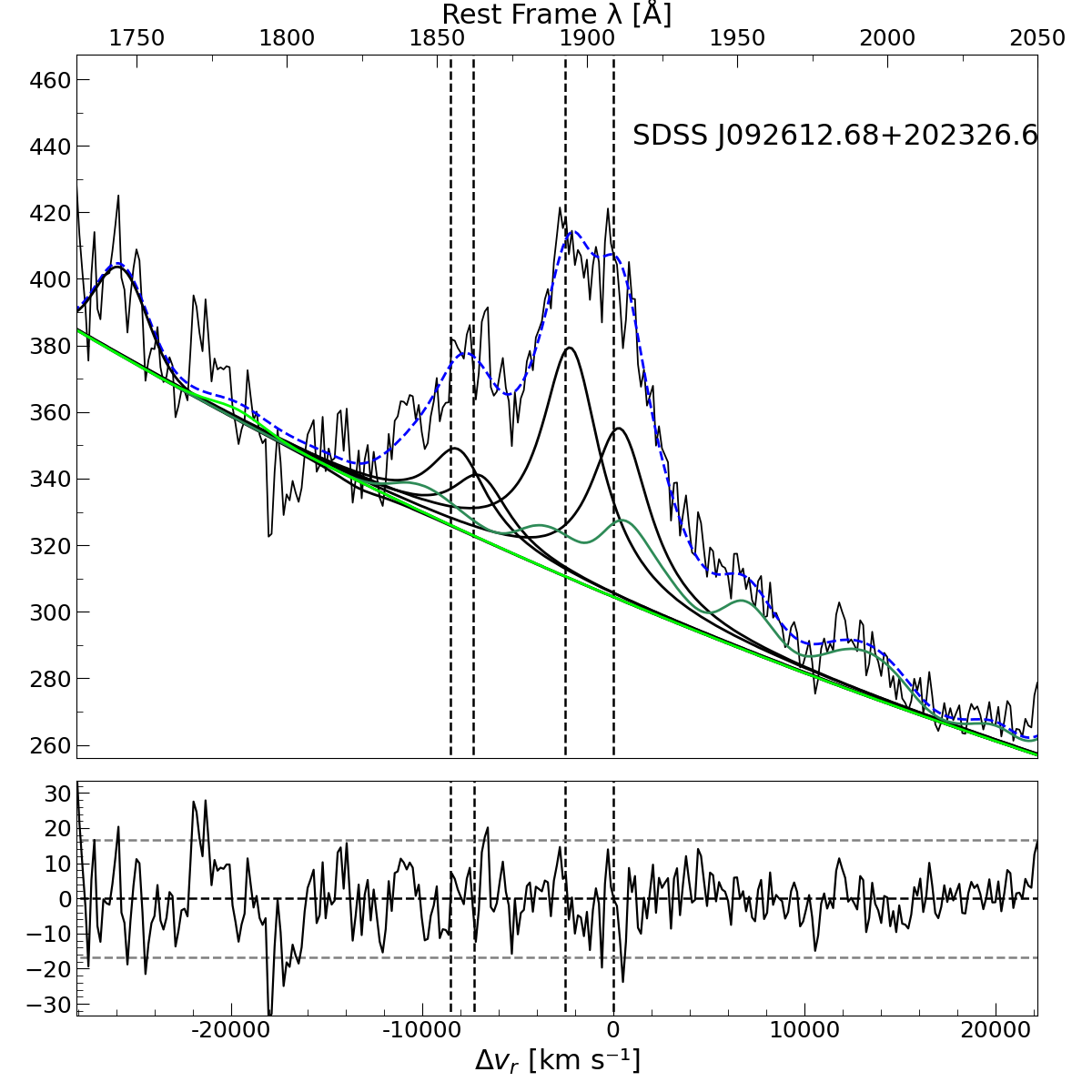}
        \includegraphics[width=0.3\textwidth]{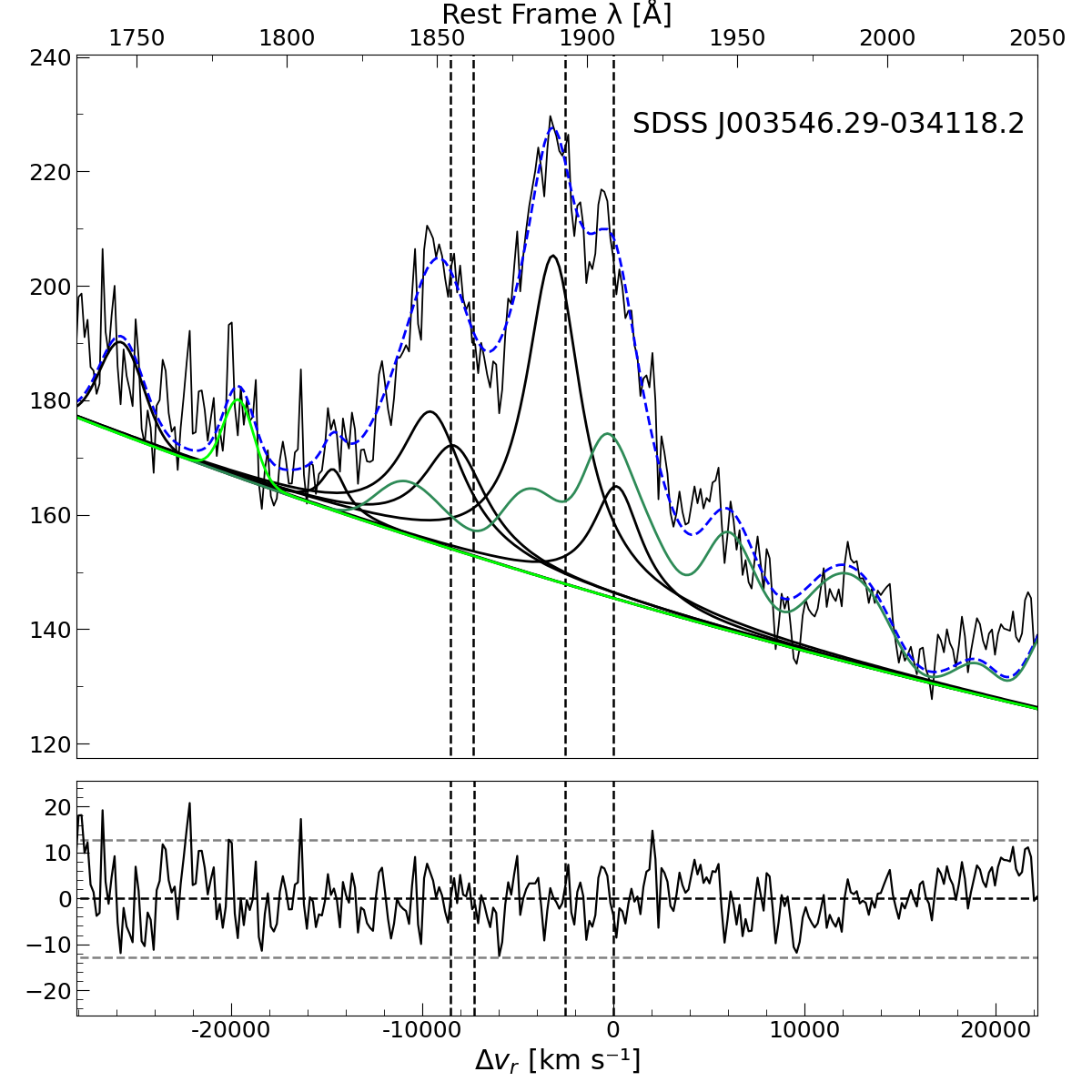}
        \caption{Analysis of the 1900\AA\ blend as described in section \ref{sec:multifitting} for the 11 xA sources in our work. Abscissa scales are rest-frame wavelength in \AA. Ordinate scale is the specific flux. Black lines identify the BC of \aliii, \siiii\ and \ciii. Dashed blue line is the multi-component model obtained by \texttt{specfit}. Green lines trace the adopted \feii\ (pale) and \feiii\ template (dark). }
        \label{fig:xA_spectra}
    \end{figure*} 
 
\section{Line fitting procedures and derived computations header table}\label{app_B} 
    Data obtained for this work and can be described as follows by number of columns in Table \ref{tab:T3}: \\
    (1) file SDSS name, \\
    (2-5) $z$ of this work obtained with \oii\ ($\Delta z = z_{[OII]} - z_{SDSS}$) and the SDSS extracted values with errors, \\ 
    (6) signal-to-noise ratio around 1700\AA, \\
    (7-10) continuum flux at 1700\AA\ and its normalisation flux with errors, \\
    (11) assigned profile of the \ciii\ line: Gaussian or Lorentzian, for xA sources it was added to the name the notation "xA" \\
    (12-13) power-law index - $\alpha$ with error for REGION 1, \\
    (14-41) flux, equivalent width, FWHM, shift of \aliii, \siiii\ and \ciii\ BC, \\
    (42-45) UV diagnostics ratios with errors, \\
    (46-69) flux and FWHM of \niii, \siii, \feii(UV191), \feiiiuv, \cnl VBC and NC with errors obtained by \texttt{specfit}, \\
    (70) power-law index - $\alpha$ for REGION 2, \\
    (71-74) flux and FWHM of \oii\ with errors, \\
    (75-80) continuum flux at 1350\AA\ and pseudo-continuum flux at 3700\AA\ with errors, \\
    (81-83) logarithmic line luminosity at 1860\AA, 1892\AA\ and 1909\AA, \\
    (84-85) logarithmic bolometric luminosity, \\
    (86) FWHM using the luminosity-dependant criterion, \\
    (87-88) logarithmic black hole mass using \ciii\ and \aliii, \\
    (89-92) eddington ratio using \ciii\ and \aliii\ with errors, \\
    (93-94) logarithmic virial luminosity for Pop. xA sources with errors

    \setcounter{table}{1}
    \longtab[1]{
        \centering
        \begin{longtable}{lllllll}
        \caption{Header description of the sample with individual measurements.}\\
        \hline\hline
          COL & Identifier & Type & Units & Description \\
        \hline
        \endfirsthead
        \caption{continued.}\\
        \hline\hline
          COL & Identifier & Type & Units & Description \\
        \hline
        \endhead
        1 & SDSS & CHAR & NULL & File name \\
        2 & z & FLOAT & NULL & z for this work, measured using \oii\ (see text)\\
        3 & z$\_$ERR & FLOAT & NULL & z (this work) error \\
        4 & z$\_$SDSS & FLOAT & NULL & z given by the SDSS database \\ 
        5 & z$\_$SDSS$\_$ERR & FLOAT & NULL & z given by the SDSS database error \\ 
        6 & SN & FLOAT & NULL & S/N Ratio measured around 1700\AA  \\ 
        7 & C1700 & FLOAT & \ergss\ cm$^{-2}$ \AA$^{-1}$ & Continuum Flux at 1700\AA	\\
        8 & C1700$\_$ERR & FLOAT & \ergss\ cm$^{-2}$ \AA$^{-1}$ & Continuum Flux at 1700\AA\ error \\
        9 & N1700 & FLOAT & NULL & Continuum normalisation at 1700\AA \\
        10 & N1700$\_$ERR & FLOAT &  NULL & Continuum normalisation at 1700\AA\ error \\	
        11 & CIII$\_$PROFILE & CHAR & NULL & \ciii\ BC Line Profile. G = Gaussian, L = Lorentzian \\
        12 & ALPHA$\_$R1 & FLOAT & NULL & Power Law Index - $\alpha$ in Region 1 (see text)\\
        13 & ALPHA$\_$R1$\_$ERR & FLOAT & NULL & Power Law Index - $\alpha$ error \\ 
        14 & FLUX$\_$FEIII & FLOAT & 10$^{-17}$\ergss\ cm$^{-2}$ \AA$^{-1}$ & \feiii\ Flux \\
        15 & FLUX$\_$FEIII$\_$ERR & FLOAT & 10$^{-17}$\ergss\ cm$^{-2}$ \AA$^{-1}$ & \feiii\ Flux error \\ 
        16 & SHIFT$\_$FEIII & FLOAT & \AA & \feiii\ shift with respect to the Rest-frame \\
        17 & SHIFT$\_$FEIII$\_$ERR & FLOAT & \AA & \feiii\ shift with respect to the Rest-frame error \\
        18 & EW$\_$CIIIBC & FLOAT & \AA & \ciii\ BC Equivalent Width\\ 
        19 & EW$\_$CIIIBC$\_$ERR & FLOAT & \AA & \ciii\ BC Equivalent Width error \\ 
        20 & FLUX$\_$CIIIBC & FLOAT & 10$^{-17}$\ergss\ cm$^{-2}$ \AA$^{-1}$ & \ciii\ BC Flux\\ 
        21 & FLUX$\_$CIIIBC$\_$ERR & FLOAT & 10$^{-17}$\ergss\ cm$^{-2}$ \AA$^{-1}$ & \ciii\ BC Flux error \\
        22 & SHIFT$\_$CIIIBC & FLOAT & \kms & \ciii\ BC shift with respect to the Rest-frame \\
        23 & SHIFT$\_$CIIIBC$\_$ERR & FLOAT & \kms & \ciii\ BC shift with respect to the Rest-frame error \\
        24 & FWHM$\_$CIIIBC & FLOAT & \kms & \ciii\ BC Full Width at Half Maximuum \\ 
        25 & FWHM$\_$CIIIBC$\_$ERR & FLOAT & \kms & \ciii\ BC Full Width at Half Maximuum error \\
        26 & EW$\_$SIIII & FLOAT & \AA & \siiii\ Equivalent Width\\ 
        27 & EW$\_$SIIII$\_$ERR & FLOAT & \AA & \siiii\ Equivalent Width error \\ 
        28 & FLUX$\_$SIIII & FLOAT &  10$^{-17}$\ergss\ cm$^{-2}$ \AA$^{-1}$ & \siiii\ Flux \\ 
        29 & FLUX$\_$SIIII$\_$ERR & FLOAT & 10$^{-17}$\ergss\ cm$^{-2}$ \AA$^{-1}$ & \siiii\ Flux error \\ 
        30 & SHIFT$\_$SIIII & FLOAT & \kms & \siiii\ shift with respect to the Rest-frame \\ 
        31 & SHIFT$\_$SIIII$\_$ERR & FLOAT & \kms & \siiii\ shift with respect to the Rest-frame error \\ 
        32 & FWHM$\_$SIIII & FLOAT & \kms & \siiii\ Full Width at Half Maximuum \\ 
        33 & FWHM$\_$SiIII$\_$ERR & FLOAT & \kms & \siiii\ Full Width at Half Maximuum error \\
        34 & EW$\_$ALIII & FLOAT & \AA & \aliii\ Equivalent Width \\ 
        35 & EW$\_$ALIII$\_$ERR & FLOAT & \AA & \aliii\ Equivalent Width error \\ 
        36 & FLUX$\_$AlIII & FLOAT & 10$^{-17}$\ergss\ cm$^{-2}$ \AA$^{-1}$ & \aliii\ Flux \\ 
        37 & FLUX$\_$AlIII$\_$ERR & FLOAT & 10$^{-17}$\ergss\ cm$^{-2}$ \AA$^{-1}$ & \aliii\ Flux error \\ 
        38 & SHIFT$\_$AlIII & FLOAT & \kms & \aliii\ shift with respect to the Rest-frame \\ 
        39 & SHIFT$\_$AlIII$\_$ERR & FLOAT & \kms & \aliii\ shift with respect to the Rest-frame error \\ 
        40 & FWHM$\_$AlIII & FLOAT & \kms & \aliii\ Full Width at Half Maximuum \\ 
        41 & FWHM$\_$AlIII$\_$ERR & FLOAT & \kms & \aliii\ Full Width at Half Maximuum error \\ 
        42 & AlIII$\_$SIIII & FLOAT & NULL & UV Diagnostic ratio \aliii/\siiii \\ 
        43 & AlIII$\_$SIIII$\_$ERR & FLOAT & NULL & UV Diagnostic ratio \aliii/\siiii\ error \\
        44 & CIII$\_$SIIII & DOUBLE & NULL & UV Diagnostic ratio \ciii/\siiii \\ 
        45 & CIII$\_$SIIII$\_$ERR & FLOAT & NULL & UV Diagnostic ratio \ciii/\siiii\ error \\ 
        46 & FLUX$\_$NIII & FLOAT & 10$^{-17}$\ergss\ cm$^{-2}$ \AA$^{-1}$ & \niii\ Flux \\
        47 & FLUX$\_$NIII$\_$ERR & FLOAT & 10$^{-17}$\ergss\ cm$^{-2}$ \AA$^{-1}$ & \niii\ Flux err \\ 
        48 & FWHM$\_$NIII & FLOAT & \kms & \niii\ Full Width at Half Maximuum \\ 
        49 & FWHM$\_$NIII$\_$ERR & FLOAT & \kms  & \niii\ Full Width at Half Maximuum error\\
        50 & FLUX$\_$SII & FLOAT & 10$^{-17}$\ergss\ cm$^{-2}$ \AA$^{-1}$ & \siii\ Flux\\ 
        51 & FLUX$\_$SII$\_$ERR & FLOAT & 10$^{-17}$\ergss\ cm$^{-2}$ \AA$^{-1}$ & \siii\ Flux error\\ 
        52 & FWHM$\_$SII & FLOAT & \kms & \siii\ Full Width at Half Maximuum\\ 
        53 & FWHM$\_$SII$\_$ERR & FLOAT & \kms  & \siii\ Full Width at Half Maximuum error \\
        54 & FLUX$\_$FEII & FLOAT & 10$^{-17}$\ergss\ cm$^{-2}$ \AA$^{-1}$ & \feii\ Flux\\ 
        55 & FLUX$\_$FEII$\_$ERR & FLOAT & 10$^{-17}$\ergss\ cm$^{-2}$ \AA$^{-1}$ & \feii\ Flux error \\ 
        56 & FWHM$\_$FEII & FLOAT & \kms & \feii\ Full Width at Half Maximuum \\ 
        57 & FWHM$\_$FEII$\_$ERR & FLOAT & \kms & \feii\ Full Width at Half Maximuum error\\
        58 & FLUX$\_$CIIINC & FLOAT & 10$^{-17}$\ergss\ cm$^{-2}$ \AA$^{-1}$ & \ciii\ NC Flux \\ 
        59 & FLUX$\_$CIIINC$\_$ERR & FLOAT & 10$^{-17}$\ergss\ cm$^{-2}$ \AA$^{-1}$ & \ciii\ NC Flux error \\
        60 & FWHM$\_$CIIINC & FLOAT & \kms & \ciii\ NC Full Width at Half Maximuum\\
        61 & FWHM$\_$CIIINC$\_$ERR & FLOAT & \kms & \ciii\ NC Full Width at Half Maximuum error \\
        62 & FLUX$\_$CIIIVBC & FLOAT & 10$^{-17}$\ergss\ cm$^{-2}$ \AA$^{-1}$ & \ciii\ VBC Flux \\
        63 & FLUX$\_$CIIIVBC$\_$ERR & FLOAT & 10$^{-17}$\ergss\ cm$^{-2}$ \AA$^{-1}$ & \ciii VBC Flux error \\
        64 & FWHM$\_$CIIIVBC & FLOAT & \kms & \ciii\ VBC Full Width at Half Maximuum\\
        65 & FWHM$\_$CIIIVBC$\_$ERR & FLOAT & \kms & \ciii\ VBC Full Width at Half Maximuum error \\
        66 & FLUX$\_$FE1914 & FLOAT & 10$^{-17}$\ergss\ cm$^{-2}$ \AA$^{-1}$ & \feiii $\lambda1914$ Flux \\ 
        67 & FLUX$\_$FE1914$\_$ERR & FLOAT & 10$^{-17}$\ergss\ cm$^{-2}$ \AA$^{-1}$ & \feiii $\lambda1914$ Flux error \\
        68 & FWHM$\_$FE1914 & FLOAT & \kms & \feiii$\lambda$1914 Full Width at Half Maximuum \\ 
        69 & FWHM$\_$FE1914$\_$ERR & FLOAT & \kms & \feiii$\lambda$1914 Full Width at Half Maximuum error \\
        70 & ALPHA$\_$R2 & FLOAT & NULL & Power Law Index - $\alpha$ in Region 2 (see text)\\
        71 & FLUX$\_$OII & FLOAT & 10$^{-17}$\ergss\ cm$^{-2}$ \AA$^{-1}$ & \oii\ Flux \\ 
        72 & FLUX$\_$OII$\_$ERR & FLOAT & 10$^{-17}$\ergss\ cm$^{-2}$ \AA$^{-1}$ & \oii\  Flux error \\
        73 & FWHM$\_$OII & FLOAT & \kms & \oii\ Full Width at Half Maximuum \\ 
        74 & FWHM$\_$OII$\_$ERR & FLOAT & \kms & \oii\ Full Width at Half Maximuum error \\
        75 & C1350 & FLOAT & 10$^{-17}$\ergss\ cm$^{-2}$ \AA$^{-1}$ & Continuum Flux at 1350 \AA\\ 
        76 & C1350$\_$ERR & FLOAT & 10$^{-17}$\ergss\ cm$^{-2}$ \AA$^{-1}$ & Continuum Flux at 1350 \AA\ error \\ 
        77 & C3700 & FLOAT & 10$^{-17}$\ergss\ cm$^{-2}$ \AA$^{-1}$ & Pseudo-continuum Flux at 3700 \AA \\
        78 & C3700$\_$ERR & FLOAT & 10$^{-17}$\ergss\ cm$^{-2}$ \AA$^{-1}$ & Pseudo-continuum Flux at 3700 \AA error \\
        79 & C5100 & FLOAT & 10$^{-17}$\ergss\ cm$^{-2}$ \AA$^{-1}$ & Continuum Flux at 5100 \AA \\
        80 & C5100$\_$ERR & FLOAT & 10$^{-17}$\ergss\ cm$^{-2}$ \AA$^{-1}$ & Continuum Flux at 5100 \AA\ error \\
        81 & LOG$\_$L1860 & DOUBLE & \ergss &  Logarithmic Line Luminosity at 1860 \AA \\
        82 & LOG$\_$L1892 & DOUBLE & \ergss &  Logarithmic Line Luminosity at 1892 \AA \\
        83 & LOG$\_$L1909 & DOUBLE & \ergss &  Logarithmic Line Luminosity at 1909 \AA \\
        84 & LOG$\_$L$\_$BOL & DOUBLE & \ergss &  Logarithmic bolometric Luminosity at 1700 \AA \\ 
        85 & LOG$\_$L$\_$BOL$\_$ERR & FLOAT & \ergss &  Logarithmic bolometric Luminosity at 1700 \AA\ error \\ 
        86 & FWHM$\_$AB & DOUBLE & \kms & FWHM$_{AB}$ using Sulentic et al. (2017) criterion \\ 
        87 & LOG$\_$MBH$\_$CIII & FLOAT & NULL & Logarithmic Black Hole Mass in solar masses of \ciii \\ 
        88 & LOG$\_$MBH$\_$AlIII & FLOAT & NULL & Logarithmic Black Hole Mass in solar masses of \aliii\\ 
        89 & REDD$\_$CIII & FLOAT & NULL & Eddington Ratio using \ciii\ line \\ 
        90 & REDD$\_$CIII$\_$ERR & FLOAT & NULL & Eddington Ratio using \ciii\ line error \\ 
        91 & REDD$\_$ALIII & FLOAT & NULL & Eddington Ratio using \aliii\ line \\ 
        92 & REDD$\_$ALIII$\_$ERR & FLOAT & NULL & Eddington Ratio using \aliii\ line error \\ 
        93 & LOG$\_$L$\_$VIR & DOUBLE & \ergss & Logarithmic Virial Luminosity for Pop. xA sources \\
        94 & LOG$\_$L$\_$VIR$\_$ERR & FLOAT & \ergss & Logarithmic Virial Luminosity for Pop. xA sources error\\
        \hline
        \label{tab:T3}
        \end{longtable}}

\end{appendix}

\end{document}